\documentclass[twocolumn,superscriptaddress,longbibliography,aps,prx,floatfix,
preprintnumbers]{revtex4-2}

\usepackage{graphicx}
\usepackage{bm}
\usepackage{color}
\usepackage{amsmath}
\usepackage{amssymb}

\usepackage[urlcolor=blue,colorlinks=true,citecolor=blue,linkcolor=blue,pdfstartview={FitH},bookmarks=false]{hyperref}
\usepackage[dvipsnames]{xcolor}

\graphicspath{{fig/}{./fig/}{.}}

\begin{document}

\title{
Correlation stabilized ferromagnetic MnRuAs with distorted kagome lattice
}

\author{Anusree C V}
\affiliation{Department of Physics, Indian Institute of Technology Hyderabad, Kandi, Sangareddy 502285, Telangana, India}

\author{Andrzej~Ptok}
\email[e-mail: ]{aptok@mmj.pl}
\affiliation{Institute of Nuclear Physics, Polish Academy of Sciences, W. E. Radzikowskiego 152, PL-31342 Krak\'{o}w, Poland}

\author{Pawe\l{}~Sobieszczyk}
\affiliation{Institute of Nuclear Physics, Polish Academy of Sciences, W. E. Radzikowskiego 152, PL-31342 Krak\'{o}w, Poland}

\author{G. Vaitheeswaran}
\email[e-mail: ]{vaithee@uohyd.ac.in} 
\affiliation{School of Physics, University of Hyderabad, Prof. C. R. Rao Road, Gachibowli, Hyderabad 500-046, Telangana, India}

\author{V. Kanchana}
\email[e-mail: ]{kanchana@iith.ac.in} 
\affiliation{Department of Physics, Indian Institute of Technology Hyderabad, Kandi, Sangareddy 502285, Telangana, India}

\begin{abstract}
We present an in-depth analysis of MnRuAs, a compound crystallizing in the \textit{P$\bar{6}2m$} symmetry with a distorted kagome lattice, revealing its distinctive structural, magnetic, and electronic properties through state-of-the-art \textit{ab initio} calculations. By incorporating strong correlation effects using the DFT+U approach, we demonstrate the stabilization of MnRuAs, transforming its inherent dynamical instability into a robust ferromagnetic state with significant coupling along the $c$ axis. The calculated magnon dispersion reveals a parabolic profile with a minimum at the $\Gamma$ point, indicative of ferromagnetic behavior. Furthermore, MnRuAs exhibits intriguing electronic properties, including quasi-one-dimensional Fermi surface and the formation of nodal sphere.
Our study also delves into the electronic surface states and constant energy contours, offering valuable insights into the complex physics of this material.
\end{abstract}

\maketitle

\section{Introduction}
The kagome lattice, characterized by its distinctive arrangement of corner-sharing triangles, exhibits fascinating electronic properties. Like the honeycomb lattice, the electronic band structure of the kagome lattice features a Dirac cone and a saddle point. However, it also uniquely hosts an ideal flat band, which has garnered significant attention for real systems incorporating kagome lattices~\cite{yin.lian.22,wang.lei.24}. Several classes of compounds with intriguing physical properties that include kagome lattices are worth highlighting: 
%%%%%%%%%%%%%%%%%%%%%%%%%% 
{\it (i)} $A$V$_{3}$Sb$_{5}$ ($A=$ alkali metals), a family of vanadium-based compounds, presents a rare coexistence of charge density wave and superconductivity~\cite{ortiz.gomes.19,ortiz.teicher.20,li.zhang.21,linag.hou.21};
%%%%%%%%%%%%%%%%%%%%%%%%%% 
{\it (ii)} $A$V$_{6}$Sb$_{6}$ ($A=$ alkali metals), a vanadium kagome bilayer system, features Dirac nodal lines near the Fermi level and superconductivity under pressure~\cite{yin.tu.21,yang.fan.21,shi.yu.22,mantravadi.gvozdetskyi.23}; 
%%%%%%%%%%%%%%%%%%%%%%%%%% 
{\it (iii)} $R$V$_{6}$Sn$_{6}$ ($R=$ rare earth), another vanadium kagome bilayer system, is notable for its charge density wave properties~\cite{hu.wu.22,arachchige.meier.22,cao.xu.23}; %%%%%%%%%%%%%%%%%%%%%%%%%% 
{\it (iv)} Co$_{3}$Sn$_{2}$S$_{2}$, a Weyl semimetal with a ferromagnetic cobalt kagome lattice, exhibits a giant anomalous Hall effect~\cite{liu.sun.18,wang.xu.18,liu.liang.19,morali.batabyal.19,yin.zhang.19,xu.zhao.20}; 
%%%%%%%%%%%%%%%%%%%%%%%%%% 
{\it (v)} topological ferrimagnetic or ferromagnetic kagome metals, such as FeSn~\cite{kang.ye.20,han.inoue.21,zhang.oli.23}, Fe$_{3}$Sn$_{2}$~\cite{wang.sun.16,ye.kang.18,lin.choi.18,yin.zhang.18}, and $R$Mn$_{6}$Sn$_{6}$ ($R=$ rare earth)~\cite{yin.ma.20,ghimire.dally.20,ma.xu.21,li.wang.21,riberolles.slade.22,mielke.ma.22,zhang.koo.22,fruhling.streeter.24}, further exemplify the noteworthy physics of kagome lattices.

However, the distortion of the kagome lattice, breaking the inversion symmetry, can enhance the topological properties~\cite{roychowdhury.samanta.24}.
Such distortion can be introduced by rotating the triangles forming the kagome lattice, for instance, during a structural phase transition from \textit{P6/mmm}
 to \textit{P$\bar{6}2m$} symmetry~\cite{ptok.kobialka.21,ptok.meier.23}.
The absence of inversion symmetry gives rise to antisymmetric \mbox{spin--orbit} coupling (SOC)~\cite{smidman.salamon.17}.
Notable examples include NbReSi~\cite{shang.tay.22,sajilesh.motla.22,nandi.sasmal.23} and $T$RuSi ($T$ = transition metal)~\cite{sajilesh.singh.21,shang.zhao.22,sharma.sajilesh.23}, which are noncentrosymmetric topological superconductors with \textit{P$\bar{6}2m$} symmetry, containing distorted kagome lattices.
Furthermore, the absence of inversion symmetry can also lead to highly frustrated magnetic states~\cite{balents.10}, as reported in Cr$T$As ($T =$ Fe, Rh) compounds~\cite{rau.kee.11,florez.vargas.13,jin.meven.19,huang.jeschke.23}.

In this paper, we investigate MnRuAs, a member of the extensive family of MM'X compounds, where M and M' represent $3d$ and $4d$ transition metals, respectively, while X is a \mbox{{\it p}-block} element~\cite{szymanski.zach.23,kanomata.kawashima.91}. These ternary compounds typically crystallize in the Fe$_{2}$P-type structure, characterized by \textit{P$\bar{6}2m$} symmetry, which includes a distorted kagome lattice. This class of materials are well-known for exhibiting magnetic order, and MnRuAs is no exception, displaying ferromagnetic order below the Curie temperature of $496$~K~\cite{kanomata.kawashima.91,kaneko.kanomata.92}. The combination of ferromagnetism and the absence of inversion symmetry in MnRuAs suggests potential topological properties worthy of further exploration.

Initial calculations suggest that MnRuAs is not stable with \textit{P$\bar{6}2m$} symmetry, which contradicts experimental observations~\cite{kanomata.kawashima.91,kaneko.kanomata.92,szymanski.zach.23}. In fact, ordinary DFT calculations reveal imaginary soft modes in the phonon spectra, indicating instability. However, incorporating correlation effects via the DFT+U approach stabilizes the system by yielding positive phonon spectra, consistent with the experimentally observed symmetry. The influence of correlation effects on system stability has been documented in the literature. For example, in kagome LaRu$_{3}$Si$_{2}$~\cite{wang.23}, enhanced electronic correlations stabilize a perfect kagome lattice and induce significant ferromagnetic fluctuations. Similarly, in kagome FeGe, correlation effects lead to phonon softening and the emergence of a charge density wave~\cite{teng.oh.23,wang.23f,ptok.basak.24}. These observations motivate a more precise investigation of MnRuAs properties using advanced {\it \textit{ab initio}} techniques.

The organization of the paper is as follows. Section~\ref{sec.comp} provides details of the numerical calculations. Our findings are presented and discussed in Section~\ref{sec.res}. Specifically, we cover the crystal structure in Section~\ref{sec.crystal}, lattice dynamics in Section~\ref{sec.phonons}, magnetic properties and spin dynamics in Section~\ref{sec.magnetic}, electronic properties, including the electronic band structure, Fermi surface, and nodal crossings features, in Section~\ref{sec.ele}. Finally, we summarize and conclude our results in Section~\ref{sec.summary}.

\section{Computational details}
\label{sec.comp}

The first-principles density functional theory (DFT) calculations were carried out using the projector augmented-wave (PAW) potentials~\cite{blochl.94} as implemented in the \textit{Vienna Ab initio Simulation Package} ({\sc Vasp})~\cite{kresse.hafner.94,kresse.furthmuller.96,kresse.joubert.99}. 
The exchange-correlation energy was treated using the generalized gradient approximation (GGA) within the Perdew--Burke--Ernzerhof (PBE) parametrization~\cite{perdew.burke.96}. 
An energy cutoff of 400~eV was applied for the plane-wave expansion. 
To account for the correlation effects in the Mn-$3d$ orbitals, the DFT+U approach proposed by Dudarev \textit{et al.}~\cite{dudarev.botton.98} was employed. 
Here, the effective Hubbard parameter $U_{\text{eff}}$ is used, where the on-site Coulomb and exchange contributions are combined and not treated separately~\cite{shishkin.sato.19}.

Lattice constants and atomic positions were optimized for magnetic unit cells with ferromagnetic (FM) order. The optimization was performed using an $8 \times 8 \times 13$ \textbf{k}-point grid within the Monkhorst--Pack scheme~\cite{monkhorst.pack.76}. The convergence criteria for the optimization loop were set to an energy change below $10^{-6}$~eV for the ionic degrees of freedom and $10^{-8}$~eV for the electronic degrees of freedom. 
Post-optimization, the symmetry of the structures was analyzed using {\sc FindSym}~\cite{stokes.hatch.05} and {\sc Spglib}~\cite{togo.tanaka.18}, while momentum space analysis was conducted with {\sc SeeK-path}~\cite{hinuma.pizzi.17}. 
The Fermi surface was investigated using {\sc FermiSurfer}~\cite{kawamura.19} and {\sc IFermi}~\cite{ganose.searle.21}.

\begin{figure}[!t]
\centering
\includegraphics[width=\linewidth]{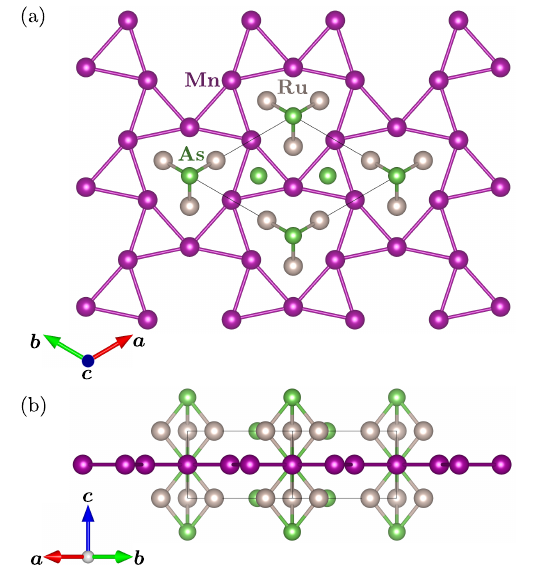}
\caption{
Crystal structure of MnRuAs with \textit{P$\bar{6}2m$} symmetry, featuring a distorted kagome lattice of Mn atoms.
(a) Top view and (b) view along the (110) direction.
\label{fig.crys}
}
\end{figure}

\begin{figure*}[t]
\centering
\includegraphics[width=\linewidth]{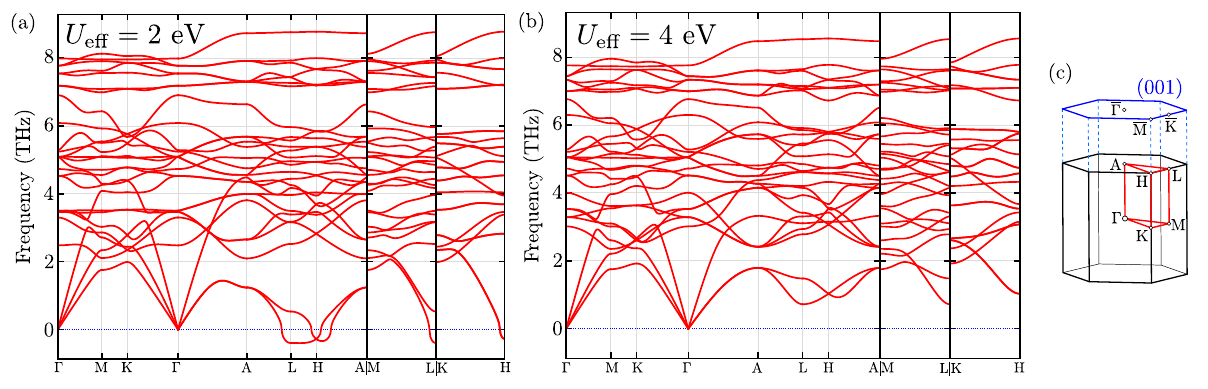}
\caption{
The influence of correlation effects on the phonon dispersion curves.
Phonon dispersion curves of MnRuAs for $U_{\text{eff}} = 2$~eV (a) and $4$~eV (b), along high symmetry directions of the Brillouin zone (c).
\label{fig.phonon}
}
\end{figure*}

Dynamical properties were calculated using the direct \textit{Parlinski--Li--Kawazoe} method~\cite{parlinski.li.97}, as implemented in the {\sc Phonopy} package~\cite{togo.chaput.23,togo.23}. In this method, the interatomic force constants are derived from the Hellmann-Feynman (HF) forces acting on atoms, following individual atomic displacements within the supercell. The calculations were performed using a supercell with dimensions corresponding to $2 \times 2 \times 2$, and a reduced $4 \times 4 \times 5$ {\bf k}-grid was employed during these calculations.

The tight-binding model in the basis of maximally localized Wannier orbitals~\cite{marzari.mostofi.12,marzari.vanderbilt.97,souza.marzari.01} was derived from the exact DFT electronic band structure using the {\sc Wannier90} software~\cite{mostofi.yates.08,mostofi.yates.14,pizzi.vitale.20}. In our calculations, a $\Gamma$-centered \textbf{k}-point grid of $12 \times 12 \times 10$ was employed, with the initial projections chosen as the $d$ orbitals for Mn and Ru, and the $p$ orbitals for As. The resulting tight-binding model, consisting of $39$ orbitals and $78$ bands. Furthermore, this model was applied to study the surface Green's function for a semi-infinite system using the {\sc WannierTools} package~\cite{sancho.sancho.85,wu.zhang.18}.

Additionally, the tight-binding model in the Wannier orbital basis, was employed to compute the magnetic interactions between atoms (here Mn atoms).
The exchange couplings were determined using {\sc TB2J} software~\cite{he.helbig.21}, which is based on the Green’s function method, treating local rigid spin rotation as a perturbation. The resulting effective exchange couplings were subsequently utilized to investigate spin dynamics.

\section{Results and discussion}
\label{sec.res}

\subsection{Crystal structure}
\label{sec.crystal}

MnRuAs crystallizes in the hexagonal Fe$_{2}$P-type structure with \textit{P$\bar{6}2m$} symmetry (space group No.~189). 
This structure lacks inversion symmetry and contains several three- and six-fold rotational axes along the $c$ axis, which is parallel to the $z$ direction.
The experimental lattice constants are reported as $a = b = 6.518$\AA\ and $c = 3.619$\AA~\cite{kanomata.kawashima.91}. The correlation effects introduced via DFT+U typically influence system parameters. However, in MnRuAs, the $U_{\text{eff}}$ does not significantly alter the lattice parameters upon optimization [see Table~\ref{tab.hubbers} in the Supplemental Material (SM)~\footnote{See Supplemental Material at [URL will be inserted by publisher] for additional theoretical results, including Ref.~\cite{cococcioni.gironcoli.05}.}]. Additionally, the atomic positions show minimal dependence on the $U_{\text{eff}}$. For  $U_{\text{eff}} = 4$eV, the optimized lattice constants are $a = b = 6.623$~\AA\ and $c = 3.671$~\AA.
The atomic positions in the unit cell are as follows: Mn occupies the Wyckoff position $3g$ $(0.5952, 0, 1/2)$, Ru occupies the $3f$ position $(0.2546, 0, 0)$, while As atoms occupy two non-equivalent positions, As(1) at $1b$ $(0, 0, 1/2)$ and As(2) at $2c$ $(1/3, 2/3, 0)$.

The crystal structure of MnRuAs is depicted in Fig.~\ref{fig.crys}.
The arrangement of atoms results in the formation of MnAs and RuAs layers, which are stacked along the $c$-axis.
Mn atoms form a characteristic distorted kagome-like lattice, decorated by As(1) atoms [see Fig.~\ref{fig.crys}(a)].
Similarly, a lattice of Ru trimers, decorated by As(2) atoms, is located above.
Due to the relatively small distances between Ru and As (around $2.5$~\AA), a chain-like structure is formed along the $c$ direction [see Fig.~\ref{fig.crys}(b)], alternating Ru trimers and As atoms.
This structure significantly influences the electronic band structure and the shape of the Fermi surface, which exhibits quasi-one-dimensional (1D) features (see Sec.~\ref{sec.ele}).

\subsection{Lattice dynamics}
\label{sec.phonons}

The phonon dispersion curves of MnRuAs are presented in Fig.~\ref{fig.phonon}. For relatively weak correlation effects (e.g., $U_{\text{eff}} = 2$~eV), the phonon spectrum exhibits imaginary soft modes along the L-H path, as shown in Fig.~\ref{fig.phonon}(a) with negative frequencies. This suggests that MnRuAs is dynamically unstable with the \textit{P$\bar{6}2m$} structure. This result contrasts with experimental observations of MnRuAs in the \textit{P$\bar{6}2m$} structure. In contrast, increasing the correlation effect (e.g., to $U_{\text{eff}} = 4$~eV) leads to a phonon spectrum with positive frequencies, as depicted in Fig.~\ref{fig.phonon}(b). Thus, for relatively strong correlation effects, the system is stable with the experimentally observed \textit{P$\bar{6}2m$} symmetry.

From this analysis, there is a critical effective Hubbard parameter $U_{\text{eff}}^{\text{c}}$ that describes the transition from an unstable to a stable \textit{P$\bar{6}2m$} structure. 
As observed, a $U_{\text{eff}}$ greater than $U_{\text{eff}}^{\text{c}}$ is necessary for the system's stabilization.
This feature is clearly visible in the evolution of the phonon dispersion curve with $U_{\text{eff}}$ (see Fig.~\ref{fig.smphband} in the SM~\cite{Note1}). At relatively weak correlation effects ($U_\text{eff} < 2.5$~eV), the imaginary soft mode appears along two acoustic branches in the A-L-H-A path. However, when the correlation effects become stronger ($U_\text{eff} > 2$~eV), these acoustic branches stabilize and become positive. A similar trend can be seen in the phonon density of states (DOS), shown in Fig.~\ref{fig.smphdos} in the SM~\cite{Note1}. As $U_\text{eff}$ increases, the main features of the phonon DOS remain mostly unchanged. The imaginary soft modes are associated with vibrations of all atoms and exhibit a shift between $U_\text{eff} = 2$~eV and $2.5$~eV.

We adopt $U_{\text{eff}} = 4$~eV for our subsequent calculations, as this value stabilizes the structure. 
This choice is consistent with values reported in the literature for Mn-based materials, such as the magnetic topological insulators MnBi$_{2}$Te$_{4}$~\cite{otrokov.klimoviskih.19, klimovskikh.otrokov.20, kobialka.sternik.22} and Mn$_{2}$Bi$_{2}$(Te/Se)$_{5}$~\cite{eremeev.otrokov.22}, where the $U_{\text{eff}}$ reasonably 
describes the electronic band structure. 
In the case of MnRuAs, $U_{\text{eff}}$ calculated using the linear response ansatz of Cococcioni {\it et al.}.~\cite{cococcioni.gironcoli.05} is $5.44$~eV (see Fig.~\ref{fig.lin_resp} in the SM~\cite{Note1}).
In our opinion this value is overestimated, and $U_{\text{eff}} = 4$~eV is a more realistic value.
This value is also consistent with the linear response studies of DFT+U correlation parameters for Mn, as discussed in Ref.~\cite{moore.horton.24}, where a $U_{\text{eff}}$ of $4.14$~eV was determined.

\begin{figure}[t]
\centering
\includegraphics[width=\linewidth]{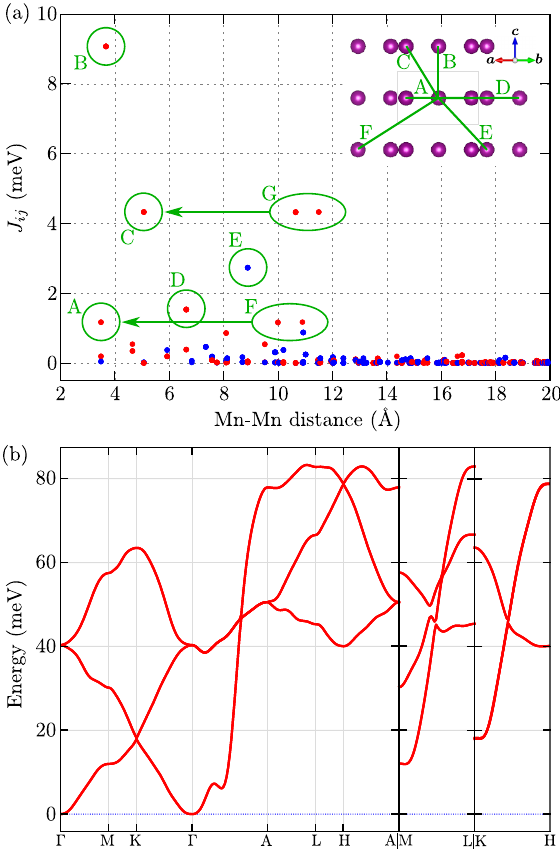}
\caption{The calculated exchange constants $J_{ij}$ as a function of the distance between Mn atoms are shown in (a), with the schematic representation provided in the inset. The color of the dots corresponds to the sign of the exchange: red indicates positive exchange, while blue represents negative exchange. The magnon dispersion curve (b) for the obtained exchange couplings along the high-symmetry directions of the Brillouin zone, as presented in Fig.~\ref{fig.phonon}(c).
\label{fig.mag}
}
\end{figure}

\subsection{Magnetic properties}
\label{sec.magnetic}

The magnetic ground state of MnRuAs was determined by considering $2 \times 2 \times 2$ supercell configurations, including both FM and antiferromagnetic (AFM) arrangements. Regardless of the value of $U_{\text{eff}}$, the FM state emerges as the magnetic ground state, consistent with experimental observations~\cite{kanomata.kawashima.91,kaneko.kanomata.92,szymanski.zach.23}. For $U_{\text{eff}} = 4$~eV, the energy of the FM state is $\sim 0.1$~eV/f.u. lower than that of the A-type AFM state (i.e., arrangement of FM layers with AFM order along the $c$ direction). Comparison of the system energies with different magnetic order can be found in Fig.~\ref{fig.ene_mag} in the SM~\cite{Note1}.
We find the magnetic moment of Mn to be $4.29$~$\mu_{B}$, while the experimentally reported value is $3.96$~$\mu_{B}$~\cite{kaneko.kanomata.92}. It is worth noting that the magnetic ground state is highly sensitive to the chemical composition; for instance, the isostructural compounds MnRhAs and MnRuP exhibit AFM order~\cite{kaneko.kanomata.92}. The introduction of SOC allows for the specification of the magneto-crystalline anisotropy energy (MAE). For the magnetic moment in the $ab$ plane ($\bm{M} \perp c$), the system energy is lower than for the case where $\bm{M} \parallel c$. In this scenario, the MAE is relatively small, with a value of $0.23~\text{meV/f.u.}$. Furthermore, the difference in energy between states with $\bm{M}$ oriented along $x \parallel a$ and $y$ is negligible. From this observation, we can conclude that $ab$ plane is the easy plane.

\begin{figure}[t]
\centering
\includegraphics[width=\linewidth]{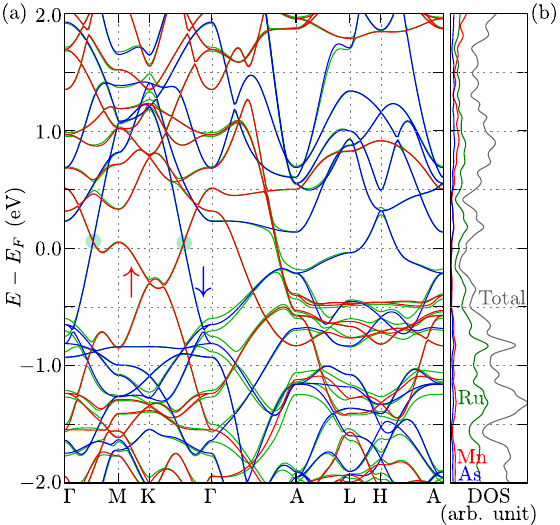}
\caption{
Electronic band structure (a) and density of states (b) of MnRuAs with \textit{P$\bar{6}2m$} symmetry. In the band structure plot, spin $\uparrow$ and spin $\downarrow$ bands are shown in red and blue colors. For comparison, the band structure including SOC is depicted by the green color. 
\label{fig.el_band}
}
\end{figure}

The magnetic properties of MnRuAs are described using the Heisenberg model, given by the Hamiltonian:
$H = - \sum J_{ij} \; {\bm e}_{i} \cdot {\bm e}_{j}$, 
where \( i \) and \( j \) represent the lattice vectors of atoms within the sublattices. The exchange couplings \( J_{ij} \) between Mn atoms were calculated using the Green's function method, which treats local rigid spin rotations as a perturbation.
The resulting exchange couplings are presented in Fig.~\ref{fig.mag}(a). As shown, the exchange couplings are grouped according to the Mn-Mn distance [see inset of Fig.~\ref{fig.mag}(a)]. For Mn pairs within the $ab$ plane (group A), the exchange coupling is $\sim 1.14$~meV, with a corresponding distance of $3.49$~\AA. More dominant couplings are observed between Mn atoms in neighboring unit cells along the $c$ direction. In this case, we identify group B with a distance of $3.67$~\AA\ and an exchange coupling of $8.79$~meV, and group C with a distance of $5.06$~\AA\ and a coupling of $4.27$~meV. Both of these groups correspond to neighboring planes, where Mn atoms are either directly above each other (group B) or exhibit a small lateral shift (group C). Further, there are couplings between Mn atoms in next-neighboring unit cells. For instance, group D exhibits a coupling of $1.49$~meV with a corresponding distance of $6.62$~\AA. Some of these couplings have an AFM character, such as in the case of group E, which has a coupling of $-2.68$~meV and a distance of $8.88$~\AA. Groups F and G correspond to the same Mn pairs as groups A and C, respectively, but with larger interatomic distances.

For the obtained exchange coupling, we also calculated the magnon dispersion curve [Fig.~\ref{fig.mag}(b)].
Around the $\Gamma$ point, we observed a parabolic-like feature characteristic of a ferromagnetic system~\cite{pajda.kudrnovsky.01,chen.mao.22,zakeri.vonfaber.24}. The absence of imaginary soft modes in the magnon spectra further supports the FM order as the ground state.
Due to the strong coupling along the $c$-axis, the magnon dispersion exhibits a pronounced $k_{z}$-dependence.
This is clearly visible along high-symmetry paths such as $\Gamma$-A, M-L, and K-H. The magnon dispersion contains three branches, corresponding to the three Mn atoms in the unit cell~\cite{zhuo.li.22}. Additionally, due to the rotational symmetry axis parallel to the $c$ direction, some branches along the $\Gamma$-A and K-H paths are doubly degenerate.
It is also noteworthy that the magnon spectrum lies within the range of $\sim 85$~meV, which is comparable to the phonon energies ($\sim 32$~meV). This suggests the possibility of strong magnon-phonon coupling in MnRuAs.

The magnon dispersion is shaped by contributions from multiple exchange interactions spanning different crystallographic directions. 
Although the dominant exchange interaction (group B) lies along the $c$-axis, it involves a relatively limited number of Mn magnetic moment pairs. 
In contrast, group C interactions, which are approximately half as strong but involve a significantly larger number of Mn pairs, introduce substantial contributions to the dispersion. 
This interplay of in-plane and out-of-plane interactions affect on the the magnon spectrum by shifting the maximum away from the A point to the A-L/A-H directions. Such behavior can be expected in systems with anisotropic exchange couplings.

\subsection{Electronic properties}
\label{sec.ele}

\begin{figure}[t]
\centering
\includegraphics[width=\linewidth]{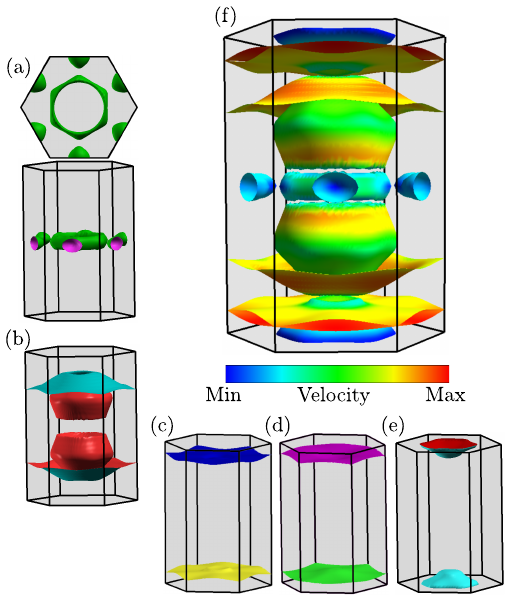}
\caption{
The Fermi surface of MnRuAs with \textit{P$\bar{6}2m$} symmetry. Panels (a) to (e) display the individual Fermi surface pockets, while panel (f) presents the complete Fermi surface, overlaid with a color map indicating the Fermi velocity. The results are obtained with the inclusion of SOC.
\label{fig.fermi}
}
\end{figure}

\begin{figure*}[t]
\centering
\includegraphics[width=\linewidth]{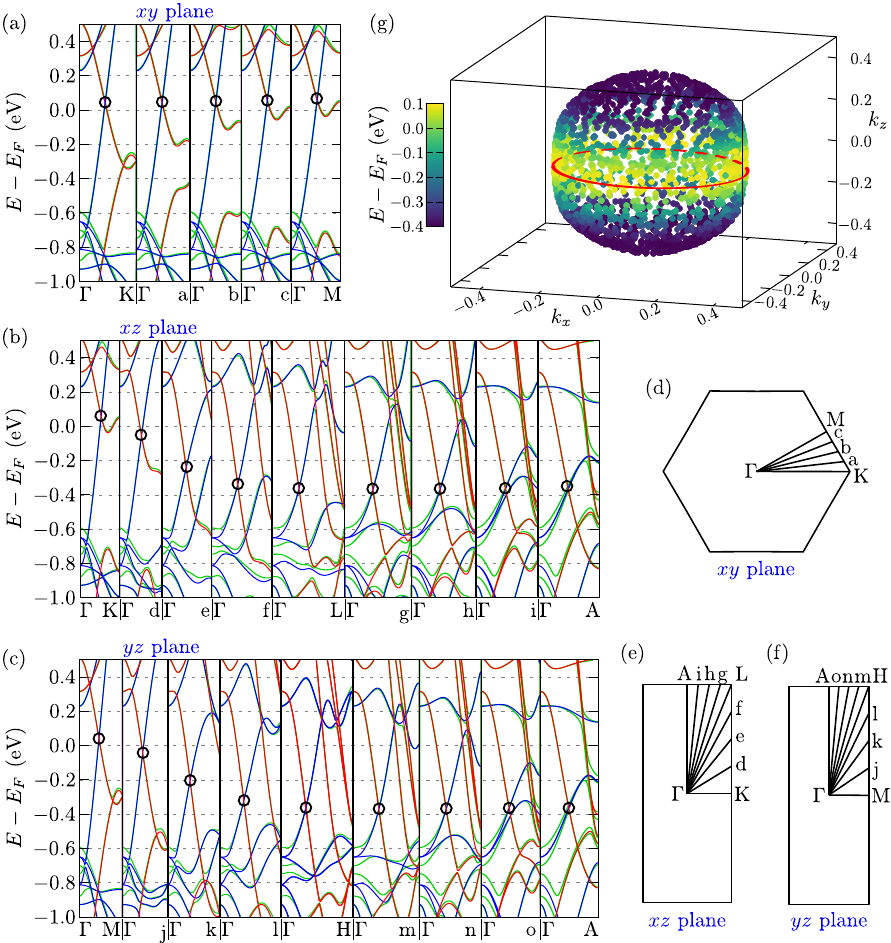}
\caption{
Realization of the nodal sphere. The electronic band structure (a)-(c), along different directions of the Brillouin zone
(d)-(f). The band structure in the absence of spin–orbit coupling with spin $\uparrow$ and spin $\downarrow$ bands presented by red and blue colors, respectively. For comparison, the band structure in the presence of the SOC is presented by green line. Crossing
of the spin $\uparrow$ and spin $\downarrow$ (marked by black circles), creates nodal surface in form of the sphere, as presented in (g). The
energy of the nodal crossing with respect of the Fermi level are presented by the point color.
\label{fig.nodal}
}
\end{figure*}

The electronic band structure and DOS of MnRuAs with \textit{P$\bar{6}2m$} symmetry are shown in Fig.~\ref{fig.el_band}. The bands corresponding to spin $\uparrow$ and spin $\downarrow$ states, in the absence of SOC, are represented by red and blue colors, respectively. The system exhibits metallic character, which is consistent with the metallic bonding between atoms (see Fig.~\ref{fig.bond} in the SM~\cite{Note1}). The splitting of the spin $\uparrow$ and $\downarrow$ bands due to FM order is clearly visible. The Fermi level is crossed by bands associated with both spin contributions. Moreover, the interplay between magnetism and correlation effects, introduced by the $U_{\text{eff}}$ parameter, causes a shift of the Mn-$d$ states to energies further from the Fermi level. Consequently, the states around the Fermi level are predominantly contributed by Ru atoms, as shown in the DOS [Fig.~\ref{fig.el_band}(b)].
This is further evident from the atomic projection of the electronic band structure (see Fig.~\ref{fig.orb} in the SM~\cite{Note1}).

The effect of SOC can manifest as the mixing of spin $\uparrow$ and $\downarrow$ bands or as the lifting of band degeneracy. For instance, the gap opening at points where bands with different spin characters cross is evident, such as along the $\Gamma$-A path [cf.~the blue and red lines with the green line in Fig.~\ref{fig.el_band}(a)]. For Mn atoms, the SOC is $\sim 4$~meV on $p$ orbitals, while it is negligibly small ($< 0.8$~meV) for $d$ orbitals. Similarly, for As-$p$ orbitals, the SOC is around $4$~meV. In contrast, the heavier Ru atom exhibits SOC values in the range of $37$~meV on $p$ orbitals and $1.5$~meV on $d$ orbitals. This is reflected in the electronic band structure, where the impact of SOC, in the form of band splitting, is clearly visible in bands with significant contributions from Ru orbitals [e.g., below $-0.5$~eV in Fig.~\ref{fig.el_band}(a)]. However, the SOC has a relatively weak effect on the electronic band structure near the Fermi level.

{\it Fermi surface.}---
The Fermi surface of MnRuAs, which crystallizes in the \textit{P$\bar{6}2m$} symmetry, consists of five distinct Fermi pockets, as depicted separately in Fig.~\ref{fig.fermi}(a)-(e). Among these, three pockets [Fig.~\ref{fig.fermi}(a),~\ref{fig.fermi}(b), and~\ref{fig.fermi}(e)] demonstrate a three-dimensional (3D) character, owing to their strong $k$-dependence. In contrast, the remaining two pockets [Fig.~\ref{fig.fermi}(c) and~\ref{fig.fermi}(d)] exhibit distinctly 1D features, manifesting as flat sheets in the Fermi surface. The Fermi velocities, shown in Fig.~\ref{fig.fermi}(f), range from $\sim 0.5 \times 10^{5}$~m/s to $5 \times 10^{5}$~m/s. Notably, the highest velocities are associated with the 1D flat sheets, whereas the 3D Fermi pockets, such as the one around the A point [see Fig.~\ref{fig.fermi}(e)], correspond to smaller velocities.

The 1D features observed in certain Fermi surface pockets are indicative of the Ru-As chains present in the crystal structure [see Fig.~\ref{fig.crys}(b)]. In materials where atomic chains or linear arrangements exist, electronic states frequently exhibit 1D characteristics due to the restricted electron motion along these chains. This manifests in the Fermi surface as flat pockets, as discussed above. Similarly, in the electronic band structure, we observe bands with strong dispersion along the $c$ axis and weak dispersion in the plane perpendicular to $c$. For instance, several flat bands around $-0.5$~eV can be identified along the A-L-H-A path in Fig.~\ref{fig.el_band}(a). Such features have been reported in materials like ${A}_{2}$Cr$_{3}$As$_{3}$ ($A$ = K, Rb, Cs)~\cite{wu.yang.15,jiang.cao.15,yang.feng.19,xu.wu.20,taddei.lei.23}, Ba$_{3}$TiTe$_{5}$~\cite{zhang.jia.19}, $R$Rh$_{3}$B$_{2}$ ($R$ = Ce, La)~\cite{okubo.yamada.03}, Tl$_{2}$Mo$_{6}$Se$_{6}$~\cite{song.li.20}, Ce$_3$TiSb$_5$~\cite{he.li.24}, and NbReSi~\cite{basak.ptok.23}. However, such characteristics are rarely observed in magnetic systems with a distorted kagome lattice. Furthermore, the presence of flat Fermi surface sheets suggests potential (charge or spin) density waves along the $c$ direction due to ideal nesting conditions~\cite{wu.yang.15,jiang.cao.15}.

{\it Nodal Lines/Sphere.}---
We now turn our attention to the spin $\uparrow$ and $\downarrow$ band crossings clearly visible in Fig.~\ref{fig.el_band}(a) along the $\Gamma$-M and K-$\Gamma$ paths (marked by green circles). In the absence of SOC, where there is no mixing between the spin $\uparrow$ and $\downarrow$ bands, these crossings occur along all directions. This can be demonstrated by examining the band structure along different directions within the $xy$, $xz$, and $yz$ planes (see Fig.~\ref{fig.nodal}). As shown, for $k_{z} = 0$, the energy of the nodal plane remains independent of the direction around the $z$ axis. In this scenario, the nodal points are located at $\sim 55$~meV, in close proximity to the Fermi level [Fig.~\ref{fig.nodal}(a)]. Similarly, a weak $k_{z}$-dependent dispersion of the nodal points is observed for $k_{z} = \pm \pi/c$ [i.e., along paths from $\Gamma$ to L, g, h, i, and A in Fig.~\ref{fig.nodal}(b) or from $\Gamma$ to H, m, n, o, and A in Fig.~\ref{fig.nodal}(c)], where the nodal points are situated around $-0.35$~eV. Although a $k_{z}$-dependent dispersion is evident from $k_{z} = 0$ to $\pm \pi/c$, the energies of the nodal points remain mostly consistent for fixed $k_{z}$. Ultimately, in the absence of SOC, all nodal points form a nodal surface resembling a sphere [Fig.~\ref{fig.nodal}(g)].

\begin{figure}[t]
\centering
\includegraphics[width=\linewidth]{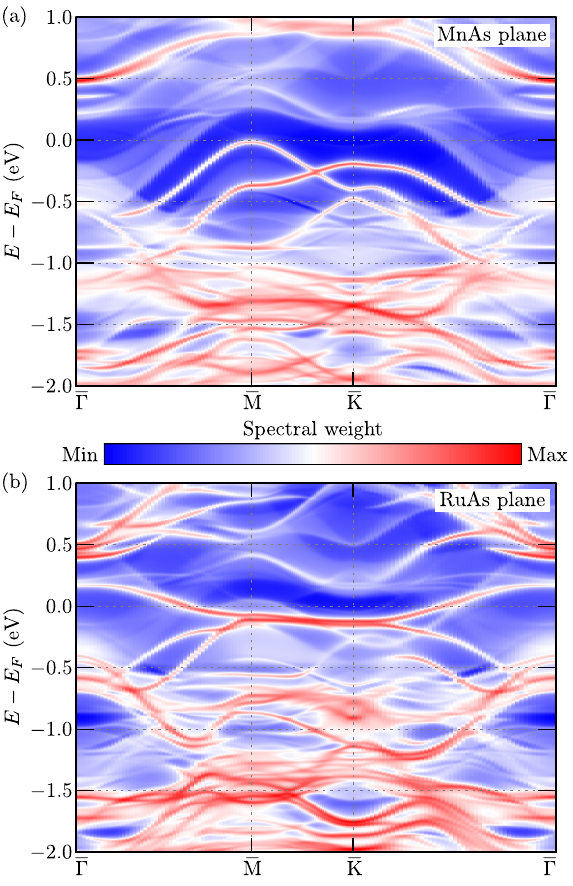}
\caption{
The surface Green function for the MnRuAs (001) surface is shown for MnAs termination in (a) and RuAs termination in (b). The two-dimensional surface Brillouin zone and high-symmetry points are illustrated in Fig.~\ref{fig.phonon}(c).
\label{fig.spec}
}
\end{figure}

The introduction of SOC leads to the opening of a gap along the entire surface. Interestingly, the band splitting at $k_{z} = 0$ remains negligibly small. 
As a result, an effective nodal ring emerges in the presence of SOC within the $xy$ plane, as depicted by the red line in Fig.~\ref{fig.nodal}(g). Indeed, in Fig.~\ref{fig.nodal}(a)-(c), the gap at the nodal points is practically invisible, with the bands in the presence of SOC shown by green color. 
This could be attributed to the significant contribution of Ru atoms (which have weak SOC) to the band structure near the Fermi level. 
Similar small gap openings along the nodal ring have been previously reported in other systems, such as ZrSiS~\cite{mofazzelhosen.dimitri.17}, Cu$_{3}$PdN~\cite{yu.weng.15}, TiB$_{2}$~\cite{liu.lou.18}, and MnGaGe~\cite{anusree.sonali.24}.

The realization of a closed nodal line protected by the mirror symmetry within the $xy$ plane aligns with previous studies on systems possessing \textit{P$\bar{6}2m$} symmetry~\cite{zhou.yang.24}.
As demonstrated in Ref.~\cite{zhou.yang.24}, the shape and size of the nodal line relative to the Brillouin zone are highly dependent on the lattice constants $a = b$ and $c$ of the system. 
For MnRuAs, the nodal line closely resembles those predicted for LiErC, LiNdC, and NaDyC, which exhibit comparable lattice constants~\cite{zhou.yang.24}.

\begin{figure*}[t]
\centering
\includegraphics[width=\linewidth]{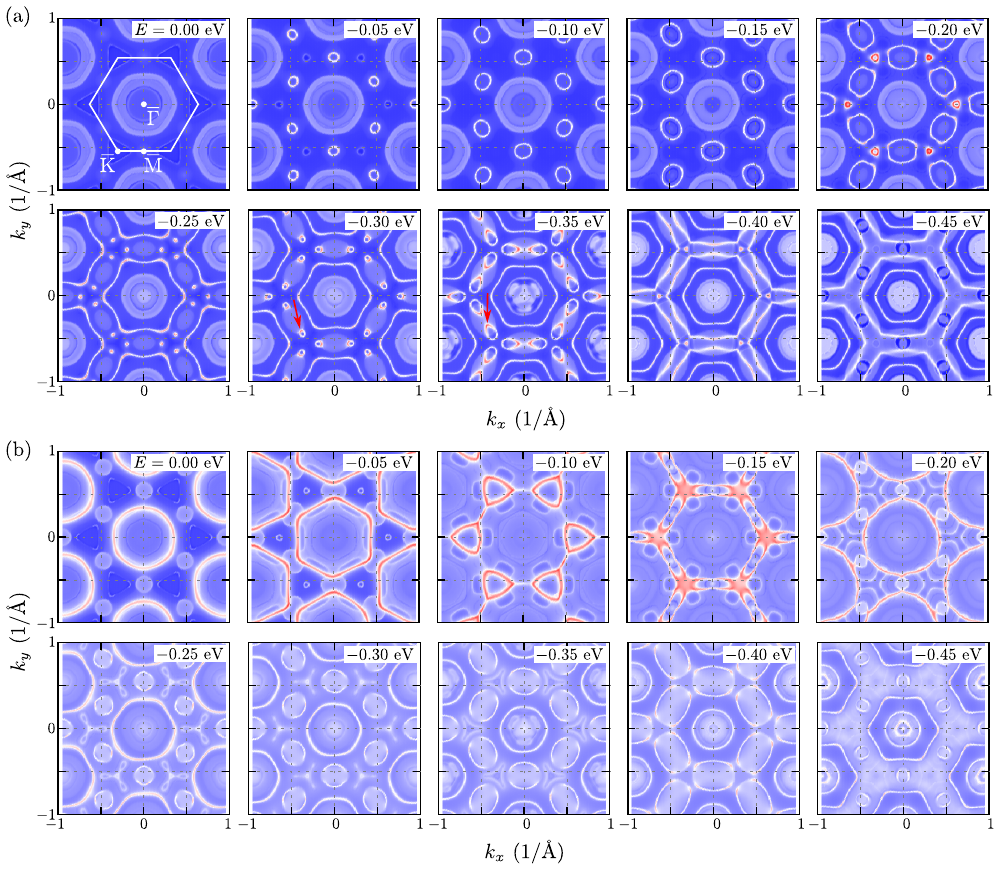}
\caption{
The constant energy contour of the surface Green function for the MnRuAs (001) surface is shown for MnAs termination in (a) and RuAs termination in (b). The top-left panel illustrates the two-dimensional surface Brillouin zone and high-symmetry points, corresponding to Fig.~\ref{fig.phonon}(c).
Color map is the same as given in Fig.~\ref{fig.spec}.
\label{fig.arc}
}
\end{figure*}

{\it Surface states.}--- 
In this part, we focus on the (001) surface, which can be realized in two terminations: MnAs and RuAs. 
The results of the surface Green function calculations are presented in Fig.~\ref{fig.spec}. 
As demonstrated, the obtained results strongly depend on the specific termination and can be used to experimentally confirm one of the terminations~\cite{rosmus.olszowska.22}. 
The surface states are readily identified by their relatively strong contribution to the spectral function (indicated by red colors).

For the (001) surface with MnAs termination [Fig.~\ref{fig.spec}(a)], the surface states initially appear at the $\bar{\text{M}}$ point, with their energy shifting below the Fermi level. 
Subsequently, the surface state becomes visible around the $\bar{\text{K}}$ point. 
Finally, a Dirac-like crossing of both surface states can be observed along the $\bar{\text{K}}$-$\bar{\text{M}}$ path. 
This crossing should also be observable in the constant energy contour (to be discussed later) around an energy of $-0.25$~eV. 
The results obtained for the (001) surface with RuAs termination, as shown in Fig.~\ref{fig.spec}(b), exhibit a much more complex behavior. 
In this case, the surface states, appearing as hole-like bands crossing the $\bar{\Gamma}$-$\bar{\text{M}}$ and $\bar{\Gamma}$-$\bar{\text{K}}$ paths, should intersect the Fermi level. 
At an energy of $\sim -0.1$~eV, these surface states are mixed with other surface states from lower energies. 
This behavior is observed as multiple nearly-flat surface states along the $\bar{\text{K}}$-$\bar{\text{M}}$ path.

MnRuAs exhibits metallic characteristics, reflected in the complex structure of its electronic band projections on the (001) surface. Despite this complexity, the main features of the band structure can be effectively captured by the spectral function. For instance, the bulk electronic band structure, obtained through exact DFT calculations and projected onto the (001) surface (see Fig.~\ref{fig.bandkz_bulk} in the SM~\cite{Note1}), is accurately reproduced by the bulk spectral function (see Fig.~\ref{fig.spec_bulk} in the SM~\cite{Note1}). By comparing the bulk and surface spectral functions, the edge modes can be easily identified.

We will now discuss the constant energy contours for the (001) surface, as presented in Fig.~\ref{fig.arc}. 
For clarity, the first panel also includes the two-dimensional surface Brillouin zone [see Fig.~\ref{fig.phonon}(c)]. 
For the MnAs termination [Fig.~\ref{fig.arc}(a)], the first surface state appears as circles around the $\bar{\text{M}}$ point, corresponding to energies from $-0.05$~eV to $-0.15$~eV. 
At $-0.20$~eV, the second surface state emerges as ``small'' circles around the $\bar{\text{K}}$ point. 
The crossing of both surface states, occurring around $-0.25$~eV, forms a group of three points around the $\bar{\text{K}}$ point, with pairs of points from neighboring $\bar{\text{K}}$ point aligned along the $\bar{\text{K}}$-$\bar{\text{M}}$ path. 
Additionally, the energy dependence of these surface states confirms the formation of Dirac cones. 
Furthermore, below $-0.25$~eV, a Fermi-arc-like structure can be observed as strong spectral weight (indicated by red color) connecting bulk states (white shapes). 
However, unlike the typical Fermi arc seen in Weyl systems, this does not connect different nodes. 
These states are indicated by red arrows at the $-0.3$~eV and $-0.35$~eV panels in Fig.~\ref{fig.arc}(a). 
At lower energies, such as $-0.4$~eV and $-0.45$~eV these states later form surface states with a hexagonal shape.

Similarly, for the RuAs termination [Fig.~\ref{fig.arc}(b)], the surface states initially appear as circular features around the $\bar{\Gamma}$ point. 
As the energy decreases, these surface states shift towards the edge of the Brillouin zone, forming a hexagonal shape (e.g., at $-0.05$~eV). 
At $-0.1$~eV, the surface states evolve into a triangle-like configuration around the $\bar{\text{K}}$ points. 
The mixing of the surface states becomes evident at $-0.15$~eV, where the surface states are predominantly located along the edge of the Brillouin zone. 
Subsequently, the surface states again adopt a circular form around the $\bar{\Gamma}$ point, with additional states appearing around the $\bar{\text{K}}$ and $\bar{\text{M}}$ points (e.g., at $-0.3$~eV to $-0.4$~eV). 
As the energy decreases further, around $-0.45$~eV, these states evolve into surface states with a hexagonal structure.

As expected for metals, the bulk states contribute significantly to the complexity of the spectral function. Not all states, however, contribute equally to the resulting spectral function. For instance, the flat Fermi pockets [Fig.~\ref{fig.fermi}(c)--(e)] projected onto the (001) surface span the entire surface Brillouin zone at the Fermi level. In this case, due to the weak dispersion within the $ab$-plane, such states provide a homogeneous contribution to the spectral function. Consequently, the regions of minimum intensity in the theoretical spectral function do not correspond to the absence of states.

The surface states observed within the energy range of the bulk states exhibit a resonant nature, as anticipated for resonant surface states~\cite{lu.ran.22}. In MnRuAs, there is no clear hybridization between the surface and bulk states, aligning with this characteristic. Notably, no direct connection exists between the observed surface states (illustrated in Fig.~\ref{fig.spec} and Fig.~\ref{fig.arc}) and the bulk spectral function contribution (displayed in Fig.~\ref{fig.spec_bulk} and Fig.~\ref{fig.arc_bulk} in the SM~\cite{Note1}).
Additionally, it is worth noting that surface relaxation and potential reconstruction could significantly alter the surface state dispersion or even introduce new surface states~\cite{souza.criivillero.23}, though these effects are beyond the scope of the method applied in calculating the surface states.

\section{Summary}
\label{sec.summary}

In this paper, we investigate MnRuAs with \textit{P$\bar{6}2m$} symmetry, featuring a distorted kagome lattice. Utilizing advanced \textit{ab initio} techniques, we explore the lattice dynamics, magnetic properties, and electronic characteristics of the material.

Our study reveals the critical role of correlation effects in determining the structural stability and unique properties of MnRuAs. We have shown that MnRuAs exhibits dynamical instability in the absence of strong correlation effects, as evidenced by the presence of imaginary frequencies in the phonon spectra, indicating that the system is unstable. In contrast, when correlation effects are stronger, the system stabilizes with \textit{P$\bar{6}2m$} symmetry, which corresponds to positive phonon spectra as reported experimentally.

The absence of inversion symmetry in the system can indicate the possibility of magnetic frustration, as observed in other ternary magnetic compounds with \textit{P$\bar{6}2m$} symmetry. Our study directly demonstrates that the ferromagnetic state is the ground state of the system. To substantiate this assertion, we explore the magnetic properties using the Heisenberg spin model. Analysis of the exchange interactions between Mn magnetic moments reveals a strong ferromagnetic coupling along the $c$ direction. Additionally, the magnon dispersion curves display a parabolic-like band structure at the $\Gamma$ point, which is characteristic of ferromagnetic systems.

Finally, we investigate the electronic properties. The electronic band structure near the Fermi level is predominantly associated with Ru-$d$ states. The Fermi surface exhibits flat pockets, indicative of quasi-1D electron character. These features can be attributed to RuAs chains within the MnRuAs crystal structure. Around the Fermi level, band crossings with spin $\uparrow$ and $\downarrow$ contributions are observed. In the absence of SOC, these band crossings form a nodal surface that is spherical in shape and centered at the $\Gamma$ point. Although SOC is relatively weak, it leads to a gap opening at these band crossings. Consequently, in the presence of SOC, a nodal ring emerges in the $ab$ plane.

We also investigate the surface states for the (001) surface with both MnAs and RuAs terminations. Our results show distinct features in the surface Green function for each termination. For the MnAs termination, surface states are observed at $\bar{\text{K}}$ and $\bar{\text{M}}$, slightly below the Fermi level. Additionally, these states can form Dirac-like crossings along the $\bar{\text{K}}$-$\bar{\text{M}}$ path. In contrast, for the RuAs termination, the surface states cross the Fermi level. In both cases, the surface states are also discernible in the constant energy contours.

We anticipate that this study will inspire additional experimental investigations into other members of this ternary family with distorted kagome lattices. Consequently, the presence of flat Fermi surface sheets renders MnRuAs particularly unique and significant within the context of density wave ordering.

\begin{acknowledgments}
Some figures in this work were rendered using {\sc Vesta}~\cite{momma.izumi.11} and {\sc XCrySDen}~\cite{kokalj.99} software. ACV and VK express their gratitude to IIT Hyderabad and the National Supercomputing Mission (NSM) for providing computational resources via 'PARAM SEVA' by CDAC. They also acknowledge the financial support from the DST-SERB project (CRG/2022/005228) and VK appreciates DST-FIST (SR/FST/PSI-215/2016) for financial assistance. G V would like to acknowledge the institute of Eminence, University of
Hyderabad (UOH-IOE-RC3-21-046)for funding and CMSD, University of Hyderabad for providing the Computational resources.
\end{acknowledgments}

%\nocite{*}
\bibliography{biblio.bib}

%apsrev4-2.bst 2019-01-14 (MD) hand-edited version of apsrev4-1.bst
%Control: key (0)
%Control: author (8) initials jnrlst
%Control: editor formatted (1) identically to author
%Control: production of article title (0) allowed
%Control: page (0) single
%Control: year (1) truncated
%Control: production of eprint (0) enabled
\begin{thebibliography}{112}%
\makeatletter
\providecommand \@ifxundefined [1]{%
 \@ifx{#1\undefined}
}%
\providecommand \@ifnum [1]{%
 \ifnum #1\expandafter \@firstoftwo
 \else \expandafter \@secondoftwo
 \fi
}%
\providecommand \@ifx [1]{%
 \ifx #1\expandafter \@firstoftwo
 \else \expandafter \@secondoftwo
 \fi
}%
\providecommand \natexlab [1]{#1}%
\providecommand \enquote  [1]{``#1''}%
\providecommand \bibnamefont  [1]{#1}%
\providecommand \bibfnamefont [1]{#1}%
\providecommand \citenamefont [1]{#1}%
\providecommand \href@noop [0]{\@secondoftwo}%
\providecommand \href [0]{\begingroup \@sanitize@url \@href}%
\providecommand \@href[1]{\@@startlink{#1}\@@href}%
\providecommand \@@href[1]{\endgroup#1\@@endlink}%
\providecommand \@sanitize@url [0]{\catcode `\\12\catcode `\$12\catcode `\&12\catcode `\#12\catcode `\^12\catcode `\_12\catcode `\%12\relax}%
\providecommand \@@startlink[1]{}%
\providecommand \@@endlink[0]{}%
\providecommand \url  [0]{\begingroup\@sanitize@url \@url }%
\providecommand \@url [1]{\endgroup\@href {#1}{\urlprefix }}%
\providecommand \urlprefix  [0]{URL }%
\providecommand \Eprint [0]{\href }%
\providecommand \doibase [0]{https://doi.org/}%
\providecommand \selectlanguage [0]{\@gobble}%
\providecommand \bibinfo  [0]{\@secondoftwo}%
\providecommand \bibfield  [0]{\@secondoftwo}%
\providecommand \translation [1]{[#1]}%
\providecommand \BibitemOpen [0]{}%
\providecommand \bibitemStop [0]{}%
\providecommand \bibitemNoStop [0]{.\EOS\space}%
\providecommand \EOS [0]{\spacefactor3000\relax}%
\providecommand \BibitemShut  [1]{\csname bibitem#1\endcsname}%
\let\auto@bib@innerbib\@empty
%</preamble>
\bibitem [{\citenamefont {Yin}\ \emph {et~al.}(2022)\citenamefont {Yin}, \citenamefont {Lian},\ and\ \citenamefont {Hasan}}]{yin.lian.22}%
  \BibitemOpen
  \bibfield  {author} {\bibinfo {author} {\bibfnamefont {J.-X.}\ \bibnamefont {Yin}}, \bibinfo {author} {\bibfnamefont {B.}~\bibnamefont {Lian}},\ and\ \bibinfo {author} {\bibfnamefont {M.~Z.}\ \bibnamefont {Hasan}},\ }\bibfield  {title} {\bibinfo {title} {Topological kagome magnets and superconductors},\ }\href {https://doi.org/10.1038/s41586-022-05516-0} {\bibfield  {journal} {\bibinfo  {journal} {Nature}\ }\textbf {\bibinfo {volume} {612}},\ \bibinfo {pages} {647} (\bibinfo {year} {2022})}\BibitemShut {NoStop}%
\bibitem [{\citenamefont {Wang}\ \emph {et~al.}(2024)\citenamefont {Wang}, \citenamefont {Lei}, \citenamefont {Qi},\ and\ \citenamefont {Felser}}]{wang.lei.24}%
  \BibitemOpen
  \bibfield  {author} {\bibinfo {author} {\bibfnamefont {Q.}~\bibnamefont {Wang}}, \bibinfo {author} {\bibfnamefont {H.}~\bibnamefont {Lei}}, \bibinfo {author} {\bibfnamefont {Y.}~\bibnamefont {Qi}},\ and\ \bibinfo {author} {\bibfnamefont {C.}~\bibnamefont {Felser}},\ }\bibfield  {title} {\bibinfo {title} {Topological quantum materials with kagome lattice},\ }\href {https://doi.org/10.1021/accountsmr.3c00291} {\bibfield  {journal} {\bibinfo  {journal} {Acc. Mater. Res.}\ }\textbf {\bibinfo {volume} {5}},\ \bibinfo {pages} {786} (\bibinfo {year} {2024})}\BibitemShut {NoStop}%
\bibitem [{\citenamefont {Ortiz}\ \emph {et~al.}(2019)\citenamefont {Ortiz}, \citenamefont {Gomes}, \citenamefont {Morey}, \citenamefont {Winiarski}, \citenamefont {Bordelon}, \citenamefont {Mangum}, \citenamefont {Oswald}, \citenamefont {Rodriguez-Rivera}, \citenamefont {Neilson}, \citenamefont {Wilson}, \citenamefont {Ertekin}, \citenamefont {McQueen},\ and\ \citenamefont {Toberer}}]{ortiz.gomes.19}%
  \BibitemOpen
  \bibfield  {author} {\bibinfo {author} {\bibfnamefont {B.~R.}\ \bibnamefont {Ortiz}}, \bibinfo {author} {\bibfnamefont {L.~C.}\ \bibnamefont {Gomes}}, \bibinfo {author} {\bibfnamefont {J.~R.}\ \bibnamefont {Morey}}, \bibinfo {author} {\bibfnamefont {M.}~\bibnamefont {Winiarski}}, \bibinfo {author} {\bibfnamefont {M.}~\bibnamefont {Bordelon}}, \bibinfo {author} {\bibfnamefont {J.~S.}\ \bibnamefont {Mangum}}, \bibinfo {author} {\bibfnamefont {I.~W.~H.}\ \bibnamefont {Oswald}}, \bibinfo {author} {\bibfnamefont {J.~A.}\ \bibnamefont {Rodriguez-Rivera}}, \bibinfo {author} {\bibfnamefont {J.~R.}\ \bibnamefont {Neilson}}, \bibinfo {author} {\bibfnamefont {S.~D.}\ \bibnamefont {Wilson}}, \bibinfo {author} {\bibfnamefont {E.}~\bibnamefont {Ertekin}}, \bibinfo {author} {\bibfnamefont {T.~M.}\ \bibnamefont {McQueen}},\ and\ \bibinfo {author} {\bibfnamefont {E.~S.}\ \bibnamefont {Toberer}},\ }\bibfield  {title} {\bibinfo {title} {New kagome prototype materials: discovery of {KV$_{3}$Sb$_{5}$}, {RbV$_{3}$Sb$_{5}$}, and
  {CsV$_{3}$Sb$_{5}$}},\ }\href {https://doi.org/10.1103/PhysRevMaterials.3.094407} {\bibfield  {journal} {\bibinfo  {journal} {Phys. Rev. Mater.}\ }\textbf {\bibinfo {volume} {3}},\ \bibinfo {pages} {094407} (\bibinfo {year} {2019})}\BibitemShut {NoStop}%
\bibitem [{\citenamefont {Ortiz}\ \emph {et~al.}(2020)\citenamefont {Ortiz}, \citenamefont {Teicher}, \citenamefont {Hu}, \citenamefont {Zuo}, \citenamefont {Sarte}, \citenamefont {Schueller}, \citenamefont {Abeykoon}, \citenamefont {Krogstad}, \citenamefont {Rosenkranz}, \citenamefont {Osborn}, \citenamefont {Seshadri}, \citenamefont {Balents}, \citenamefont {He},\ and\ \citenamefont {Wilson}}]{ortiz.teicher.20}%
  \BibitemOpen
  \bibfield  {author} {\bibinfo {author} {\bibfnamefont {B.~R.}\ \bibnamefont {Ortiz}}, \bibinfo {author} {\bibfnamefont {S.~M.~L.}\ \bibnamefont {Teicher}}, \bibinfo {author} {\bibfnamefont {Y.}~\bibnamefont {Hu}}, \bibinfo {author} {\bibfnamefont {J.~L.}\ \bibnamefont {Zuo}}, \bibinfo {author} {\bibfnamefont {P.~M.}\ \bibnamefont {Sarte}}, \bibinfo {author} {\bibfnamefont {E.~C.}\ \bibnamefont {Schueller}}, \bibinfo {author} {\bibfnamefont {A.~M.~M.}\ \bibnamefont {Abeykoon}}, \bibinfo {author} {\bibfnamefont {M.~J.}\ \bibnamefont {Krogstad}}, \bibinfo {author} {\bibfnamefont {S.}~\bibnamefont {Rosenkranz}}, \bibinfo {author} {\bibfnamefont {R.}~\bibnamefont {Osborn}}, \bibinfo {author} {\bibfnamefont {R.}~\bibnamefont {Seshadri}}, \bibinfo {author} {\bibfnamefont {L.}~\bibnamefont {Balents}}, \bibinfo {author} {\bibfnamefont {J.}~\bibnamefont {He}},\ and\ \bibinfo {author} {\bibfnamefont {S.~D.}\ \bibnamefont {Wilson}},\ }\bibfield  {title} {\bibinfo {title} {{CsV$_{3}$Sb$_{5}$}: A {$\mathbb{Z}_{2}$}
  topological kagome metal with a superconducting ground state},\ }\href {https://doi.org/10.1103/PhysRevLett.125.247002} {\bibfield  {journal} {\bibinfo  {journal} {Phys. Rev. Lett.}\ }\textbf {\bibinfo {volume} {125}},\ \bibinfo {pages} {247002} (\bibinfo {year} {2020})}\BibitemShut {NoStop}%
\bibitem [{\citenamefont {Li}\ \emph {et~al.}(2021{\natexlab{a}})\citenamefont {Li}, \citenamefont {Zhang}, \citenamefont {Yilmaz}, \citenamefont {Pai}, \citenamefont {Marvinney}, \citenamefont {Said}, \citenamefont {Yin}, \citenamefont {Gong}, \citenamefont {Tu}, \citenamefont {Vescovo}, \citenamefont {Nelson}, \citenamefont {Moore}, \citenamefont {Murakami}, \citenamefont {Lei}, \citenamefont {Lee}, \citenamefont {Lawrie},\ and\ \citenamefont {Miao}}]{li.zhang.21}%
  \BibitemOpen
  \bibfield  {author} {\bibinfo {author} {\bibfnamefont {H.}~\bibnamefont {Li}}, \bibinfo {author} {\bibfnamefont {T.~T.}\ \bibnamefont {Zhang}}, \bibinfo {author} {\bibfnamefont {T.}~\bibnamefont {Yilmaz}}, \bibinfo {author} {\bibfnamefont {Y.~Y.}\ \bibnamefont {Pai}}, \bibinfo {author} {\bibfnamefont {C.~E.}\ \bibnamefont {Marvinney}}, \bibinfo {author} {\bibfnamefont {A.}~\bibnamefont {Said}}, \bibinfo {author} {\bibfnamefont {Q.~W.}\ \bibnamefont {Yin}}, \bibinfo {author} {\bibfnamefont {C.~S.}\ \bibnamefont {Gong}}, \bibinfo {author} {\bibfnamefont {Z.~J.}\ \bibnamefont {Tu}}, \bibinfo {author} {\bibfnamefont {E.}~\bibnamefont {Vescovo}}, \bibinfo {author} {\bibfnamefont {C.~S.}\ \bibnamefont {Nelson}}, \bibinfo {author} {\bibfnamefont {R.~G.}\ \bibnamefont {Moore}}, \bibinfo {author} {\bibfnamefont {S.}~\bibnamefont {Murakami}}, \bibinfo {author} {\bibfnamefont {H.~C.}\ \bibnamefont {Lei}}, \bibinfo {author} {\bibfnamefont {H.~N.}\ \bibnamefont {Lee}}, \bibinfo {author} {\bibfnamefont {B.~J.}\
  \bibnamefont {Lawrie}},\ and\ \bibinfo {author} {\bibfnamefont {H.}~\bibnamefont {Miao}},\ }\bibfield  {title} {\bibinfo {title} {Observation of unconventional charge density wave without acoustic phonon anomaly in kagome superconductors {$A$V$_{3}$Sb$_{5}$} ({$A=$Rb, Cs})},\ }\href {https://doi.org/10.1103/PhysRevX.11.031050} {\bibfield  {journal} {\bibinfo  {journal} {Phys. Rev. X}\ }\textbf {\bibinfo {volume} {11}},\ \bibinfo {pages} {031050} (\bibinfo {year} {2021}{\natexlab{a}})}\BibitemShut {NoStop}%
\bibitem [{\citenamefont {Liang}\ \emph {et~al.}(2021)\citenamefont {Liang}, \citenamefont {Hou}, \citenamefont {Zhang}, \citenamefont {Ma}, \citenamefont {Wu}, \citenamefont {Zhang}, \citenamefont {Yu}, \citenamefont {Ying}, \citenamefont {Jiang}, \citenamefont {Shan}, \citenamefont {Wang},\ and\ \citenamefont {Chen}}]{linag.hou.21}%
  \BibitemOpen
  \bibfield  {author} {\bibinfo {author} {\bibfnamefont {Z.}~\bibnamefont {Liang}}, \bibinfo {author} {\bibfnamefont {X.}~\bibnamefont {Hou}}, \bibinfo {author} {\bibfnamefont {F.}~\bibnamefont {Zhang}}, \bibinfo {author} {\bibfnamefont {W.}~\bibnamefont {Ma}}, \bibinfo {author} {\bibfnamefont {P.}~\bibnamefont {Wu}}, \bibinfo {author} {\bibfnamefont {Z.}~\bibnamefont {Zhang}}, \bibinfo {author} {\bibfnamefont {F.}~\bibnamefont {Yu}}, \bibinfo {author} {\bibfnamefont {J.-J.}\ \bibnamefont {Ying}}, \bibinfo {author} {\bibfnamefont {K.}~\bibnamefont {Jiang}}, \bibinfo {author} {\bibfnamefont {L.}~\bibnamefont {Shan}}, \bibinfo {author} {\bibfnamefont {Z.}~\bibnamefont {Wang}},\ and\ \bibinfo {author} {\bibfnamefont {X.-H.}\ \bibnamefont {Chen}},\ }\bibfield  {title} {\bibinfo {title} {Three-dimensional charge density wave and surface-dependent vortex-core states in a kagome superconductor {CsV$_{3}$Sb$_{5}$}},\ }\href {https://doi.org/10.1103/PhysRevX.11.031026} {\bibfield  {journal} {\bibinfo  {journal} {Phys.
  Rev. X}\ }\textbf {\bibinfo {volume} {11}},\ \bibinfo {pages} {031026} (\bibinfo {year} {2021})}\BibitemShut {NoStop}%
\bibitem [{\citenamefont {Yin}\ \emph {et~al.}(2021)\citenamefont {Yin}, \citenamefont {Tu}, \citenamefont {Gong}, \citenamefont {Tian},\ and\ \citenamefont {Lei}}]{yin.tu.21}%
  \BibitemOpen
  \bibfield  {author} {\bibinfo {author} {\bibfnamefont {Q.}~\bibnamefont {Yin}}, \bibinfo {author} {\bibfnamefont {Z.}~\bibnamefont {Tu}}, \bibinfo {author} {\bibfnamefont {C.}~\bibnamefont {Gong}}, \bibinfo {author} {\bibfnamefont {S.}~\bibnamefont {Tian}},\ and\ \bibinfo {author} {\bibfnamefont {H.}~\bibnamefont {Lei}},\ }\bibfield  {title} {\bibinfo {title} {Structures and physical properties of {V}-based kagome metals {CsV$_{6}$Sb$_{6}$} and {CsV$_{8}$Sb$_{12}$}},\ }\href {https://doi.org/10.1088/0256-307X/38/12/127401} {\bibfield  {journal} {\bibinfo  {journal} {Chinese Phys. Lett.}\ }\textbf {\bibinfo {volume} {38}},\ \bibinfo {pages} {127401} (\bibinfo {year} {2021})}\BibitemShut {NoStop}%
\bibitem [{\citenamefont {Yang}\ \emph {et~al.}(2021)\citenamefont {Yang}, \citenamefont {Fan}, \citenamefont {Zhang}, \citenamefont {Chen}, \citenamefont {Chen}, \citenamefont {Ying}, \citenamefont {Wu}, \citenamefont {Yang}, \citenamefont {Meng}, \citenamefont {Li}, \citenamefont {Li}, \citenamefont {Gu}, \citenamefont {Qian}, \citenamefont {Schnyder}, \citenamefont {gang Guo},\ and\ \citenamefont {Chen}}]{yang.fan.21}%
  \BibitemOpen
  \bibfield  {author} {\bibinfo {author} {\bibfnamefont {Y.}~\bibnamefont {Yang}}, \bibinfo {author} {\bibfnamefont {W.}~\bibnamefont {Fan}}, \bibinfo {author} {\bibfnamefont {Q.}~\bibnamefont {Zhang}}, \bibinfo {author} {\bibfnamefont {Z.}~\bibnamefont {Chen}}, \bibinfo {author} {\bibfnamefont {X.}~\bibnamefont {Chen}}, \bibinfo {author} {\bibfnamefont {T.}~\bibnamefont {Ying}}, \bibinfo {author} {\bibfnamefont {X.}~\bibnamefont {Wu}}, \bibinfo {author} {\bibfnamefont {X.}~\bibnamefont {Yang}}, \bibinfo {author} {\bibfnamefont {F.}~\bibnamefont {Meng}}, \bibinfo {author} {\bibfnamefont {G.}~\bibnamefont {Li}}, \bibinfo {author} {\bibfnamefont {S.}~\bibnamefont {Li}}, \bibinfo {author} {\bibfnamefont {L.}~\bibnamefont {Gu}}, \bibinfo {author} {\bibfnamefont {T.}~\bibnamefont {Qian}}, \bibinfo {author} {\bibfnamefont {A.~P.}\ \bibnamefont {Schnyder}}, \bibinfo {author} {\bibfnamefont {J.}~\bibnamefont {gang Guo}},\ and\ \bibinfo {author} {\bibfnamefont {X.}~\bibnamefont {Chen}},\ }\bibfield  {title} {\bibinfo
  {title} {Discovery of two families of {VSb}-based compounds with {V}-kagome lattice},\ }\href {https://doi.org/10.1088/0256-307X/38/12/127102} {\bibfield  {journal} {\bibinfo  {journal} {Chinese Phys. Lett.}\ }\textbf {\bibinfo {volume} {38}},\ \bibinfo {pages} {127102} (\bibinfo {year} {2021})}\BibitemShut {NoStop}%
\bibitem [{\citenamefont {Shi}\ \emph {et~al.}(2022)\citenamefont {Shi}, \citenamefont {Yu}, \citenamefont {Yang}, \citenamefont {Meng}, \citenamefont {Lei}, \citenamefont {Luo}, \citenamefont {Sun}, \citenamefont {He}, \citenamefont {Wang}, \citenamefont {Jiang}, \citenamefont {Liu}, \citenamefont {Shen}, \citenamefont {Wu}, \citenamefont {Wang}, \citenamefont {Xiang}, \citenamefont {Ying},\ and\ \citenamefont {Chen}}]{shi.yu.22}%
  \BibitemOpen
  \bibfield  {author} {\bibinfo {author} {\bibfnamefont {M.}~\bibnamefont {Shi}}, \bibinfo {author} {\bibfnamefont {F.}~\bibnamefont {Yu}}, \bibinfo {author} {\bibfnamefont {Y.}~\bibnamefont {Yang}}, \bibinfo {author} {\bibfnamefont {F.}~\bibnamefont {Meng}}, \bibinfo {author} {\bibfnamefont {B.}~\bibnamefont {Lei}}, \bibinfo {author} {\bibfnamefont {Y.}~\bibnamefont {Luo}}, \bibinfo {author} {\bibfnamefont {Z.}~\bibnamefont {Sun}}, \bibinfo {author} {\bibfnamefont {J.}~\bibnamefont {He}}, \bibinfo {author} {\bibfnamefont {R.}~\bibnamefont {Wang}}, \bibinfo {author} {\bibfnamefont {Z.}~\bibnamefont {Jiang}}, \bibinfo {author} {\bibfnamefont {Z.}~\bibnamefont {Liu}}, \bibinfo {author} {\bibfnamefont {D.}~\bibnamefont {Shen}}, \bibinfo {author} {\bibfnamefont {T.}~\bibnamefont {Wu}}, \bibinfo {author} {\bibfnamefont {Z.}~\bibnamefont {Wang}}, \bibinfo {author} {\bibfnamefont {Z.}~\bibnamefont {Xiang}}, \bibinfo {author} {\bibfnamefont {J.}~\bibnamefont {Ying}},\ and\ \bibinfo {author} {\bibfnamefont
  {X.}~\bibnamefont {Chen}},\ }\bibfield  {title} {\bibinfo {title} {A new class of bilayer kagome lattice compounds with {Dirac} nodal lines and pressure-induced superconductivity},\ }\href {https://doi.org/10.1038/s41467-022-30442-0} {\bibfield  {journal} {\bibinfo  {journal} {Nat. Commun.}\ }\textbf {\bibinfo {volume} {13}},\ \bibinfo {pages} {2773} (\bibinfo {year} {2022})}\BibitemShut {NoStop}%
\bibitem [{\citenamefont {Mantravadi}\ \emph {et~al.}(2023)\citenamefont {Mantravadi}, \citenamefont {Gvozdetskyi}, \citenamefont {Sarkar}, \citenamefont {Mudryk},\ and\ \citenamefont {Zaikina}}]{mantravadi.gvozdetskyi.23}%
  \BibitemOpen
  \bibfield  {author} {\bibinfo {author} {\bibfnamefont {A.}~\bibnamefont {Mantravadi}}, \bibinfo {author} {\bibfnamefont {V.}~\bibnamefont {Gvozdetskyi}}, \bibinfo {author} {\bibfnamefont {A.}~\bibnamefont {Sarkar}}, \bibinfo {author} {\bibfnamefont {Y.}~\bibnamefont {Mudryk}},\ and\ \bibinfo {author} {\bibfnamefont {J.~V.}\ \bibnamefont {Zaikina}},\ }\bibfield  {title} {\bibinfo {title} {Exploring the {$A$-V-Sb} landscape beyond {$A$V$_{3}$Sb$_{5}$}: A case study on the {KV$_{6}$Sb$_{6}$} kagome compound},\ }\href {https://doi.org/10.1103/PhysRevMaterials.7.115002} {\bibfield  {journal} {\bibinfo  {journal} {Phys. Rev. Mater.}\ }\textbf {\bibinfo {volume} {7}},\ \bibinfo {pages} {115002} (\bibinfo {year} {2023})}\BibitemShut {NoStop}%
\bibitem [{\citenamefont {Hu}\ \emph {et~al.}(2022)\citenamefont {Hu}, \citenamefont {Wu}, \citenamefont {Yang}, \citenamefont {Gao}, \citenamefont {Plumb}, \citenamefont {Schnyder}, \citenamefont {Xie}, \citenamefont {Ma},\ and\ \citenamefont {Shi}}]{hu.wu.22}%
  \BibitemOpen
  \bibfield  {author} {\bibinfo {author} {\bibfnamefont {Y.}~\bibnamefont {Hu}}, \bibinfo {author} {\bibfnamefont {X.}~\bibnamefont {Wu}}, \bibinfo {author} {\bibfnamefont {Y.}~\bibnamefont {Yang}}, \bibinfo {author} {\bibfnamefont {S.}~\bibnamefont {Gao}}, \bibinfo {author} {\bibfnamefont {N.~C.}\ \bibnamefont {Plumb}}, \bibinfo {author} {\bibfnamefont {A.~P.}\ \bibnamefont {Schnyder}}, \bibinfo {author} {\bibfnamefont {W.}~\bibnamefont {Xie}}, \bibinfo {author} {\bibfnamefont {J.}~\bibnamefont {Ma}},\ and\ \bibinfo {author} {\bibfnamefont {M.}~\bibnamefont {Shi}},\ }\bibfield  {title} {\bibinfo {title} {Tunable topological {Dirac} surface states and van hove singularities in kagome metal {GdV$_{6}$Sn$_{6}$}},\ }\href {https://doi.org/10.1126/sciadv.add2024} {\bibfield  {journal} {\bibinfo  {journal} {Sci. Adv.}\ }\textbf {\bibinfo {volume} {8}},\ \bibinfo {pages} {eadd2024} (\bibinfo {year} {2022})}\BibitemShut {NoStop}%
\bibitem [{\citenamefont {Arachchige}\ \emph {et~al.}(2022)\citenamefont {Arachchige}, \citenamefont {Meier}, \citenamefont {Marshall}, \citenamefont {Matsuoka}, \citenamefont {Xue}, \citenamefont {McGuire}, \citenamefont {Hermann}, \citenamefont {Cao},\ and\ \citenamefont {Mandrus}}]{arachchige.meier.22}%
  \BibitemOpen
  \bibfield  {author} {\bibinfo {author} {\bibfnamefont {H.~W.~S.}\ \bibnamefont {Arachchige}}, \bibinfo {author} {\bibfnamefont {W.~R.}\ \bibnamefont {Meier}}, \bibinfo {author} {\bibfnamefont {M.}~\bibnamefont {Marshall}}, \bibinfo {author} {\bibfnamefont {T.}~\bibnamefont {Matsuoka}}, \bibinfo {author} {\bibfnamefont {R.}~\bibnamefont {Xue}}, \bibinfo {author} {\bibfnamefont {M.~A.}\ \bibnamefont {McGuire}}, \bibinfo {author} {\bibfnamefont {R.~P.}\ \bibnamefont {Hermann}}, \bibinfo {author} {\bibfnamefont {H.}~\bibnamefont {Cao}},\ and\ \bibinfo {author} {\bibfnamefont {D.}~\bibnamefont {Mandrus}},\ }\bibfield  {title} {\bibinfo {title} {Charge density wave in kagome lattice intermetallic {ScV$_{6}$Sn$_{6}$}},\ }\href {https://doi.org/10.1103/PhysRevLett.129.216402} {\bibfield  {journal} {\bibinfo  {journal} {Phys. Rev. Lett.}\ }\textbf {\bibinfo {volume} {129}},\ \bibinfo {pages} {216402} (\bibinfo {year} {2022})}\BibitemShut {NoStop}%
\bibitem [{\citenamefont {Cao}\ \emph {et~al.}(2023)\citenamefont {Cao}, \citenamefont {Xu}, \citenamefont {Fukui}, \citenamefont {Manjo}, \citenamefont {Dong}, \citenamefont {Shi}, \citenamefont {Liu}, \citenamefont {Cao},\ and\ \citenamefont {Song}}]{cao.xu.23}%
  \BibitemOpen
  \bibfield  {author} {\bibinfo {author} {\bibfnamefont {S.}~\bibnamefont {Cao}}, \bibinfo {author} {\bibfnamefont {C.}~\bibnamefont {Xu}}, \bibinfo {author} {\bibfnamefont {H.}~\bibnamefont {Fukui}}, \bibinfo {author} {\bibfnamefont {T.}~\bibnamefont {Manjo}}, \bibinfo {author} {\bibfnamefont {Y.}~\bibnamefont {Dong}}, \bibinfo {author} {\bibfnamefont {M.}~\bibnamefont {Shi}}, \bibinfo {author} {\bibfnamefont {Y.}~\bibnamefont {Liu}}, \bibinfo {author} {\bibfnamefont {C.}~\bibnamefont {Cao}},\ and\ \bibinfo {author} {\bibfnamefont {Y.}~\bibnamefont {Song}},\ }\bibfield  {title} {\bibinfo {title} {Competing charge-density wave instabilities in the kagome metal {ScV$_{6}$Sn$_{6}$}},\ }\href {https://doi.org/10.1038/s41467-023-43454-1} {\bibfield  {journal} {\bibinfo  {journal} {Nat. Commun.}\ }\textbf {\bibinfo {volume} {14}},\ \bibinfo {pages} {7671} (\bibinfo {year} {2023})}\BibitemShut {NoStop}%
\bibitem [{\citenamefont {Liu}\ \emph {et~al.}(2018{\natexlab{a}})\citenamefont {Liu}, \citenamefont {Sun}, \citenamefont {Kumar}, \citenamefont {Muechler}, \citenamefont {Sun}, \citenamefont {Jiao}, \citenamefont {Yang}, \citenamefont {Liu}, \citenamefont {Liang}, \citenamefont {Xu}, \citenamefont {Kroder}, \citenamefont {S{\"u}{\ss}}, \citenamefont {Borrmann}, \citenamefont {Shekhar}, \citenamefont {Wang}, \citenamefont {Xi}, \citenamefont {Wang}, \citenamefont {Schnelle}, \citenamefont {Wirth}, \citenamefont {Chen}, \citenamefont {Goennenwein},\ and\ \citenamefont {Felser}}]{liu.sun.18}%
  \BibitemOpen
  \bibfield  {author} {\bibinfo {author} {\bibfnamefont {E.}~\bibnamefont {Liu}}, \bibinfo {author} {\bibfnamefont {Y.}~\bibnamefont {Sun}}, \bibinfo {author} {\bibfnamefont {N.}~\bibnamefont {Kumar}}, \bibinfo {author} {\bibfnamefont {L.}~\bibnamefont {Muechler}}, \bibinfo {author} {\bibfnamefont {A.}~\bibnamefont {Sun}}, \bibinfo {author} {\bibfnamefont {L.}~\bibnamefont {Jiao}}, \bibinfo {author} {\bibfnamefont {S.-Y.}\ \bibnamefont {Yang}}, \bibinfo {author} {\bibfnamefont {D.}~\bibnamefont {Liu}}, \bibinfo {author} {\bibfnamefont {A.}~\bibnamefont {Liang}}, \bibinfo {author} {\bibfnamefont {Q.}~\bibnamefont {Xu}}, \bibinfo {author} {\bibfnamefont {J.}~\bibnamefont {Kroder}}, \bibinfo {author} {\bibfnamefont {V.}~\bibnamefont {S{\"u}{\ss}}}, \bibinfo {author} {\bibfnamefont {H.}~\bibnamefont {Borrmann}}, \bibinfo {author} {\bibfnamefont {C.}~\bibnamefont {Shekhar}}, \bibinfo {author} {\bibfnamefont {Z.}~\bibnamefont {Wang}}, \bibinfo {author} {\bibfnamefont {C.}~\bibnamefont {Xi}}, \bibinfo {author}
  {\bibfnamefont {W.}~\bibnamefont {Wang}}, \bibinfo {author} {\bibfnamefont {W.}~\bibnamefont {Schnelle}}, \bibinfo {author} {\bibfnamefont {S.}~\bibnamefont {Wirth}}, \bibinfo {author} {\bibfnamefont {Y.}~\bibnamefont {Chen}}, \bibinfo {author} {\bibfnamefont {S.~T.~B.}\ \bibnamefont {Goennenwein}},\ and\ \bibinfo {author} {\bibfnamefont {C.}~\bibnamefont {Felser}},\ }\bibfield  {title} {\bibinfo {title} {Giant anomalous {Hall} effect in a ferromagnetic kagome-lattice semimetal},\ }\href {https://doi.org/10.1038/s41567-018-0234-5} {\bibfield  {journal} {\bibinfo  {journal} {Nat. Phys.}\ }\textbf {\bibinfo {volume} {14}},\ \bibinfo {pages} {1125} (\bibinfo {year} {2018}{\natexlab{a}})}\BibitemShut {NoStop}%
\bibitem [{\citenamefont {Wang}\ \emph {et~al.}(2018)\citenamefont {Wang}, \citenamefont {Xu}, \citenamefont {Lou}, \citenamefont {Liu}, \citenamefont {Li}, \citenamefont {Huang}, \citenamefont {Shen}, \citenamefont {Weng}, \citenamefont {Wang},\ and\ \citenamefont {Lei}}]{wang.xu.18}%
  \BibitemOpen
  \bibfield  {author} {\bibinfo {author} {\bibfnamefont {Q.}~\bibnamefont {Wang}}, \bibinfo {author} {\bibfnamefont {Y.}~\bibnamefont {Xu}}, \bibinfo {author} {\bibfnamefont {R.}~\bibnamefont {Lou}}, \bibinfo {author} {\bibfnamefont {Z.}~\bibnamefont {Liu}}, \bibinfo {author} {\bibfnamefont {M.}~\bibnamefont {Li}}, \bibinfo {author} {\bibfnamefont {Y.}~\bibnamefont {Huang}}, \bibinfo {author} {\bibfnamefont {D.}~\bibnamefont {Shen}}, \bibinfo {author} {\bibfnamefont {H.}~\bibnamefont {Weng}}, \bibinfo {author} {\bibfnamefont {S.}~\bibnamefont {Wang}},\ and\ \bibinfo {author} {\bibfnamefont {H.}~\bibnamefont {Lei}},\ }\bibfield  {title} {\bibinfo {title} {Large intrinsic anomalous {Hall} effect in half-metallic ferromagnet {Co$_{3}$Sn$_{2}$S$_{2}$} with magnetic {Weyl} fermions},\ }\href {https://doi.org/10.1038/s41467-018-06088-2} {\bibfield  {journal} {\bibinfo  {journal} {Nat. Commun.}\ }\textbf {\bibinfo {volume} {9}},\ \bibinfo {pages} {3681} (\bibinfo {year} {2018})}\BibitemShut {NoStop}%
\bibitem [{\citenamefont {Liu}\ \emph {et~al.}(2019)\citenamefont {Liu}, \citenamefont {Liang}, \citenamefont {Liu}, \citenamefont {Xu}, \citenamefont {Li}, \citenamefont {Chen}, \citenamefont {Pei}, \citenamefont {Shi}, \citenamefont {Mo}, \citenamefont {Dudin}, \citenamefont {Kim}, \citenamefont {Cacho}, \citenamefont {Li}, \citenamefont {Sun}, \citenamefont {Yang}, \citenamefont {Liu}, \citenamefont {Parkin}, \citenamefont {Felser},\ and\ \citenamefont {Chen}}]{liu.liang.19}%
  \BibitemOpen
  \bibfield  {author} {\bibinfo {author} {\bibfnamefont {D.~F.}\ \bibnamefont {Liu}}, \bibinfo {author} {\bibfnamefont {A.~J.}\ \bibnamefont {Liang}}, \bibinfo {author} {\bibfnamefont {E.~K.}\ \bibnamefont {Liu}}, \bibinfo {author} {\bibfnamefont {Q.~N.}\ \bibnamefont {Xu}}, \bibinfo {author} {\bibfnamefont {Y.~W.}\ \bibnamefont {Li}}, \bibinfo {author} {\bibfnamefont {C.}~\bibnamefont {Chen}}, \bibinfo {author} {\bibfnamefont {D.}~\bibnamefont {Pei}}, \bibinfo {author} {\bibfnamefont {W.~J.}\ \bibnamefont {Shi}}, \bibinfo {author} {\bibfnamefont {S.~K.}\ \bibnamefont {Mo}}, \bibinfo {author} {\bibfnamefont {P.}~\bibnamefont {Dudin}}, \bibinfo {author} {\bibfnamefont {T.}~\bibnamefont {Kim}}, \bibinfo {author} {\bibfnamefont {C.}~\bibnamefont {Cacho}}, \bibinfo {author} {\bibfnamefont {G.}~\bibnamefont {Li}}, \bibinfo {author} {\bibfnamefont {Y.}~\bibnamefont {Sun}}, \bibinfo {author} {\bibfnamefont {L.~X.}\ \bibnamefont {Yang}}, \bibinfo {author} {\bibfnamefont {Z.~K.}\ \bibnamefont {Liu}}, \bibinfo {author}
  {\bibfnamefont {S.~S.~P.}\ \bibnamefont {Parkin}}, \bibinfo {author} {\bibfnamefont {C.}~\bibnamefont {Felser}},\ and\ \bibinfo {author} {\bibfnamefont {Y.~L.}\ \bibnamefont {Chen}},\ }\bibfield  {title} {\bibinfo {title} {Magnetic {Weyl} semimetal phase in a kagom\'{e} crystal},\ }\href {https://doi.org/10.1126/science.aav2873} {\bibfield  {journal} {\bibinfo  {journal} {Science}\ }\textbf {\bibinfo {volume} {365}},\ \bibinfo {pages} {1282} (\bibinfo {year} {2019})}\BibitemShut {NoStop}%
\bibitem [{\citenamefont {Morali}\ \emph {et~al.}(2019)\citenamefont {Morali}, \citenamefont {Batabyal}, \citenamefont {Nag}, \citenamefont {Liu}, \citenamefont {Xu}, \citenamefont {Sun}, \citenamefont {Yan}, \citenamefont {Felser}, \citenamefont {Avraham},\ and\ \citenamefont {Beidenkopf}}]{morali.batabyal.19}%
  \BibitemOpen
  \bibfield  {author} {\bibinfo {author} {\bibfnamefont {N.}~\bibnamefont {Morali}}, \bibinfo {author} {\bibfnamefont {R.}~\bibnamefont {Batabyal}}, \bibinfo {author} {\bibfnamefont {P.~K.}\ \bibnamefont {Nag}}, \bibinfo {author} {\bibfnamefont {E.}~\bibnamefont {Liu}}, \bibinfo {author} {\bibfnamefont {Q.}~\bibnamefont {Xu}}, \bibinfo {author} {\bibfnamefont {Y.}~\bibnamefont {Sun}}, \bibinfo {author} {\bibfnamefont {B.}~\bibnamefont {Yan}}, \bibinfo {author} {\bibfnamefont {C.}~\bibnamefont {Felser}}, \bibinfo {author} {\bibfnamefont {N.}~\bibnamefont {Avraham}},\ and\ \bibinfo {author} {\bibfnamefont {H.}~\bibnamefont {Beidenkopf}},\ }\bibfield  {title} {\bibinfo {title} {Fermi-arc diversity on surface terminations of the magnetic {Weyl} semimetal {Co$_{3}$Sn$_{2}$S$_{2}$}},\ }\href {https://doi.org/10.1126/science.aav2334} {\bibfield  {journal} {\bibinfo  {journal} {Science}\ }\textbf {\bibinfo {volume} {365}},\ \bibinfo {pages} {1286} (\bibinfo {year} {2019})}\BibitemShut {NoStop}%
\bibitem [{\citenamefont {Yin}\ \emph {et~al.}(2019)\citenamefont {Yin}, \citenamefont {Zhang}, \citenamefont {Chang}, \citenamefont {Wang}, \citenamefont {Tsirkin}, \citenamefont {Guguchia}, \citenamefont {Lian}, \citenamefont {Zhou}, \citenamefont {Jiang}, \citenamefont {Belopolski}, \citenamefont {Shumiya}, \citenamefont {Multer}, \citenamefont {Litskevich}, \citenamefont {Cochran}, \citenamefont {Lin}, \citenamefont {Wang}, \citenamefont {Neupert}, \citenamefont {Jia}, \citenamefont {Lei},\ and\ \citenamefont {Hasan}}]{yin.zhang.19}%
  \BibitemOpen
  \bibfield  {author} {\bibinfo {author} {\bibfnamefont {J.-X.}\ \bibnamefont {Yin}}, \bibinfo {author} {\bibfnamefont {S.~S.}\ \bibnamefont {Zhang}}, \bibinfo {author} {\bibfnamefont {G.}~\bibnamefont {Chang}}, \bibinfo {author} {\bibfnamefont {Q.}~\bibnamefont {Wang}}, \bibinfo {author} {\bibfnamefont {S.~S.}\ \bibnamefont {Tsirkin}}, \bibinfo {author} {\bibfnamefont {Z.}~\bibnamefont {Guguchia}}, \bibinfo {author} {\bibfnamefont {B.}~\bibnamefont {Lian}}, \bibinfo {author} {\bibfnamefont {H.}~\bibnamefont {Zhou}}, \bibinfo {author} {\bibfnamefont {K.}~\bibnamefont {Jiang}}, \bibinfo {author} {\bibfnamefont {I.}~\bibnamefont {Belopolski}}, \bibinfo {author} {\bibfnamefont {N.}~\bibnamefont {Shumiya}}, \bibinfo {author} {\bibfnamefont {D.}~\bibnamefont {Multer}}, \bibinfo {author} {\bibfnamefont {M.}~\bibnamefont {Litskevich}}, \bibinfo {author} {\bibfnamefont {T.~A.}\ \bibnamefont {Cochran}}, \bibinfo {author} {\bibfnamefont {H.}~\bibnamefont {Lin}}, \bibinfo {author} {\bibfnamefont {Z.}~\bibnamefont
  {Wang}}, \bibinfo {author} {\bibfnamefont {T.}~\bibnamefont {Neupert}}, \bibinfo {author} {\bibfnamefont {S.}~\bibnamefont {Jia}}, \bibinfo {author} {\bibfnamefont {H.}~\bibnamefont {Lei}},\ and\ \bibinfo {author} {\bibfnamefont {M.~Z.}\ \bibnamefont {Hasan}},\ }\bibfield  {title} {\bibinfo {title} {Negative flat band magnetism in a spin--orbit-coupled correlated kagome magnet},\ }\href {https://doi.org/10.1038/s41567-019-0426-7} {\bibfield  {journal} {\bibinfo  {journal} {Nat. Phys.}\ }\textbf {\bibinfo {volume} {15}},\ \bibinfo {pages} {443} (\bibinfo {year} {2019})}\BibitemShut {NoStop}%
\bibitem [{\citenamefont {Xu}\ \emph {et~al.}(2020{\natexlab{a}})\citenamefont {Xu}, \citenamefont {Zhao}, \citenamefont {Yi}, \citenamefont {Wang}, \citenamefont {Yin}, \citenamefont {Wang}, \citenamefont {Hu}, \citenamefont {Wang}, \citenamefont {Liu}, \citenamefont {Xu}, \citenamefont {Lu}, \citenamefont {Soluyanov}, \citenamefont {Lei}, \citenamefont {Shi}, \citenamefont {Luo},\ and\ \citenamefont {Chen}}]{xu.zhao.20}%
  \BibitemOpen
  \bibfield  {author} {\bibinfo {author} {\bibfnamefont {Y.}~\bibnamefont {Xu}}, \bibinfo {author} {\bibfnamefont {J.}~\bibnamefont {Zhao}}, \bibinfo {author} {\bibfnamefont {C.}~\bibnamefont {Yi}}, \bibinfo {author} {\bibfnamefont {Q.}~\bibnamefont {Wang}}, \bibinfo {author} {\bibfnamefont {Q.}~\bibnamefont {Yin}}, \bibinfo {author} {\bibfnamefont {Y.}~\bibnamefont {Wang}}, \bibinfo {author} {\bibfnamefont {X.}~\bibnamefont {Hu}}, \bibinfo {author} {\bibfnamefont {L.}~\bibnamefont {Wang}}, \bibinfo {author} {\bibfnamefont {E.}~\bibnamefont {Liu}}, \bibinfo {author} {\bibfnamefont {G.}~\bibnamefont {Xu}}, \bibinfo {author} {\bibfnamefont {L.}~\bibnamefont {Lu}}, \bibinfo {author} {\bibfnamefont {A.~A.}\ \bibnamefont {Soluyanov}}, \bibinfo {author} {\bibfnamefont {H.}~\bibnamefont {Lei}}, \bibinfo {author} {\bibfnamefont {Y.}~\bibnamefont {Shi}}, \bibinfo {author} {\bibfnamefont {J.}~\bibnamefont {Luo}},\ and\ \bibinfo {author} {\bibfnamefont {Z.-G.}\ \bibnamefont {Chen}},\ }\bibfield  {title} {\bibinfo
  {title} {Electronic correlations and flattened band in magnetic {Weyl} semimetal candidate {Co$_{3}$Sn$_{2}$S$_{2}$}},\ }\href {https://doi.org/10.1038/s41467-020-17234-0} {\bibfield  {journal} {\bibinfo  {journal} {Nat. Commun.}\ }\textbf {\bibinfo {volume} {11}},\ \bibinfo {pages} {3985} (\bibinfo {year} {2020}{\natexlab{a}})}\BibitemShut {NoStop}%
\bibitem [{\citenamefont {Kang}\ \emph {et~al.}(2020)\citenamefont {Kang}, \citenamefont {Ye}, \citenamefont {Fang}, \citenamefont {You}, \citenamefont {Levitan}, \citenamefont {Han}, \citenamefont {Facio}, \citenamefont {Jozwiak}, \citenamefont {Bostwick}, \citenamefont {Rotenberg}, \citenamefont {Chan}, \citenamefont {McDonald}, \citenamefont {Graf}, \citenamefont {Kaznatcheev}, \citenamefont {Vescovo}, \citenamefont {Bell}, \citenamefont {Kaxiras}, \citenamefont {van~den Brink}, \citenamefont {Richter}, \citenamefont {Prasad~Ghimire}, \citenamefont {Checkelsky},\ and\ \citenamefont {Comin}}]{kang.ye.20}%
  \BibitemOpen
  \bibfield  {author} {\bibinfo {author} {\bibfnamefont {M.}~\bibnamefont {Kang}}, \bibinfo {author} {\bibfnamefont {L.}~\bibnamefont {Ye}}, \bibinfo {author} {\bibfnamefont {S.}~\bibnamefont {Fang}}, \bibinfo {author} {\bibfnamefont {J.-S.}\ \bibnamefont {You}}, \bibinfo {author} {\bibfnamefont {A.}~\bibnamefont {Levitan}}, \bibinfo {author} {\bibfnamefont {M.}~\bibnamefont {Han}}, \bibinfo {author} {\bibfnamefont {J.~I.}\ \bibnamefont {Facio}}, \bibinfo {author} {\bibfnamefont {C.}~\bibnamefont {Jozwiak}}, \bibinfo {author} {\bibfnamefont {A.}~\bibnamefont {Bostwick}}, \bibinfo {author} {\bibfnamefont {E.}~\bibnamefont {Rotenberg}}, \bibinfo {author} {\bibfnamefont {M.~K.}\ \bibnamefont {Chan}}, \bibinfo {author} {\bibfnamefont {R.~D.}\ \bibnamefont {McDonald}}, \bibinfo {author} {\bibfnamefont {D.}~\bibnamefont {Graf}}, \bibinfo {author} {\bibfnamefont {K.}~\bibnamefont {Kaznatcheev}}, \bibinfo {author} {\bibfnamefont {E.}~\bibnamefont {Vescovo}}, \bibinfo {author} {\bibfnamefont {D.~C.}\ \bibnamefont
  {Bell}}, \bibinfo {author} {\bibfnamefont {E.}~\bibnamefont {Kaxiras}}, \bibinfo {author} {\bibfnamefont {J.}~\bibnamefont {van~den Brink}}, \bibinfo {author} {\bibfnamefont {M.}~\bibnamefont {Richter}}, \bibinfo {author} {\bibfnamefont {M.}~\bibnamefont {Prasad~Ghimire}}, \bibinfo {author} {\bibfnamefont {J.~G.}\ \bibnamefont {Checkelsky}},\ and\ \bibinfo {author} {\bibfnamefont {R.}~\bibnamefont {Comin}},\ }\bibfield  {title} {\bibinfo {title} {{Dirac} fermions and flat bands in the ideal kagome metal {FeSn}},\ }\href {https://doi.org/10.1038/s41563-019-0531-0} {\bibfield  {journal} {\bibinfo  {journal} {Nat. Mater.}\ }\textbf {\bibinfo {volume} {19}},\ \bibinfo {pages} {163} (\bibinfo {year} {2020})}\BibitemShut {NoStop}%
\bibitem [{\citenamefont {Han}\ \emph {et~al.}(2021)\citenamefont {Han}, \citenamefont {Inoue}, \citenamefont {Fang}, \citenamefont {John}, \citenamefont {Ye}, \citenamefont {Chan}, \citenamefont {Graf}, \citenamefont {Suzuki}, \citenamefont {Ghimire}, \citenamefont {Cho}, \citenamefont {Kaxiras},\ and\ \citenamefont {Checkelsky}}]{han.inoue.21}%
  \BibitemOpen
  \bibfield  {author} {\bibinfo {author} {\bibfnamefont {M.}~\bibnamefont {Han}}, \bibinfo {author} {\bibfnamefont {H.}~\bibnamefont {Inoue}}, \bibinfo {author} {\bibfnamefont {S.}~\bibnamefont {Fang}}, \bibinfo {author} {\bibfnamefont {C.}~\bibnamefont {John}}, \bibinfo {author} {\bibfnamefont {L.}~\bibnamefont {Ye}}, \bibinfo {author} {\bibfnamefont {M.~K.}\ \bibnamefont {Chan}}, \bibinfo {author} {\bibfnamefont {D.}~\bibnamefont {Graf}}, \bibinfo {author} {\bibfnamefont {T.}~\bibnamefont {Suzuki}}, \bibinfo {author} {\bibfnamefont {M.~P.}\ \bibnamefont {Ghimire}}, \bibinfo {author} {\bibfnamefont {W.~J.}\ \bibnamefont {Cho}}, \bibinfo {author} {\bibfnamefont {E.}~\bibnamefont {Kaxiras}},\ and\ \bibinfo {author} {\bibfnamefont {J.~G.}\ \bibnamefont {Checkelsky}},\ }\bibfield  {title} {\bibinfo {title} {Evidence of two-dimensional flat band at the surface of antiferromagnetic kagome metal {FeSn}},\ }\href {https://doi.org/10.1038/s41467-021-25705-1} {\bibfield  {journal} {\bibinfo  {journal} {Nat. Commun.}\
  }\textbf {\bibinfo {volume} {12}},\ \bibinfo {pages} {5345} (\bibinfo {year} {2021})}\BibitemShut {NoStop}%
\bibitem [{\citenamefont {Zhang}\ \emph {et~al.}(2023)\citenamefont {Zhang}, \citenamefont {Oli}, \citenamefont {Zou}, \citenamefont {Guo}, \citenamefont {Wang},\ and\ \citenamefont {Li}}]{zhang.oli.23}%
  \BibitemOpen
  \bibfield  {author} {\bibinfo {author} {\bibfnamefont {H.}~\bibnamefont {Zhang}}, \bibinfo {author} {\bibfnamefont {B.~D.}\ \bibnamefont {Oli}}, \bibinfo {author} {\bibfnamefont {Q.}~\bibnamefont {Zou}}, \bibinfo {author} {\bibfnamefont {X.}~\bibnamefont {Guo}}, \bibinfo {author} {\bibfnamefont {Z.}~\bibnamefont {Wang}},\ and\ \bibinfo {author} {\bibfnamefont {L.}~\bibnamefont {Li}},\ }\bibfield  {title} {\bibinfo {title} {Visualizing symmetry-breaking electronic orders in epitaxial kagome magnet {FeSn} films},\ }\href {https://doi.org/10.1038/s41467-023-41831-4} {\bibfield  {journal} {\bibinfo  {journal} {Nat. Commun.}\ }\textbf {\bibinfo {volume} {14}},\ \bibinfo {pages} {6167} (\bibinfo {year} {2023})}\BibitemShut {NoStop}%
\bibitem [{\citenamefont {Wang}\ \emph {et~al.}(2016)\citenamefont {Wang}, \citenamefont {Sun}, \citenamefont {Zhang}, \citenamefont {Pang},\ and\ \citenamefont {Lei}}]{wang.sun.16}%
  \BibitemOpen
  \bibfield  {author} {\bibinfo {author} {\bibfnamefont {Q.}~\bibnamefont {Wang}}, \bibinfo {author} {\bibfnamefont {S.}~\bibnamefont {Sun}}, \bibinfo {author} {\bibfnamefont {X.}~\bibnamefont {Zhang}}, \bibinfo {author} {\bibfnamefont {F.}~\bibnamefont {Pang}},\ and\ \bibinfo {author} {\bibfnamefont {H.}~\bibnamefont {Lei}},\ }\bibfield  {title} {\bibinfo {title} {Anomalous {Hall} effect in a ferromagnetic {Fe$_{3}$Sn$_{2}$} single crystal with a geometrically frustrated {Fe} bilayer kagome lattice},\ }\href {https://doi.org/10.1103/PhysRevB.94.075135} {\bibfield  {journal} {\bibinfo  {journal} {Phys. Rev. B}\ }\textbf {\bibinfo {volume} {94}},\ \bibinfo {pages} {075135} (\bibinfo {year} {2016})}\BibitemShut {NoStop}%
\bibitem [{\citenamefont {Ye}\ \emph {et~al.}(2018)\citenamefont {Ye}, \citenamefont {Kang}, \citenamefont {Liu}, \citenamefont {von Cube}, \citenamefont {Wicker}, \citenamefont {Suzuki}, \citenamefont {Jozwiak}, \citenamefont {Bostwick}, \citenamefont {Rotenberg}, \citenamefont {Bell}, \citenamefont {Fu}, \citenamefont {Comin},\ and\ \citenamefont {Checkelsky}}]{ye.kang.18}%
  \BibitemOpen
  \bibfield  {author} {\bibinfo {author} {\bibfnamefont {L.}~\bibnamefont {Ye}}, \bibinfo {author} {\bibfnamefont {M.}~\bibnamefont {Kang}}, \bibinfo {author} {\bibfnamefont {J.}~\bibnamefont {Liu}}, \bibinfo {author} {\bibfnamefont {F.}~\bibnamefont {von Cube}}, \bibinfo {author} {\bibfnamefont {C.~R.}\ \bibnamefont {Wicker}}, \bibinfo {author} {\bibfnamefont {T.}~\bibnamefont {Suzuki}}, \bibinfo {author} {\bibfnamefont {C.}~\bibnamefont {Jozwiak}}, \bibinfo {author} {\bibfnamefont {A.}~\bibnamefont {Bostwick}}, \bibinfo {author} {\bibfnamefont {E.}~\bibnamefont {Rotenberg}}, \bibinfo {author} {\bibfnamefont {D.~C.}\ \bibnamefont {Bell}}, \bibinfo {author} {\bibfnamefont {L.}~\bibnamefont {Fu}}, \bibinfo {author} {\bibfnamefont {R.}~\bibnamefont {Comin}},\ and\ \bibinfo {author} {\bibfnamefont {J.~G.}\ \bibnamefont {Checkelsky}},\ }\bibfield  {title} {\bibinfo {title} {Massive {Dirac} fermions in a ferromagnetic kagome metal},\ }\href {https://doi.org/10.1038/nature25987} {\bibfield  {journal} {\bibinfo
  {journal} {Nature}\ }\textbf {\bibinfo {volume} {555}},\ \bibinfo {pages} {638} (\bibinfo {year} {2018})}\BibitemShut {NoStop}%
\bibitem [{\citenamefont {Lin}\ \emph {et~al.}(2018)\citenamefont {Lin}, \citenamefont {Choi}, \citenamefont {Zhang}, \citenamefont {Qin}, \citenamefont {Yi}, \citenamefont {Wang}, \citenamefont {Li}, \citenamefont {Wang}, \citenamefont {Zhang}, \citenamefont {Sun}, \citenamefont {Wei}, \citenamefont {Zhang}, \citenamefont {Guo}, \citenamefont {Lu}, \citenamefont {Cho}, \citenamefont {Zeng},\ and\ \citenamefont {Zhang}}]{lin.choi.18}%
  \BibitemOpen
  \bibfield  {author} {\bibinfo {author} {\bibfnamefont {Z.}~\bibnamefont {Lin}}, \bibinfo {author} {\bibfnamefont {J.-H.}\ \bibnamefont {Choi}}, \bibinfo {author} {\bibfnamefont {Q.}~\bibnamefont {Zhang}}, \bibinfo {author} {\bibfnamefont {W.}~\bibnamefont {Qin}}, \bibinfo {author} {\bibfnamefont {S.}~\bibnamefont {Yi}}, \bibinfo {author} {\bibfnamefont {P.}~\bibnamefont {Wang}}, \bibinfo {author} {\bibfnamefont {L.}~\bibnamefont {Li}}, \bibinfo {author} {\bibfnamefont {Y.}~\bibnamefont {Wang}}, \bibinfo {author} {\bibfnamefont {H.}~\bibnamefont {Zhang}}, \bibinfo {author} {\bibfnamefont {Z.}~\bibnamefont {Sun}}, \bibinfo {author} {\bibfnamefont {L.}~\bibnamefont {Wei}}, \bibinfo {author} {\bibfnamefont {S.}~\bibnamefont {Zhang}}, \bibinfo {author} {\bibfnamefont {T.}~\bibnamefont {Guo}}, \bibinfo {author} {\bibfnamefont {Q.}~\bibnamefont {Lu}}, \bibinfo {author} {\bibfnamefont {J.-H.}\ \bibnamefont {Cho}}, \bibinfo {author} {\bibfnamefont {C.}~\bibnamefont {Zeng}},\ and\ \bibinfo {author} {\bibfnamefont
  {Z.}~\bibnamefont {Zhang}},\ }\bibfield  {title} {\bibinfo {title} {Flatbands and emergent ferromagnetic ordering in {Fe$_{3}$Sn$_{2}$} kagome lattices},\ }\href {https://doi.org/10.1103/PhysRevLett.121.096401} {\bibfield  {journal} {\bibinfo  {journal} {Phys. Rev. Lett.}\ }\textbf {\bibinfo {volume} {121}},\ \bibinfo {pages} {096401} (\bibinfo {year} {2018})}\BibitemShut {NoStop}%
\bibitem [{\citenamefont {Yin}\ \emph {et~al.}(2018)\citenamefont {Yin}, \citenamefont {Zhang}, \citenamefont {Li}, \citenamefont {Jiang}, \citenamefont {Chang}, \citenamefont {Zhang}, \citenamefont {Lian}, \citenamefont {Xiang}, \citenamefont {Belopolski}, \citenamefont {Zheng}, \citenamefont {Cochran}, \citenamefont {Xu}, \citenamefont {Bian}, \citenamefont {Liu}, \citenamefont {Chang}, \citenamefont {Lin}, \citenamefont {Lu}, \citenamefont {Wang}, \citenamefont {Jia}, \citenamefont {Wang},\ and\ \citenamefont {Hasan}}]{yin.zhang.18}%
  \BibitemOpen
  \bibfield  {author} {\bibinfo {author} {\bibfnamefont {J.-X.}\ \bibnamefont {Yin}}, \bibinfo {author} {\bibfnamefont {S.~S.}\ \bibnamefont {Zhang}}, \bibinfo {author} {\bibfnamefont {H.}~\bibnamefont {Li}}, \bibinfo {author} {\bibfnamefont {K.}~\bibnamefont {Jiang}}, \bibinfo {author} {\bibfnamefont {G.}~\bibnamefont {Chang}}, \bibinfo {author} {\bibfnamefont {B.}~\bibnamefont {Zhang}}, \bibinfo {author} {\bibfnamefont {B.}~\bibnamefont {Lian}}, \bibinfo {author} {\bibfnamefont {C.}~\bibnamefont {Xiang}}, \bibinfo {author} {\bibfnamefont {I.}~\bibnamefont {Belopolski}}, \bibinfo {author} {\bibfnamefont {H.}~\bibnamefont {Zheng}}, \bibinfo {author} {\bibfnamefont {T.~A.}\ \bibnamefont {Cochran}}, \bibinfo {author} {\bibfnamefont {S.-Y.}\ \bibnamefont {Xu}}, \bibinfo {author} {\bibfnamefont {G.}~\bibnamefont {Bian}}, \bibinfo {author} {\bibfnamefont {K.}~\bibnamefont {Liu}}, \bibinfo {author} {\bibfnamefont {T.-R.}\ \bibnamefont {Chang}}, \bibinfo {author} {\bibfnamefont {H.}~\bibnamefont {Lin}}, \bibinfo
  {author} {\bibfnamefont {Z.-Y.}\ \bibnamefont {Lu}}, \bibinfo {author} {\bibfnamefont {Z.}~\bibnamefont {Wang}}, \bibinfo {author} {\bibfnamefont {S.}~\bibnamefont {Jia}}, \bibinfo {author} {\bibfnamefont {W.}~\bibnamefont {Wang}},\ and\ \bibinfo {author} {\bibfnamefont {M.~Z.}\ \bibnamefont {Hasan}},\ }\bibfield  {title} {\bibinfo {title} {Giant and anisotropic many-body spin--orbit tunability in a strongly correlated kagome magnet},\ }\href {https://doi.org/10.1038/s41586-018-0502-7} {\bibfield  {journal} {\bibinfo  {journal} {Nature}\ }\textbf {\bibinfo {volume} {562}},\ \bibinfo {pages} {91} (\bibinfo {year} {2018})}\BibitemShut {NoStop}%
\bibitem [{\citenamefont {Yin}\ \emph {et~al.}(2020)\citenamefont {Yin}, \citenamefont {Ma}, \citenamefont {Cochran}, \citenamefont {Xu}, \citenamefont {Zhang}, \citenamefont {Tien}, \citenamefont {Shumiya}, \citenamefont {Cheng}, \citenamefont {Jiang}, \citenamefont {Lian}, \citenamefont {Song}, \citenamefont {Chang}, \citenamefont {Belopolski}, \citenamefont {Multer}, \citenamefont {Litskevich}, \citenamefont {Cheng}, \citenamefont {Yang}, \citenamefont {Swidler}, \citenamefont {Zhou}, \citenamefont {Lin}, \citenamefont {Neupert}, \citenamefont {Wang}, \citenamefont {Yao}, \citenamefont {Chang}, \citenamefont {Jia},\ and\ \citenamefont {Zahid~Hasan}}]{yin.ma.20}%
  \BibitemOpen
  \bibfield  {author} {\bibinfo {author} {\bibfnamefont {J.-X.}\ \bibnamefont {Yin}}, \bibinfo {author} {\bibfnamefont {W.}~\bibnamefont {Ma}}, \bibinfo {author} {\bibfnamefont {T.~A.}\ \bibnamefont {Cochran}}, \bibinfo {author} {\bibfnamefont {X.}~\bibnamefont {Xu}}, \bibinfo {author} {\bibfnamefont {S.~S.}\ \bibnamefont {Zhang}}, \bibinfo {author} {\bibfnamefont {H.-J.}\ \bibnamefont {Tien}}, \bibinfo {author} {\bibfnamefont {N.}~\bibnamefont {Shumiya}}, \bibinfo {author} {\bibfnamefont {G.}~\bibnamefont {Cheng}}, \bibinfo {author} {\bibfnamefont {K.}~\bibnamefont {Jiang}}, \bibinfo {author} {\bibfnamefont {B.}~\bibnamefont {Lian}}, \bibinfo {author} {\bibfnamefont {Z.}~\bibnamefont {Song}}, \bibinfo {author} {\bibfnamefont {G.}~\bibnamefont {Chang}}, \bibinfo {author} {\bibfnamefont {I.}~\bibnamefont {Belopolski}}, \bibinfo {author} {\bibfnamefont {D.}~\bibnamefont {Multer}}, \bibinfo {author} {\bibfnamefont {M.}~\bibnamefont {Litskevich}}, \bibinfo {author} {\bibfnamefont {Z.-J.}\ \bibnamefont {Cheng}},
  \bibinfo {author} {\bibfnamefont {X.~P.}\ \bibnamefont {Yang}}, \bibinfo {author} {\bibfnamefont {B.}~\bibnamefont {Swidler}}, \bibinfo {author} {\bibfnamefont {H.}~\bibnamefont {Zhou}}, \bibinfo {author} {\bibfnamefont {H.}~\bibnamefont {Lin}}, \bibinfo {author} {\bibfnamefont {T.}~\bibnamefont {Neupert}}, \bibinfo {author} {\bibfnamefont {Z.}~\bibnamefont {Wang}}, \bibinfo {author} {\bibfnamefont {N.}~\bibnamefont {Yao}}, \bibinfo {author} {\bibfnamefont {T.-R.}\ \bibnamefont {Chang}}, \bibinfo {author} {\bibfnamefont {S.}~\bibnamefont {Jia}},\ and\ \bibinfo {author} {\bibfnamefont {M.}~\bibnamefont {Zahid~Hasan}},\ }\bibfield  {title} {\bibinfo {title} {Quantum-limit {Chern} topological magnetism in {TbMn$_{6}$Sn$_{6}$}},\ }\href {https://doi.org/10.1038/s41586-020-2482-7} {\bibfield  {journal} {\bibinfo  {journal} {Nature}\ }\textbf {\bibinfo {volume} {583}},\ \bibinfo {pages} {533} (\bibinfo {year} {2020})}\BibitemShut {NoStop}%
\bibitem [{\citenamefont {Ghimire}\ \emph {et~al.}(2020)\citenamefont {Ghimire}, \citenamefont {Dally}, \citenamefont {Poudel}, \citenamefont {Jones}, \citenamefont {Michel}, \citenamefont {Magar}, \citenamefont {Bleuel}, \citenamefont {McGuire}, \citenamefont {Jiang}, \citenamefont {Mitchell}, \citenamefont {Lynn},\ and\ \citenamefont {Mazin}}]{ghimire.dally.20}%
  \BibitemOpen
  \bibfield  {author} {\bibinfo {author} {\bibfnamefont {N.~J.}\ \bibnamefont {Ghimire}}, \bibinfo {author} {\bibfnamefont {R.~L.}\ \bibnamefont {Dally}}, \bibinfo {author} {\bibfnamefont {L.}~\bibnamefont {Poudel}}, \bibinfo {author} {\bibfnamefont {D.~C.}\ \bibnamefont {Jones}}, \bibinfo {author} {\bibfnamefont {D.}~\bibnamefont {Michel}}, \bibinfo {author} {\bibfnamefont {N.~T.}\ \bibnamefont {Magar}}, \bibinfo {author} {\bibfnamefont {M.}~\bibnamefont {Bleuel}}, \bibinfo {author} {\bibfnamefont {M.~A.}\ \bibnamefont {McGuire}}, \bibinfo {author} {\bibfnamefont {J.~S.}\ \bibnamefont {Jiang}}, \bibinfo {author} {\bibfnamefont {J.~F.}\ \bibnamefont {Mitchell}}, \bibinfo {author} {\bibfnamefont {J.~W.}\ \bibnamefont {Lynn}},\ and\ \bibinfo {author} {\bibfnamefont {I.~I.}\ \bibnamefont {Mazin}},\ }\bibfield  {title} {\bibinfo {title} {Competing magnetic phases and fluctuation-driven scalar spin chirality in the kagome metal {YMn$_{6}$Sn$_{6}$}},\ }\href {https://doi.org/10.1126/sciadv.abe2680} {\bibfield
  {journal} {\bibinfo  {journal} {Sci. Adv.}\ }\textbf {\bibinfo {volume} {6}},\ \bibinfo {pages} {eabe2680} (\bibinfo {year} {2020})}\BibitemShut {NoStop}%
\bibitem [{\citenamefont {Ma}\ \emph {et~al.}(2021)\citenamefont {Ma}, \citenamefont {Xu}, \citenamefont {Yin}, \citenamefont {Yang}, \citenamefont {Zhou}, \citenamefont {Cheng}, \citenamefont {Huang}, \citenamefont {Qu}, \citenamefont {Wang}, \citenamefont {Hasan},\ and\ \citenamefont {Jia}}]{ma.xu.21}%
  \BibitemOpen
  \bibfield  {author} {\bibinfo {author} {\bibfnamefont {W.}~\bibnamefont {Ma}}, \bibinfo {author} {\bibfnamefont {X.}~\bibnamefont {Xu}}, \bibinfo {author} {\bibfnamefont {J.-X.}\ \bibnamefont {Yin}}, \bibinfo {author} {\bibfnamefont {H.}~\bibnamefont {Yang}}, \bibinfo {author} {\bibfnamefont {H.}~\bibnamefont {Zhou}}, \bibinfo {author} {\bibfnamefont {Z.-J.}\ \bibnamefont {Cheng}}, \bibinfo {author} {\bibfnamefont {Y.}~\bibnamefont {Huang}}, \bibinfo {author} {\bibfnamefont {Z.}~\bibnamefont {Qu}}, \bibinfo {author} {\bibfnamefont {F.}~\bibnamefont {Wang}}, \bibinfo {author} {\bibfnamefont {M.~Z.}\ \bibnamefont {Hasan}},\ and\ \bibinfo {author} {\bibfnamefont {S.}~\bibnamefont {Jia}},\ }\bibfield  {title} {\bibinfo {title} {Rare earth engineering in {$R$Mn$_{6}$Sn$_{6}$} ({$R=$Gd--Tm, Lu}) topological kagome magnets},\ }\href {https://doi.org/10.1103/PhysRevLett.126.246602} {\bibfield  {journal} {\bibinfo  {journal} {Phys. Rev. Lett.}\ }\textbf {\bibinfo {volume} {126}},\ \bibinfo {pages} {246602} (\bibinfo
  {year} {2021})}\BibitemShut {NoStop}%
\bibitem [{\citenamefont {Li}\ \emph {et~al.}(2021{\natexlab{b}})\citenamefont {Li}, \citenamefont {Wang}, \citenamefont {Wang}, \citenamefont {Yuan}, \citenamefont {Song}, \citenamefont {Lou}, \citenamefont {Liu}, \citenamefont {Huang}, \citenamefont {Liu}, \citenamefont {Lei}, \citenamefont {Yin},\ and\ \citenamefont {Wang}}]{li.wang.21}%
  \BibitemOpen
  \bibfield  {author} {\bibinfo {author} {\bibfnamefont {M.}~\bibnamefont {Li}}, \bibinfo {author} {\bibfnamefont {Q.}~\bibnamefont {Wang}}, \bibinfo {author} {\bibfnamefont {G.}~\bibnamefont {Wang}}, \bibinfo {author} {\bibfnamefont {Z.}~\bibnamefont {Yuan}}, \bibinfo {author} {\bibfnamefont {W.}~\bibnamefont {Song}}, \bibinfo {author} {\bibfnamefont {R.}~\bibnamefont {Lou}}, \bibinfo {author} {\bibfnamefont {Z.}~\bibnamefont {Liu}}, \bibinfo {author} {\bibfnamefont {Y.}~\bibnamefont {Huang}}, \bibinfo {author} {\bibfnamefont {Z.}~\bibnamefont {Liu}}, \bibinfo {author} {\bibfnamefont {H.}~\bibnamefont {Lei}}, \bibinfo {author} {\bibfnamefont {Z.}~\bibnamefont {Yin}},\ and\ \bibinfo {author} {\bibfnamefont {S.}~\bibnamefont {Wang}},\ }\bibfield  {title} {\bibinfo {title} {Dirac cone, flat band and saddle point in kagome magnet {YMn$_{6}$Sn$_{6}$}},\ }\href {https://doi.org/10.1038/s41467-021-23536-8} {\bibfield  {journal} {\bibinfo  {journal} {Nat. Commun.}\ }\textbf {\bibinfo {volume} {12}},\ \bibinfo
  {pages} {3129} (\bibinfo {year} {2021}{\natexlab{b}})}\BibitemShut {NoStop}%
\bibitem [{\citenamefont {Riberolles}\ \emph {et~al.}(2022)\citenamefont {Riberolles}, \citenamefont {Slade}, \citenamefont {Abernathy}, \citenamefont {Granroth}, \citenamefont {Li}, \citenamefont {Lee}, \citenamefont {Canfield}, \citenamefont {Ueland}, \citenamefont {Ke},\ and\ \citenamefont {McQueeney}}]{riberolles.slade.22}%
  \BibitemOpen
  \bibfield  {author} {\bibinfo {author} {\bibfnamefont {S.~X.~M.}\ \bibnamefont {Riberolles}}, \bibinfo {author} {\bibfnamefont {T.~J.}\ \bibnamefont {Slade}}, \bibinfo {author} {\bibfnamefont {D.~L.}\ \bibnamefont {Abernathy}}, \bibinfo {author} {\bibfnamefont {G.~E.}\ \bibnamefont {Granroth}}, \bibinfo {author} {\bibfnamefont {B.}~\bibnamefont {Li}}, \bibinfo {author} {\bibfnamefont {Y.}~\bibnamefont {Lee}}, \bibinfo {author} {\bibfnamefont {P.~C.}\ \bibnamefont {Canfield}}, \bibinfo {author} {\bibfnamefont {B.~G.}\ \bibnamefont {Ueland}}, \bibinfo {author} {\bibfnamefont {L.}~\bibnamefont {Ke}},\ and\ \bibinfo {author} {\bibfnamefont {R.~J.}\ \bibnamefont {McQueeney}},\ }\bibfield  {title} {\bibinfo {title} {Low-temperature competing magnetic energy scales in the topological ferrimagnet {TbMn$_{6}$Sn$_{6}$}},\ }\href {https://doi.org/10.1103/PhysRevX.12.021043} {\bibfield  {journal} {\bibinfo  {journal} {Phys. Rev. X}\ }\textbf {\bibinfo {volume} {12}},\ \bibinfo {pages} {021043} (\bibinfo {year}
  {2022})}\BibitemShut {NoStop}%
\bibitem [{\citenamefont {Mielke~III}\ \emph {et~al.}(2022)\citenamefont {Mielke~III}, \citenamefont {Ma}, \citenamefont {Pomjakushin}, \citenamefont {Zaharko}, \citenamefont {Sturniolo}, \citenamefont {Liu}, \citenamefont {Ukleev}, \citenamefont {White}, \citenamefont {Yin}, \citenamefont {Tsirkin}, \citenamefont {Larsen}, \citenamefont {Cochran}, \citenamefont {Medarde}, \citenamefont {Das}, \citenamefont {Gupta}, \citenamefont {Chang}, \citenamefont {Wang}, \citenamefont {Khasanov}, \citenamefont {Neupert}, \citenamefont {Amato}, \citenamefont {Liborio}, \citenamefont {Jia}, \citenamefont {Hasan}, \citenamefont {Luetkens},\ and\ \citenamefont {Guguchia}}]{mielke.ma.22}%
  \BibitemOpen
  \bibfield  {author} {\bibinfo {author} {\bibfnamefont {C.}~\bibnamefont {Mielke~III}}, \bibinfo {author} {\bibfnamefont {W.~L.}\ \bibnamefont {Ma}}, \bibinfo {author} {\bibfnamefont {V.}~\bibnamefont {Pomjakushin}}, \bibinfo {author} {\bibfnamefont {O.}~\bibnamefont {Zaharko}}, \bibinfo {author} {\bibfnamefont {S.}~\bibnamefont {Sturniolo}}, \bibinfo {author} {\bibfnamefont {X.}~\bibnamefont {Liu}}, \bibinfo {author} {\bibfnamefont {V.}~\bibnamefont {Ukleev}}, \bibinfo {author} {\bibfnamefont {J.~S.}\ \bibnamefont {White}}, \bibinfo {author} {\bibfnamefont {J.-X.}\ \bibnamefont {Yin}}, \bibinfo {author} {\bibfnamefont {S.~S.}\ \bibnamefont {Tsirkin}}, \bibinfo {author} {\bibfnamefont {C.~B.}\ \bibnamefont {Larsen}}, \bibinfo {author} {\bibfnamefont {T.~A.}\ \bibnamefont {Cochran}}, \bibinfo {author} {\bibfnamefont {V.}~\bibnamefont {Medarde}, \bibfnamefont {M.and~Por{\'e}e}}, \bibinfo {author} {\bibfnamefont {D.}~\bibnamefont {Das}}, \bibinfo {author} {\bibfnamefont {C.~N.}\ \bibnamefont {Gupta},
  \bibfnamefont {R.and~Wang}}, \bibinfo {author} {\bibfnamefont {J.}~\bibnamefont {Chang}}, \bibinfo {author} {\bibfnamefont {Z.~Q.}\ \bibnamefont {Wang}}, \bibinfo {author} {\bibfnamefont {R.}~\bibnamefont {Khasanov}}, \bibinfo {author} {\bibfnamefont {T.}~\bibnamefont {Neupert}}, \bibinfo {author} {\bibfnamefont {A.}~\bibnamefont {Amato}}, \bibinfo {author} {\bibfnamefont {L.}~\bibnamefont {Liborio}}, \bibinfo {author} {\bibfnamefont {S.}~\bibnamefont {Jia}}, \bibinfo {author} {\bibfnamefont {M.~Z.}\ \bibnamefont {Hasan}}, \bibinfo {author} {\bibfnamefont {H.}~\bibnamefont {Luetkens}},\ and\ \bibinfo {author} {\bibfnamefont {Z.}~\bibnamefont {Guguchia}},\ }\bibfield  {title} {\bibinfo {title} {Low-temperature magnetic crossover in the topological kagome magnet {TbMn$_{6}$Sn$_{6}$}},\ }\href {https://doi.org/10.1038/s42005-022-00885-4} {\bibfield  {journal} {\bibinfo  {journal} {Commun. Phys.}\ }\textbf {\bibinfo {volume} {5}},\ \bibinfo {pages} {107} (\bibinfo {year} {2022})}\BibitemShut {NoStop}%
\bibitem [{\citenamefont {Zhang}\ \emph {et~al.}(2022)\citenamefont {Zhang}, \citenamefont {Koo}, \citenamefont {Xu}, \citenamefont {Sretenovic}, \citenamefont {Yan},\ and\ \citenamefont {Ke}}]{zhang.koo.22}%
  \BibitemOpen
  \bibfield  {author} {\bibinfo {author} {\bibfnamefont {H.}~\bibnamefont {Zhang}}, \bibinfo {author} {\bibfnamefont {J.}~\bibnamefont {Koo}}, \bibinfo {author} {\bibfnamefont {C.}~\bibnamefont {Xu}}, \bibinfo {author} {\bibfnamefont {M.}~\bibnamefont {Sretenovic}}, \bibinfo {author} {\bibfnamefont {B.}~\bibnamefont {Yan}},\ and\ \bibinfo {author} {\bibfnamefont {X.}~\bibnamefont {Ke}},\ }\bibfield  {title} {\bibinfo {title} {Exchange-biased topological transverse thermoelectric effects in a kagome ferrimagnet},\ }\href {https://doi.org/10.1038/s41467-022-28733-7} {\bibfield  {journal} {\bibinfo  {journal} {Nat. Commun.}\ }\textbf {\bibinfo {volume} {13}},\ \bibinfo {pages} {1091} (\bibinfo {year} {2022})}\BibitemShut {NoStop}%
\bibitem [{\citenamefont {Fruhling}\ \emph {et~al.}(2024)\citenamefont {Fruhling}, \citenamefont {Streeter}, \citenamefont {Mardanya}, \citenamefont {Wang}, \citenamefont {Baral}, \citenamefont {Zaharko}, \citenamefont {Mazin}, \citenamefont {Chowdhury}, \citenamefont {Ratcliff},\ and\ \citenamefont {Tafti}}]{fruhling.streeter.24}%
  \BibitemOpen
  \bibfield  {author} {\bibinfo {author} {\bibfnamefont {K.}~\bibnamefont {Fruhling}}, \bibinfo {author} {\bibfnamefont {A.}~\bibnamefont {Streeter}}, \bibinfo {author} {\bibfnamefont {S.}~\bibnamefont {Mardanya}}, \bibinfo {author} {\bibfnamefont {X.}~\bibnamefont {Wang}}, \bibinfo {author} {\bibfnamefont {P.}~\bibnamefont {Baral}}, \bibinfo {author} {\bibfnamefont {O.}~\bibnamefont {Zaharko}}, \bibinfo {author} {\bibfnamefont {I.~I.}\ \bibnamefont {Mazin}}, \bibinfo {author} {\bibfnamefont {S.}~\bibnamefont {Chowdhury}}, \bibinfo {author} {\bibfnamefont {W.~D.}\ \bibnamefont {Ratcliff}},\ and\ \bibinfo {author} {\bibfnamefont {F.}~\bibnamefont {Tafti}},\ }\bibfield  {title} {\bibinfo {title} {Topological hall effect induced by chiral fluctuations in {ErMn$_{6}$Sn$_{6}$}},\ }\href {https://doi.org/10.1103/PhysRevMaterials.8.094411} {\bibfield  {journal} {\bibinfo  {journal} {Phys. Rev. Mater.}\ }\textbf {\bibinfo {volume} {8}},\ \bibinfo {pages} {094411} (\bibinfo {year} {2024})}\BibitemShut {NoStop}%
\bibitem [{\citenamefont {Roychowdhury}\ \emph {et~al.}(2024)\citenamefont {Roychowdhury}, \citenamefont {Samanta}, \citenamefont {Singh}, \citenamefont {Schnelle}, \citenamefont {Zhang}, \citenamefont {Noky}, \citenamefont {Vergniory}, \citenamefont {Shekhar},\ and\ \citenamefont {Felser}}]{roychowdhury.samanta.24}%
  \BibitemOpen
  \bibfield  {author} {\bibinfo {author} {\bibfnamefont {S.}~\bibnamefont {Roychowdhury}}, \bibinfo {author} {\bibfnamefont {K.}~\bibnamefont {Samanta}}, \bibinfo {author} {\bibfnamefont {S.}~\bibnamefont {Singh}}, \bibinfo {author} {\bibfnamefont {W.}~\bibnamefont {Schnelle}}, \bibinfo {author} {\bibfnamefont {Y.}~\bibnamefont {Zhang}}, \bibinfo {author} {\bibfnamefont {J.}~\bibnamefont {Noky}}, \bibinfo {author} {\bibfnamefont {M.~G.}\ \bibnamefont {Vergniory}}, \bibinfo {author} {\bibfnamefont {C.}~\bibnamefont {Shekhar}},\ and\ \bibinfo {author} {\bibfnamefont {C.}~\bibnamefont {Felser}},\ }\bibfield  {title} {\bibinfo {title} {Enhancement of the anomalous {Hall} effect by distorting the kagome lattice in an antiferromagnetic material},\ }\href {https://doi.org/10.1073/pnas.2401970121} {\bibfield  {journal} {\bibinfo  {journal} {PNAS}\ }\textbf {\bibinfo {volume} {121}},\ \bibinfo {pages} {e2401970121} (\bibinfo {year} {2024})}\BibitemShut {NoStop}%
\bibitem [{\citenamefont {Ptok}\ \emph {et~al.}(2021)\citenamefont {Ptok}, \citenamefont {Kobia\l{}ka}, \citenamefont {Sternik}, \citenamefont {\L{}a\.{z}ewski}, \citenamefont {Jochym}, \citenamefont {Ole\'{s}}, \citenamefont {Stankov},\ and\ \citenamefont {Piekarz}}]{ptok.kobialka.21}%
  \BibitemOpen
  \bibfield  {author} {\bibinfo {author} {\bibfnamefont {A.}~\bibnamefont {Ptok}}, \bibinfo {author} {\bibfnamefont {A.}~\bibnamefont {Kobia\l{}ka}}, \bibinfo {author} {\bibfnamefont {M.}~\bibnamefont {Sternik}}, \bibinfo {author} {\bibfnamefont {J.}~\bibnamefont {\L{}a\.{z}ewski}}, \bibinfo {author} {\bibfnamefont {P.~T.}\ \bibnamefont {Jochym}}, \bibinfo {author} {\bibfnamefont {A.~M.}\ \bibnamefont {Ole\'{s}}}, \bibinfo {author} {\bibfnamefont {S.}~\bibnamefont {Stankov}},\ and\ \bibinfo {author} {\bibfnamefont {P.}~\bibnamefont {Piekarz}},\ }\bibfield  {title} {\bibinfo {title} {Chiral phonons in the honeycomb sublattice of layered {CoSn}-like compounds},\ }\href {https://doi.org/10.1103/PhysRevB.104.054305} {\bibfield  {journal} {\bibinfo  {journal} {Phys. Rev. B}\ }\textbf {\bibinfo {volume} {104}},\ \bibinfo {pages} {054305} (\bibinfo {year} {2021})}\BibitemShut {NoStop}%
\bibitem [{\citenamefont {Ptok}\ \emph {et~al.}(2023)\citenamefont {Ptok}, \citenamefont {Meier}, \citenamefont {Kobia\l{}ka}, \citenamefont {Basak}, \citenamefont {Sternik}, \citenamefont {\L{}a\.{z}ewski}, \citenamefont {Jochym}, \citenamefont {McGuire}, \citenamefont {Sales}, \citenamefont {Miao}, \citenamefont {Piekarz},\ and\ \citenamefont {Ole\'{s}}}]{ptok.meier.23}%
  \BibitemOpen
  \bibfield  {author} {\bibinfo {author} {\bibfnamefont {A.}~\bibnamefont {Ptok}}, \bibinfo {author} {\bibfnamefont {W.~R.}\ \bibnamefont {Meier}}, \bibinfo {author} {\bibfnamefont {A.}~\bibnamefont {Kobia\l{}ka}}, \bibinfo {author} {\bibfnamefont {S.}~\bibnamefont {Basak}}, \bibinfo {author} {\bibfnamefont {M.}~\bibnamefont {Sternik}}, \bibinfo {author} {\bibfnamefont {J.}~\bibnamefont {\L{}a\.{z}ewski}}, \bibinfo {author} {\bibfnamefont {P.~T.}\ \bibnamefont {Jochym}}, \bibinfo {author} {\bibfnamefont {M.~A.}\ \bibnamefont {McGuire}}, \bibinfo {author} {\bibfnamefont {B.~C.}\ \bibnamefont {Sales}}, \bibinfo {author} {\bibfnamefont {H.}~\bibnamefont {Miao}}, \bibinfo {author} {\bibfnamefont {P.}~\bibnamefont {Piekarz}},\ and\ \bibinfo {author} {\bibfnamefont {A.~M.}\ \bibnamefont {Ole\'{s}}},\ }\bibfield  {title} {\bibinfo {title} {Phononic drumhead surface state in the distorted kagome compound rhpb},\ }\href {https://doi.org/10.1103/PhysRevResearch.5.043231} {\bibfield  {journal} {\bibinfo  {journal}
  {Phys. Rev. Res.}\ }\textbf {\bibinfo {volume} {5}},\ \bibinfo {pages} {043231} (\bibinfo {year} {2023})}\BibitemShut {NoStop}%
\bibitem [{\citenamefont {Smidman}\ \emph {et~al.}(2017)\citenamefont {Smidman}, \citenamefont {Salamon}, \citenamefont {Yuan},\ and\ \citenamefont {Agterberg}}]{smidman.salamon.17}%
  \BibitemOpen
  \bibfield  {author} {\bibinfo {author} {\bibfnamefont {M.}~\bibnamefont {Smidman}}, \bibinfo {author} {\bibfnamefont {M.~B.}\ \bibnamefont {Salamon}}, \bibinfo {author} {\bibfnamefont {H.~Q.}\ \bibnamefont {Yuan}},\ and\ \bibinfo {author} {\bibfnamefont {D.~F.}\ \bibnamefont {Agterberg}},\ }\bibfield  {title} {\bibinfo {title} {Superconductivity and spin--orbit coupling in non-centrosymmetric materials: a review},\ }\href {https://doi.org/10.1088/1361-6633/80/3/036501} {\bibfield  {journal} {\bibinfo  {journal} {Rep. Prog. Phys.}\ }\textbf {\bibinfo {volume} {80}},\ \bibinfo {pages} {036501} (\bibinfo {year} {2017})}\BibitemShut {NoStop}%
\bibitem [{\citenamefont {Shang}\ \emph {et~al.}(2022{\natexlab{a}})\citenamefont {Shang}, \citenamefont {Tay}, \citenamefont {Su}, \citenamefont {Yuan},\ and\ \citenamefont {Shiroka}}]{shang.tay.22}%
  \BibitemOpen
  \bibfield  {author} {\bibinfo {author} {\bibfnamefont {T.}~\bibnamefont {Shang}}, \bibinfo {author} {\bibfnamefont {D.}~\bibnamefont {Tay}}, \bibinfo {author} {\bibfnamefont {H.}~\bibnamefont {Su}}, \bibinfo {author} {\bibfnamefont {H.~Q.}\ \bibnamefont {Yuan}},\ and\ \bibinfo {author} {\bibfnamefont {T.}~\bibnamefont {Shiroka}},\ }\bibfield  {title} {\bibinfo {title} {Evidence of fully gapped superconductivity in {NbReSi}: A combined {$\mu$SR} and {NMR} study},\ }\href {https://doi.org/10.1103/PhysRevB.105.144506} {\bibfield  {journal} {\bibinfo  {journal} {Phys. Rev. B}\ }\textbf {\bibinfo {volume} {105}},\ \bibinfo {pages} {144506} (\bibinfo {year} {2022}{\natexlab{a}})}\BibitemShut {NoStop}%
\bibitem [{\citenamefont {P.}\ \emph {et~al.}(2022)\citenamefont {P.}, \citenamefont {Motla}, \citenamefont {Meena}, \citenamefont {Kataria}, \citenamefont {Patra}, \citenamefont {K.}, \citenamefont {Hillier},\ and\ \citenamefont {Singh}}]{sajilesh.motla.22}%
  \BibitemOpen
  \bibfield  {author} {\bibinfo {author} {\bibfnamefont {S.~K.}\ \bibnamefont {P.}}, \bibinfo {author} {\bibfnamefont {K.}~\bibnamefont {Motla}}, \bibinfo {author} {\bibfnamefont {P.~K.}\ \bibnamefont {Meena}}, \bibinfo {author} {\bibfnamefont {A.}~\bibnamefont {Kataria}}, \bibinfo {author} {\bibfnamefont {C.}~\bibnamefont {Patra}}, \bibinfo {author} {\bibfnamefont {S.}~\bibnamefont {K.}}, \bibinfo {author} {\bibfnamefont {A.~D.}\ \bibnamefont {Hillier}},\ and\ \bibinfo {author} {\bibfnamefont {R.~P.}\ \bibnamefont {Singh}},\ }\bibfield  {title} {\bibinfo {title} {Superconductivity in noncentrosymmetric {NbReSi} investigated by muon spin rotation and relaxation},\ }\href {https://doi.org/10.1103/PhysRevB.105.094523} {\bibfield  {journal} {\bibinfo  {journal} {Phys. Rev. B}\ }\textbf {\bibinfo {volume} {105}},\ \bibinfo {pages} {094523} (\bibinfo {year} {2022})}\BibitemShut {NoStop}%
\bibitem [{\citenamefont {Nandi}\ \emph {et~al.}(2023)\citenamefont {Nandi}, \citenamefont {Sasmal}, \citenamefont {Maity}, \citenamefont {Sharma}, \citenamefont {Dwari}, \citenamefont {Kulkarni},\ and\ \citenamefont {Thamizhavel}}]{nandi.sasmal.23}%
  \BibitemOpen
  \bibfield  {author} {\bibinfo {author} {\bibfnamefont {S.}~\bibnamefont {Nandi}}, \bibinfo {author} {\bibfnamefont {S.}~\bibnamefont {Sasmal}}, \bibinfo {author} {\bibfnamefont {B.~B.}\ \bibnamefont {Maity}}, \bibinfo {author} {\bibfnamefont {V.}~\bibnamefont {Sharma}}, \bibinfo {author} {\bibfnamefont {G.}~\bibnamefont {Dwari}}, \bibinfo {author} {\bibfnamefont {R.}~\bibnamefont {Kulkarni}},\ and\ \bibinfo {author} {\bibfnamefont {A.}~\bibnamefont {Thamizhavel}},\ }\bibfield  {title} {\bibinfo {title} {Anisotropic properties of a noncentrosymmetric {NbReSi} superconducting single crystal},\ }\href {https://doi.org/10.1103/PhysRevB.107.134518} {\bibfield  {journal} {\bibinfo  {journal} {Phys. Rev. B}\ }\textbf {\bibinfo {volume} {107}},\ \bibinfo {pages} {134518} (\bibinfo {year} {2023})}\BibitemShut {NoStop}%
\bibitem [{\citenamefont {P.}\ and\ \citenamefont {Singh}(2021)}]{sajilesh.singh.21}%
  \BibitemOpen
  \bibfield  {author} {\bibinfo {author} {\bibfnamefont {S.~K.}\ \bibnamefont {P.}}\ and\ \bibinfo {author} {\bibfnamefont {R.~P.}\ \bibnamefont {Singh}},\ }\bibfield  {title} {\bibinfo {title} {Superconducting properties of the non-centrosymmetric superconductors {Ta$X$Si} ({$X=$Re, Ru})},\ }\href {https://doi.org/10.1088/1361-6668/abe4b7} {\bibfield  {journal} {\bibinfo  {journal} {Supercond. Sci. Technol.}\ }\textbf {\bibinfo {volume} {34}},\ \bibinfo {pages} {055003} (\bibinfo {year} {2021})}\BibitemShut {NoStop}%
\bibitem [{\citenamefont {Shang}\ \emph {et~al.}(2022{\natexlab{b}})\citenamefont {Shang}, \citenamefont {Zhao}, \citenamefont {Hu}, \citenamefont {Ma}, \citenamefont {Gawryluk}, \citenamefont {Zhu}, \citenamefont {Zhang}, \citenamefont {Zhen}, \citenamefont {Yu}, \citenamefont {Xu}, \citenamefont {Zhan}, \citenamefont {Pomjakushina}, \citenamefont {Shi},\ and\ \citenamefont {Shiroka}}]{shang.zhao.22}%
  \BibitemOpen
  \bibfield  {author} {\bibinfo {author} {\bibfnamefont {T.}~\bibnamefont {Shang}}, \bibinfo {author} {\bibfnamefont {J.}~\bibnamefont {Zhao}}, \bibinfo {author} {\bibfnamefont {L.-H.}\ \bibnamefont {Hu}}, \bibinfo {author} {\bibfnamefont {J.}~\bibnamefont {Ma}}, \bibinfo {author} {\bibfnamefont {D.~J.}\ \bibnamefont {Gawryluk}}, \bibinfo {author} {\bibfnamefont {X.}~\bibnamefont {Zhu}}, \bibinfo {author} {\bibfnamefont {H.}~\bibnamefont {Zhang}}, \bibinfo {author} {\bibfnamefont {Z.}~\bibnamefont {Zhen}}, \bibinfo {author} {\bibfnamefont {B.}~\bibnamefont {Yu}}, \bibinfo {author} {\bibfnamefont {Y.}~\bibnamefont {Xu}}, \bibinfo {author} {\bibfnamefont {Q.}~\bibnamefont {Zhan}}, \bibinfo {author} {\bibfnamefont {E.}~\bibnamefont {Pomjakushina}}, \bibinfo {author} {\bibfnamefont {M.}~\bibnamefont {Shi}},\ and\ \bibinfo {author} {\bibfnamefont {T.}~\bibnamefont {Shiroka}},\ }\bibfield  {title} {\bibinfo {title} {Unconventional superconductivity in topological {Kramers} nodal-line semimetals},\ }\href
  {https://doi.org/10.1126/sciadv.abq6589} {\bibfield  {journal} {\bibinfo  {journal} {Sci. Adv.}\ }\textbf {\bibinfo {volume} {8}},\ \bibinfo {pages} {eabq6589} (\bibinfo {year} {2022}{\natexlab{b}})}\BibitemShut {NoStop}%
\bibitem [{\citenamefont {Sharma}\ \emph {et~al.}(2023)\citenamefont {Sharma}, \citenamefont {K.~P.}, \citenamefont {Richards}, \citenamefont {Gautreau}, \citenamefont {Pula}, \citenamefont {Beare}, \citenamefont {Kojima}, \citenamefont {Yoon}, \citenamefont {Cai}, \citenamefont {Kushwaha}, \citenamefont {Agarwal}, \citenamefont {S\o{}rensen}, \citenamefont {Singh},\ and\ \citenamefont {Luke}}]{sharma.sajilesh.23}%
  \BibitemOpen
  \bibfield  {author} {\bibinfo {author} {\bibfnamefont {S.}~\bibnamefont {Sharma}}, \bibinfo {author} {\bibfnamefont {S.}~\bibnamefont {K.~P.}}, \bibinfo {author} {\bibfnamefont {A.~D.~S.}\ \bibnamefont {Richards}}, \bibinfo {author} {\bibfnamefont {J.}~\bibnamefont {Gautreau}}, \bibinfo {author} {\bibfnamefont {M.}~\bibnamefont {Pula}}, \bibinfo {author} {\bibfnamefont {J.}~\bibnamefont {Beare}}, \bibinfo {author} {\bibfnamefont {K.~M.}\ \bibnamefont {Kojima}}, \bibinfo {author} {\bibfnamefont {S.}~\bibnamefont {Yoon}}, \bibinfo {author} {\bibfnamefont {Y.}~\bibnamefont {Cai}}, \bibinfo {author} {\bibfnamefont {R.~K.}\ \bibnamefont {Kushwaha}}, \bibinfo {author} {\bibfnamefont {T.}~\bibnamefont {Agarwal}}, \bibinfo {author} {\bibfnamefont {E.~S.}\ \bibnamefont {S\o{}rensen}}, \bibinfo {author} {\bibfnamefont {R.~P.}\ \bibnamefont {Singh}},\ and\ \bibinfo {author} {\bibfnamefont {G.~M.}\ \bibnamefont {Luke}},\ }\bibfield  {title} {\bibinfo {title} {Evidence for nonunitary triplet-pairing superconductivity in
  noncentrosymmetric {TaRuSi} and comparison with isostructural {TaReSi}},\ }\href {https://doi.org/10.1103/PhysRevB.108.144510} {\bibfield  {journal} {\bibinfo  {journal} {Phys. Rev. B}\ }\textbf {\bibinfo {volume} {108}},\ \bibinfo {pages} {144510} (\bibinfo {year} {2023})}\BibitemShut {NoStop}%
\bibitem [{\citenamefont {Balents}(2010)}]{balents.10}%
  \BibitemOpen
  \bibfield  {author} {\bibinfo {author} {\bibfnamefont {L.}~\bibnamefont {Balents}},\ }\bibfield  {title} {\bibinfo {title} {Spin liquids in frustrated magnets},\ }\href {https://doi.org/10.1038/nature08917} {\bibfield  {journal} {\bibinfo  {journal} {Nature}\ }\textbf {\bibinfo {volume} {464}},\ \bibinfo {pages} {199} (\bibinfo {year} {2010})}\BibitemShut {NoStop}%
\bibitem [{\citenamefont {Rau}\ and\ \citenamefont {Kee}(2011)}]{rau.kee.11}%
  \BibitemOpen
  \bibfield  {author} {\bibinfo {author} {\bibfnamefont {J.~G.}\ \bibnamefont {Rau}}\ and\ \bibinfo {author} {\bibfnamefont {H.-Y.}\ \bibnamefont {Kee}},\ }\bibfield  {title} {\bibinfo {title} {Hidden spin liquid in an antiferromagnet: Applications to {FeCrAs}},\ }\href {https://doi.org/10.1103/PhysRevB.84.104448} {\bibfield  {journal} {\bibinfo  {journal} {Phys. Rev. B}\ }\textbf {\bibinfo {volume} {84}},\ \bibinfo {pages} {104448} (\bibinfo {year} {2011})}\BibitemShut {NoStop}%
\bibitem [{\citenamefont {Florez}\ \emph {et~al.}(2013)\citenamefont {Florez}, \citenamefont {Vargas}, \citenamefont {Garcia},\ and\ \citenamefont {Ross}}]{florez.vargas.13}%
  \BibitemOpen
  \bibfield  {author} {\bibinfo {author} {\bibfnamefont {J.~M.}\ \bibnamefont {Florez}}, \bibinfo {author} {\bibfnamefont {P.}~\bibnamefont {Vargas}}, \bibinfo {author} {\bibfnamefont {C.}~\bibnamefont {Garcia}},\ and\ \bibinfo {author} {\bibfnamefont {C.~A.}\ \bibnamefont {Ross}},\ }\bibfield  {title} {\bibinfo {title} {Magnetic entropy change plateau in a geometrically frustrated layered system: {FeCrAs}-like iron-pnictide structure as a magnetocaloric prototype},\ }\href {https://doi.org/10.1088/0953-8984/25/22/226004} {\bibfield  {journal} {\bibinfo  {journal} {J. Phys.: Condens. Matter}\ }\textbf {\bibinfo {volume} {25}},\ \bibinfo {pages} {226004} (\bibinfo {year} {2013})}\BibitemShut {NoStop}%
\bibitem [{\citenamefont {Jin}\ \emph {et~al.}(2019)\citenamefont {Jin}, \citenamefont {Meven}, \citenamefont {Deng}, \citenamefont {Su}, \citenamefont {Wu}, \citenamefont {Julian},\ and\ \citenamefont {Kim}}]{jin.meven.19}%
  \BibitemOpen
  \bibfield  {author} {\bibinfo {author} {\bibfnamefont {W.~T.}\ \bibnamefont {Jin}}, \bibinfo {author} {\bibfnamefont {M.}~\bibnamefont {Meven}}, \bibinfo {author} {\bibfnamefont {H.}~\bibnamefont {Deng}}, \bibinfo {author} {\bibfnamefont {Y.}~\bibnamefont {Su}}, \bibinfo {author} {\bibfnamefont {W.}~\bibnamefont {Wu}}, \bibinfo {author} {\bibfnamefont {S.~R.}\ \bibnamefont {Julian}},\ and\ \bibinfo {author} {\bibfnamefont {Y.-J.}\ \bibnamefont {Kim}},\ }\bibfield  {title} {\bibinfo {title} {Spin reorientation in {FeCrAs} revealed by single-crystal neutron diffraction},\ }\href {https://doi.org/10.1103/PhysRevB.100.174421} {\bibfield  {journal} {\bibinfo  {journal} {Phys. Rev. B}\ }\textbf {\bibinfo {volume} {100}},\ \bibinfo {pages} {174421} (\bibinfo {year} {2019})}\BibitemShut {NoStop}%
\bibitem [{\citenamefont {Huang}\ \emph {et~al.}(2023)\citenamefont {Huang}, \citenamefont {Jeschke},\ and\ \citenamefont {Mazin}}]{huang.jeschke.23}%
  \BibitemOpen
  \bibfield  {author} {\bibinfo {author} {\bibfnamefont {Y.~N.}\ \bibnamefont {Huang}}, \bibinfo {author} {\bibfnamefont {H.~O.}\ \bibnamefont {Jeschke}},\ and\ \bibinfo {author} {\bibfnamefont {I.~I.}\ \bibnamefont {Mazin}},\ }\bibfield  {title} {\bibinfo {title} {{CrRhAs}: a member of a large family of metallic kagome antiferromagnets},\ }\href {https://doi.org/10.1038/s41535-023-00562-x} {\bibfield  {journal} {\bibinfo  {journal} {npj Quantum Mater.}\ }\textbf {\bibinfo {volume} {8}},\ \bibinfo {pages} {32} (\bibinfo {year} {2023})}\BibitemShut {NoStop}%
\bibitem [{\citenamefont {Szyma\'{n}ski}\ \emph {et~al.}(2023)\citenamefont {Szyma\'{n}ski}, \citenamefont {Zach}, \citenamefont {Tobola}, \citenamefont {Chajec}, \citenamefont {Duraj}, \citenamefont {Gondek}, \citenamefont {Baran}, \citenamefont {Michalec}, \citenamefont {Chaudouet}, \citenamefont {Haj-Khlifa}, \citenamefont {Hlil},\ and\ \citenamefont {Fruchart}}]{szymanski.zach.23}%
  \BibitemOpen
  \bibfield  {author} {\bibinfo {author} {\bibfnamefont {D.}~\bibnamefont {Szyma\'{n}ski}}, \bibinfo {author} {\bibfnamefont {R.}~\bibnamefont {Zach}}, \bibinfo {author} {\bibfnamefont {J.}~\bibnamefont {Tobola}}, \bibinfo {author} {\bibfnamefont {W.}~\bibnamefont {Chajec}}, \bibinfo {author} {\bibfnamefont {R.}~\bibnamefont {Duraj}}, \bibinfo {author} {\bibfnamefont {L.}~\bibnamefont {Gondek}}, \bibinfo {author} {\bibfnamefont {S.}~\bibnamefont {Baran}}, \bibinfo {author} {\bibfnamefont {M.}~\bibnamefont {Michalec}}, \bibinfo {author} {\bibfnamefont {P.}~\bibnamefont {Chaudouet}}, \bibinfo {author} {\bibfnamefont {S.}~\bibnamefont {Haj-Khlifa}}, \bibinfo {author} {\bibfnamefont {E.}~\bibnamefont {Hlil}},\ and\ \bibinfo {author} {\bibfnamefont {D.}~\bibnamefont {Fruchart}},\ }\bibfield  {title} {\bibinfo {title} {Review: On the complex magnetic phase diagram of the {MnRu$_{x}$Rh$_{1-x}$As} system,crystal, a.c. susceptibility, magnetization and electronic structure characterizations},\ }\href
  {https://doi.org/10.1016/j.jallcom.2022.168602} {\bibfield  {journal} {\bibinfo  {journal} {J. Alloys Compd.}\ }\textbf {\bibinfo {volume} {938}},\ \bibinfo {pages} {168602} (\bibinfo {year} {2023})}\BibitemShut {NoStop}%
\bibitem [{\citenamefont {Kanomata}\ \emph {et~al.}(1991)\citenamefont {Kanomata}, \citenamefont {Kawashima}, \citenamefont {Utsugi}, \citenamefont {Goto}, \citenamefont {Hasegawa},\ and\ \citenamefont {Kaneko}}]{kanomata.kawashima.91}%
  \BibitemOpen
  \bibfield  {author} {\bibinfo {author} {\bibfnamefont {T.}~\bibnamefont {Kanomata}}, \bibinfo {author} {\bibfnamefont {T.}~\bibnamefont {Kawashima}}, \bibinfo {author} {\bibfnamefont {H.}~\bibnamefont {Utsugi}}, \bibinfo {author} {\bibfnamefont {T.}~\bibnamefont {Goto}}, \bibinfo {author} {\bibfnamefont {H.}~\bibnamefont {Hasegawa}},\ and\ \bibinfo {author} {\bibfnamefont {T.}~\bibnamefont {Kaneko}},\ }\bibfield  {title} {\bibinfo {title} {{Magnetic properties of the intermetallic compounds {MM'X} ({M=Cr,Mn}, {M'=Ru,Rh,Pd}, and {X=P,As})}},\ }\href {https://doi.org/10.1063/1.348281} {\bibfield  {journal} {\bibinfo  {journal} {J. Appl. Phys.}\ }\textbf {\bibinfo {volume} {69}},\ \bibinfo {pages} {4639} (\bibinfo {year} {1991})}\BibitemShut {NoStop}%
\bibitem [{\citenamefont {Kaneko}\ \emph {et~al.}(1992)\citenamefont {Kaneko}, \citenamefont {Kanomata}, \citenamefont {Kawashima}, \citenamefont {Mori}, \citenamefont {Miura},\ and\ \citenamefont {Nakagawa}}]{kaneko.kanomata.92}%
  \BibitemOpen
  \bibfield  {author} {\bibinfo {author} {\bibfnamefont {T.}~\bibnamefont {Kaneko}}, \bibinfo {author} {\bibfnamefont {T.}~\bibnamefont {Kanomata}}, \bibinfo {author} {\bibfnamefont {T.}~\bibnamefont {Kawashima}}, \bibinfo {author} {\bibfnamefont {S.}~\bibnamefont {Mori}}, \bibinfo {author} {\bibfnamefont {S.}~\bibnamefont {Miura}},\ and\ \bibinfo {author} {\bibfnamefont {Y.}~\bibnamefont {Nakagawa}},\ }\bibfield  {title} {\bibinfo {title} {High-field magnetization in intermetallic compounds {MM'X} ({M=Mn, Cr};{M'=Ru, Rh, Pd}; {X=As, P})},\ }\href {https://doi.org/10.1016/0921-4526(92)90080-C} {\bibfield  {journal} {\bibinfo  {journal} {Phys. B}\ }\textbf {\bibinfo {volume} {177}},\ \bibinfo {pages} {123} (\bibinfo {year} {1992})}\BibitemShut {NoStop}%
\bibitem [{\citenamefont {Wang}(2023{\natexlab{a}})}]{wang.23}%
  \BibitemOpen
  \bibfield  {author} {\bibinfo {author} {\bibfnamefont {Y.}~\bibnamefont {Wang}},\ }\bibfield  {title} {\bibinfo {title} {Electronic correlation effects on stabilizing a perfect kagome lattice and ferromagnetic fluctuation in {LaRu$_3$Si$_2$}},\ }\href {https://doi.org/10.52396/JUSTC-2022-0182} {\bibfield  {journal} {\bibinfo  {journal} {JUSTC}\ }\textbf {\bibinfo {volume} {53}},\ \bibinfo {pages} {0702} (\bibinfo {year} {2023}{\natexlab{a}})}\BibitemShut {NoStop}%
\bibitem [{\citenamefont {Teng}\ \emph {et~al.}(2023)\citenamefont {Teng}, \citenamefont {Oh}, \citenamefont {Tan}, \citenamefont {Chen}, \citenamefont {Huang}, \citenamefont {Gao}, \citenamefont {Yin}, \citenamefont {Chu}, \citenamefont {Hashimoto}, \citenamefont {Lu}, \citenamefont {Jozwiak}, \citenamefont {Bostwick}, \citenamefont {Rotenberg}, \citenamefont {Granroth}, \citenamefont {Yan}, \citenamefont {Birgeneau}, \citenamefont {Dai},\ and\ \citenamefont {Yi}}]{teng.oh.23}%
  \BibitemOpen
  \bibfield  {author} {\bibinfo {author} {\bibfnamefont {X.}~\bibnamefont {Teng}}, \bibinfo {author} {\bibfnamefont {J.~S.}\ \bibnamefont {Oh}}, \bibinfo {author} {\bibfnamefont {H.}~\bibnamefont {Tan}}, \bibinfo {author} {\bibfnamefont {L.}~\bibnamefont {Chen}}, \bibinfo {author} {\bibfnamefont {J.}~\bibnamefont {Huang}}, \bibinfo {author} {\bibfnamefont {B.}~\bibnamefont {Gao}}, \bibinfo {author} {\bibfnamefont {J.-X.}\ \bibnamefont {Yin}}, \bibinfo {author} {\bibfnamefont {J.-H.}\ \bibnamefont {Chu}}, \bibinfo {author} {\bibfnamefont {M.}~\bibnamefont {Hashimoto}}, \bibinfo {author} {\bibfnamefont {D.}~\bibnamefont {Lu}}, \bibinfo {author} {\bibfnamefont {C.}~\bibnamefont {Jozwiak}}, \bibinfo {author} {\bibfnamefont {A.}~\bibnamefont {Bostwick}}, \bibinfo {author} {\bibfnamefont {E.}~\bibnamefont {Rotenberg}}, \bibinfo {author} {\bibfnamefont {G.~E.}\ \bibnamefont {Granroth}}, \bibinfo {author} {\bibfnamefont {B.}~\bibnamefont {Yan}}, \bibinfo {author} {\bibfnamefont {R.~J.}\ \bibnamefont {Birgeneau}},
  \bibinfo {author} {\bibfnamefont {P.}~\bibnamefont {Dai}},\ and\ \bibinfo {author} {\bibfnamefont {M.}~\bibnamefont {Yi}},\ }\bibfield  {title} {\bibinfo {title} {Magnetism and charge density wave order in kagome {FeGe}},\ }\href {https://doi.org/10.1038/s41567-023-01985-w} {\bibfield  {journal} {\bibinfo  {journal} {Nat. Phys.}\ }\textbf {\bibinfo {volume} {19}},\ \bibinfo {pages} {814} (\bibinfo {year} {2023})}\BibitemShut {NoStop}%
\bibitem [{\citenamefont {Wang}(2023{\natexlab{b}})}]{wang.23f}%
  \BibitemOpen
  \bibfield  {author} {\bibinfo {author} {\bibfnamefont {Y.}~\bibnamefont {Wang}},\ }\bibfield  {title} {\bibinfo {title} {Enhanced spin-polarization via partial {Ge}-dimerization as the driving force of the charge density wave in {FeGe}},\ }\href {https://doi.org/10.1103/PhysRevMaterials.7.104006} {\bibfield  {journal} {\bibinfo  {journal} {Phys. Rev. Mater.}\ }\textbf {\bibinfo {volume} {7}},\ \bibinfo {pages} {104006} (\bibinfo {year} {2023}{\natexlab{b}})}\BibitemShut {NoStop}%
\bibitem [{\citenamefont {Ptok}\ \emph {et~al.}(2024)\citenamefont {Ptok}, \citenamefont {Basak}, \citenamefont {Kobia\l{}ka}, \citenamefont {Sternik}, \citenamefont {\L{}a\.{z}ewski}, \citenamefont {Jochym}, \citenamefont {Ole\'{s}},\ and\ \citenamefont {Piekarz}}]{ptok.basak.24}%
  \BibitemOpen
  \bibfield  {author} {\bibinfo {author} {\bibfnamefont {A.}~\bibnamefont {Ptok}}, \bibinfo {author} {\bibfnamefont {S.}~\bibnamefont {Basak}}, \bibinfo {author} {\bibfnamefont {A.}~\bibnamefont {Kobia\l{}ka}}, \bibinfo {author} {\bibfnamefont {M.}~\bibnamefont {Sternik}}, \bibinfo {author} {\bibfnamefont {J.}~\bibnamefont {\L{}a\.{z}ewski}}, \bibinfo {author} {\bibfnamefont {P.~T.}\ \bibnamefont {Jochym}}, \bibinfo {author} {\bibfnamefont {A.~M.}\ \bibnamefont {Ole\'{s}}},\ and\ \bibinfo {author} {\bibfnamefont {P.}~\bibnamefont {Piekarz}},\ }\bibfield  {title} {\bibinfo {title} {Lattice dynamics study of electron-correlation-induced charge density wave in antiferromagnetic kagome metal {FeGe}},\ }\href {https://doi.org/10.1103/PhysRevMaterials.8.L080601} {\bibfield  {journal} {\bibinfo  {journal} {Phys. Rev. Mater.}\ }\textbf {\bibinfo {volume} {8}},\ \bibinfo {pages} {L080601} (\bibinfo {year} {2024})}\BibitemShut {NoStop}%
\bibitem [{\citenamefont {Bl\"ochl}(1994)}]{blochl.94}%
  \BibitemOpen
  \bibfield  {author} {\bibinfo {author} {\bibfnamefont {P.~E.}\ \bibnamefont {Bl\"ochl}},\ }\bibfield  {title} {\bibinfo {title} {Projector augmented-wave method},\ }\href {https://doi.org/10.1103/PhysRevB.50.17953} {\bibfield  {journal} {\bibinfo  {journal} {Phys. Rev. B}\ }\textbf {\bibinfo {volume} {50}},\ \bibinfo {pages} {17953} (\bibinfo {year} {1994})}\BibitemShut {NoStop}%
\bibitem [{\citenamefont {Kresse}\ and\ \citenamefont {Hafner}(1994)}]{kresse.hafner.94}%
  \BibitemOpen
  \bibfield  {author} {\bibinfo {author} {\bibfnamefont {G.}~\bibnamefont {Kresse}}\ and\ \bibinfo {author} {\bibfnamefont {J.}~\bibnamefont {Hafner}},\ }\bibfield  {title} {\bibinfo {title} {Ab initio molecular-dynamics simulation of the liquid-metal--amorphous-semiconductor transition in germanium},\ }\href {https://doi.org/10.1103/PhysRevB.49.14251} {\bibfield  {journal} {\bibinfo  {journal} {Phys. Rev. B}\ }\textbf {\bibinfo {volume} {49}},\ \bibinfo {pages} {14251} (\bibinfo {year} {1994})}\BibitemShut {NoStop}%
\bibitem [{\citenamefont {Kresse}\ and\ \citenamefont {Furthm\"uller}(1996)}]{kresse.furthmuller.96}%
  \BibitemOpen
  \bibfield  {author} {\bibinfo {author} {\bibfnamefont {G.}~\bibnamefont {Kresse}}\ and\ \bibinfo {author} {\bibfnamefont {J.}~\bibnamefont {Furthm\"uller}},\ }\bibfield  {title} {\bibinfo {title} {Efficient iterative schemes for ab initio total-energy calculations using a plane-wave basis set},\ }\href {https://doi.org/10.1103/PhysRevB.54.11169} {\bibfield  {journal} {\bibinfo  {journal} {Phys. Rev. B}\ }\textbf {\bibinfo {volume} {54}},\ \bibinfo {pages} {11169} (\bibinfo {year} {1996})}\BibitemShut {NoStop}%
\bibitem [{\citenamefont {Kresse}\ and\ \citenamefont {Joubert}(1999)}]{kresse.joubert.99}%
  \BibitemOpen
  \bibfield  {author} {\bibinfo {author} {\bibfnamefont {G.}~\bibnamefont {Kresse}}\ and\ \bibinfo {author} {\bibfnamefont {D.}~\bibnamefont {Joubert}},\ }\bibfield  {title} {\bibinfo {title} {From ultrasoft pseudopotentials to the projector augmented-wave method},\ }\href {https://doi.org/10.1103/PhysRevB.59.1758} {\bibfield  {journal} {\bibinfo  {journal} {Phys. Rev. B}\ }\textbf {\bibinfo {volume} {59}},\ \bibinfo {pages} {1758} (\bibinfo {year} {1999})}\BibitemShut {NoStop}%
\bibitem [{\citenamefont {Perdew}\ \emph {et~al.}(1996)\citenamefont {Perdew}, \citenamefont {Burke},\ and\ \citenamefont {Ernzerhof}}]{perdew.burke.96}%
  \BibitemOpen
  \bibfield  {author} {\bibinfo {author} {\bibfnamefont {J.~P.}\ \bibnamefont {Perdew}}, \bibinfo {author} {\bibfnamefont {K.}~\bibnamefont {Burke}},\ and\ \bibinfo {author} {\bibfnamefont {M.}~\bibnamefont {Ernzerhof}},\ }\bibfield  {title} {\bibinfo {title} {Generalized gradient approximation made simple},\ }\href {https://doi.org/10.1103/PhysRevLett.77.3865} {\bibfield  {journal} {\bibinfo  {journal} {Phys. Rev. Lett.}\ }\textbf {\bibinfo {volume} {77}},\ \bibinfo {pages} {3865} (\bibinfo {year} {1996})}\BibitemShut {NoStop}%
\bibitem [{\citenamefont {Dudarev}\ \emph {et~al.}(1998)\citenamefont {Dudarev}, \citenamefont {Botton}, \citenamefont {Savrasov}, \citenamefont {Humphreys},\ and\ \citenamefont {Sutton}}]{dudarev.botton.98}%
  \BibitemOpen
  \bibfield  {author} {\bibinfo {author} {\bibfnamefont {S.~L.}\ \bibnamefont {Dudarev}}, \bibinfo {author} {\bibfnamefont {G.~A.}\ \bibnamefont {Botton}}, \bibinfo {author} {\bibfnamefont {S.~Y.}\ \bibnamefont {Savrasov}}, \bibinfo {author} {\bibfnamefont {C.~J.}\ \bibnamefont {Humphreys}},\ and\ \bibinfo {author} {\bibfnamefont {A.~P.}\ \bibnamefont {Sutton}},\ }\bibfield  {title} {\bibinfo {title} {Electron-energy-loss spectra and the structural stability of nickel oxide: An {LSDA+U} study},\ }\href {https://doi.org/10.1103/PhysRevB.57.1505} {\bibfield  {journal} {\bibinfo  {journal} {Phys. Rev. B}\ }\textbf {\bibinfo {volume} {57}},\ \bibinfo {pages} {1505} (\bibinfo {year} {1998})}\BibitemShut {NoStop}%
\bibitem [{\citenamefont {Shishkin}\ and\ \citenamefont {Sato}(2019)}]{shishkin.sato.19}%
  \BibitemOpen
  \bibfield  {author} {\bibinfo {author} {\bibfnamefont {M.}~\bibnamefont {Shishkin}}\ and\ \bibinfo {author} {\bibfnamefont {H.}~\bibnamefont {Sato}},\ }\bibfield  {title} {\bibinfo {title} {{DFT+U in Dudarev’s formulation with corrected interactions between the electrons with opposite spins: The form of Hamiltonian, calculation of forces, and bandgap adjustments}},\ }\href {https://doi.org/10.1063/1.5090445} {\bibfield  {journal} {\bibinfo  {journal} {The Journal of Chemical Physics}\ }\textbf {\bibinfo {volume} {151}},\ \bibinfo {pages} {024102} (\bibinfo {year} {2019})}\BibitemShut {NoStop}%
\bibitem [{\citenamefont {Monkhorst}\ and\ \citenamefont {Pack}(1976)}]{monkhorst.pack.76}%
  \BibitemOpen
  \bibfield  {author} {\bibinfo {author} {\bibfnamefont {H.~J.}\ \bibnamefont {Monkhorst}}\ and\ \bibinfo {author} {\bibfnamefont {J.~D.}\ \bibnamefont {Pack}},\ }\bibfield  {title} {\bibinfo {title} {Special points for {Brillouin}-zone integrations},\ }\href {https://doi.org/10.1103/PhysRevB.13.5188} {\bibfield  {journal} {\bibinfo  {journal} {Phys. Rev. B}\ }\textbf {\bibinfo {volume} {13}},\ \bibinfo {pages} {5188} (\bibinfo {year} {1976})}\BibitemShut {NoStop}%
\bibitem [{\citenamefont {Stokes}\ and\ \citenamefont {Hatch}(2005)}]{stokes.hatch.05}%
  \BibitemOpen
  \bibfield  {author} {\bibinfo {author} {\bibfnamefont {H.~T.}\ \bibnamefont {Stokes}}\ and\ \bibinfo {author} {\bibfnamefont {D.~M.}\ \bibnamefont {Hatch}},\ }\bibfield  {title} {\bibinfo {title} {{{\sc FindSym}: program for identifying the space-group symmetry of a crystal}},\ }\href {https://doi.org/10.1107/S0021889804031528} {\bibfield  {journal} {\bibinfo  {journal} {J. Appl. Cryst.}\ }\textbf {\bibinfo {volume} {38}},\ \bibinfo {pages} {237} (\bibinfo {year} {2005})}\BibitemShut {NoStop}%
\bibitem [{\citenamefont {Togo}\ and\ \citenamefont {Tanaka}(2018)}]{togo.tanaka.18}%
  \BibitemOpen
  \bibfield  {author} {\bibinfo {author} {\bibfnamefont {A.}~\bibnamefont {Togo}}\ and\ \bibinfo {author} {\bibfnamefont {I.}~\bibnamefont {Tanaka}},\ }\href@noop {} {\bibinfo {title} {{\sc Spglib}: a software library for crystal symmetry search}} (\bibinfo {year} {2018}),\ \Eprint {https://arxiv.org/abs/arXiv:1808.01590} {arXiv:1808.01590} \BibitemShut {NoStop}%
\bibitem [{\citenamefont {Hinuma}\ \emph {et~al.}(2017)\citenamefont {Hinuma}, \citenamefont {Pizzi}, \citenamefont {Kumagai}, \citenamefont {Oba},\ and\ \citenamefont {Tanaka}}]{hinuma.pizzi.17}%
  \BibitemOpen
  \bibfield  {author} {\bibinfo {author} {\bibfnamefont {Y.}~\bibnamefont {Hinuma}}, \bibinfo {author} {\bibfnamefont {G.}~\bibnamefont {Pizzi}}, \bibinfo {author} {\bibfnamefont {Y.}~\bibnamefont {Kumagai}}, \bibinfo {author} {\bibfnamefont {F.}~\bibnamefont {Oba}},\ and\ \bibinfo {author} {\bibfnamefont {I.}~\bibnamefont {Tanaka}},\ }\bibfield  {title} {\bibinfo {title} {Band structure diagram paths based on crystallography},\ }\href {https://doi.org/10.1016/j.commatsci.2016.10.015} {\bibfield  {journal} {\bibinfo  {journal} {Comput. Mater. Sci.}\ }\textbf {\bibinfo {volume} {128}},\ \bibinfo {pages} {140} (\bibinfo {year} {2017})}\BibitemShut {NoStop}%
\bibitem [{\citenamefont {Kawamura}(2019)}]{kawamura.19}%
  \BibitemOpen
  \bibfield  {author} {\bibinfo {author} {\bibfnamefont {M.}~\bibnamefont {Kawamura}},\ }\bibfield  {title} {\bibinfo {title} {{FermiSurfer}: {Fermi}-surface viewer providing multiple representation schemes},\ }\href {https://doi.org/10.1016/j.cpc.2019.01.017} {\bibfield  {journal} {\bibinfo  {journal} {Comput. Phys. Commun.}\ }\textbf {\bibinfo {volume} {239}},\ \bibinfo {pages} {197} (\bibinfo {year} {2019})}\BibitemShut {NoStop}%
\bibitem [{\citenamefont {Ganose}\ \emph {et~al.}(2021)\citenamefont {Ganose}, \citenamefont {Searle}, \citenamefont {Jain},\ and\ \citenamefont {Griffin}}]{ganose.searle.21}%
  \BibitemOpen
  \bibfield  {author} {\bibinfo {author} {\bibfnamefont {A.~M.}\ \bibnamefont {Ganose}}, \bibinfo {author} {\bibfnamefont {A.}~\bibnamefont {Searle}}, \bibinfo {author} {\bibfnamefont {A.}~\bibnamefont {Jain}},\ and\ \bibinfo {author} {\bibfnamefont {S.~M.}\ \bibnamefont {Griffin}},\ }\bibfield  {title} {\bibinfo {title} {{IFermi}: A python library for {Fermi} surface generation and analysis},\ }\href {https://doi.org/10.21105/joss.03089} {\bibfield  {journal} {\bibinfo  {journal} {J. Open Source Softw.}\ }\textbf {\bibinfo {volume} {6}},\ \bibinfo {pages} {3089} (\bibinfo {year} {2021})}\BibitemShut {NoStop}%
\bibitem [{\citenamefont {Parlinski}\ \emph {et~al.}(1997)\citenamefont {Parlinski}, \citenamefont {Li},\ and\ \citenamefont {Kawazoe}}]{parlinski.li.97}%
  \BibitemOpen
  \bibfield  {author} {\bibinfo {author} {\bibfnamefont {K.}~\bibnamefont {Parlinski}}, \bibinfo {author} {\bibfnamefont {Z.~Q.}\ \bibnamefont {Li}},\ and\ \bibinfo {author} {\bibfnamefont {Y.}~\bibnamefont {Kawazoe}},\ }\bibfield  {title} {\bibinfo {title} {First-principles determination of the soft mode in cubic {ZrO$_{2}$}},\ }\href {https://doi.org/10.1103/PhysRevLett.78.4063} {\bibfield  {journal} {\bibinfo  {journal} {Phys. Rev. Lett.}\ }\textbf {\bibinfo {volume} {78}},\ \bibinfo {pages} {4063} (\bibinfo {year} {1997})}\BibitemShut {NoStop}%
\bibitem [{\citenamefont {Togo}\ \emph {et~al.}(2023)\citenamefont {Togo}, \citenamefont {Chaput}, \citenamefont {Tadano},\ and\ \citenamefont {Tanaka}}]{togo.chaput.23}%
  \BibitemOpen
  \bibfield  {author} {\bibinfo {author} {\bibfnamefont {A.}~\bibnamefont {Togo}}, \bibinfo {author} {\bibfnamefont {L.}~\bibnamefont {Chaput}}, \bibinfo {author} {\bibfnamefont {T.}~\bibnamefont {Tadano}},\ and\ \bibinfo {author} {\bibfnamefont {I.}~\bibnamefont {Tanaka}},\ }\bibfield  {title} {\bibinfo {title} {Implementation strategies in phonopy and phono3py},\ }\href {https://doi.org/10.1088/1361-648X/acd831} {\bibfield  {journal} {\bibinfo  {journal} {J. Phys. Condens. Matter}\ }\textbf {\bibinfo {volume} {35}},\ \bibinfo {pages} {353001} (\bibinfo {year} {2023})}\BibitemShut {NoStop}%
\bibitem [{\citenamefont {Togo}(2023)}]{togo.23}%
  \BibitemOpen
  \bibfield  {author} {\bibinfo {author} {\bibfnamefont {A.}~\bibnamefont {Togo}},\ }\bibfield  {title} {\bibinfo {title} {First-principles phonon calculations with phonopy and phono3py},\ }\href {https://doi.org/10.7566/JPSJ.92.012001} {\bibfield  {journal} {\bibinfo  {journal} {J. Phys. Soc. Jpn.}\ }\textbf {\bibinfo {volume} {92}},\ \bibinfo {pages} {012001} (\bibinfo {year} {2023})}\BibitemShut {NoStop}%
\bibitem [{\citenamefont {Marzari}\ \emph {et~al.}(2012)\citenamefont {Marzari}, \citenamefont {Mostofi}, \citenamefont {Yates}, \citenamefont {Souza},\ and\ \citenamefont {Vanderbilt}}]{marzari.mostofi.12}%
  \BibitemOpen
  \bibfield  {author} {\bibinfo {author} {\bibfnamefont {N.}~\bibnamefont {Marzari}}, \bibinfo {author} {\bibfnamefont {A.~A.}\ \bibnamefont {Mostofi}}, \bibinfo {author} {\bibfnamefont {J.~R.}\ \bibnamefont {Yates}}, \bibinfo {author} {\bibfnamefont {I.}~\bibnamefont {Souza}},\ and\ \bibinfo {author} {\bibfnamefont {D.}~\bibnamefont {Vanderbilt}},\ }\bibfield  {title} {\bibinfo {title} {Maximally localized {Wannier} functions: Theory and applications},\ }\href {https://doi.org/10.1103/RevModPhys.84.1419} {\bibfield  {journal} {\bibinfo  {journal} {Rev. Mod. Phys.}\ }\textbf {\bibinfo {volume} {84}},\ \bibinfo {pages} {1419} (\bibinfo {year} {2012})}\BibitemShut {NoStop}%
\bibitem [{\citenamefont {Marzari}\ and\ \citenamefont {Vanderbilt}(1997)}]{marzari.vanderbilt.97}%
  \BibitemOpen
  \bibfield  {author} {\bibinfo {author} {\bibfnamefont {N.}~\bibnamefont {Marzari}}\ and\ \bibinfo {author} {\bibfnamefont {D.}~\bibnamefont {Vanderbilt}},\ }\bibfield  {title} {\bibinfo {title} {Maximally localized generalized {Wannier} functions for composite energy bands},\ }\href {https://doi.org/10.1103/PhysRevB.56.12847} {\bibfield  {journal} {\bibinfo  {journal} {Phys. Rev. B}\ }\textbf {\bibinfo {volume} {56}},\ \bibinfo {pages} {12847} (\bibinfo {year} {1997})}\BibitemShut {NoStop}%
\bibitem [{\citenamefont {Souza}\ \emph {et~al.}(2001)\citenamefont {Souza}, \citenamefont {Marzari},\ and\ \citenamefont {Vanderbilt}}]{souza.marzari.01}%
  \BibitemOpen
  \bibfield  {author} {\bibinfo {author} {\bibfnamefont {I.}~\bibnamefont {Souza}}, \bibinfo {author} {\bibfnamefont {N.}~\bibnamefont {Marzari}},\ and\ \bibinfo {author} {\bibfnamefont {D.}~\bibnamefont {Vanderbilt}},\ }\bibfield  {title} {\bibinfo {title} {Maximally localized {Wannier} functions for entangled energy bands},\ }\href {https://doi.org/10.1103/PhysRevB.65.035109} {\bibfield  {journal} {\bibinfo  {journal} {Phys. Rev. B}\ }\textbf {\bibinfo {volume} {65}},\ \bibinfo {pages} {035109} (\bibinfo {year} {2001})}\BibitemShut {NoStop}%
\bibitem [{\citenamefont {Mostofi}\ \emph {et~al.}(2008)\citenamefont {Mostofi}, \citenamefont {Yates}, \citenamefont {Lee}, \citenamefont {Souza}, \citenamefont {Vanderbilt},\ and\ \citenamefont {Marzari}}]{mostofi.yates.08}%
  \BibitemOpen
  \bibfield  {author} {\bibinfo {author} {\bibfnamefont {A.~A.}\ \bibnamefont {Mostofi}}, \bibinfo {author} {\bibfnamefont {J.~R.}\ \bibnamefont {Yates}}, \bibinfo {author} {\bibfnamefont {Y.-S.}\ \bibnamefont {Lee}}, \bibinfo {author} {\bibfnamefont {I.}~\bibnamefont {Souza}}, \bibinfo {author} {\bibfnamefont {D.}~\bibnamefont {Vanderbilt}},\ and\ \bibinfo {author} {\bibfnamefont {N.}~\bibnamefont {Marzari}},\ }\bibfield  {title} {\bibinfo {title} {{\sc wannier90}: A tool for obtaining maximally-localised wannier functions},\ }\href {https://doi.org/10.1016/j.cpc.2007.11.016} {\bibfield  {journal} {\bibinfo  {journal} {Comput. Phys. Commun.}\ }\textbf {\bibinfo {volume} {178}},\ \bibinfo {pages} {685} (\bibinfo {year} {2008})}\BibitemShut {NoStop}%
\bibitem [{\citenamefont {Mostofi}\ \emph {et~al.}(2014)\citenamefont {Mostofi}, \citenamefont {Yates}, \citenamefont {Pizzi}, \citenamefont {Lee}, \citenamefont {Souza}, \citenamefont {Vanderbilt},\ and\ \citenamefont {Marzari}}]{mostofi.yates.14}%
  \BibitemOpen
  \bibfield  {author} {\bibinfo {author} {\bibfnamefont {A.~A.}\ \bibnamefont {Mostofi}}, \bibinfo {author} {\bibfnamefont {J.~R.}\ \bibnamefont {Yates}}, \bibinfo {author} {\bibfnamefont {G.}~\bibnamefont {Pizzi}}, \bibinfo {author} {\bibfnamefont {Y.-S.}\ \bibnamefont {Lee}}, \bibinfo {author} {\bibfnamefont {I.}~\bibnamefont {Souza}}, \bibinfo {author} {\bibfnamefont {D.}~\bibnamefont {Vanderbilt}},\ and\ \bibinfo {author} {\bibfnamefont {N.}~\bibnamefont {Marzari}},\ }\bibfield  {title} {\bibinfo {title} {An updated version of {\sc wannier90}: A tool for obtaining maximally-localised wannier functions},\ }\href {https://doi.org/10.1016/j.cpc.2014.05.003} {\bibfield  {journal} {\bibinfo  {journal} {Comput. Phys. Commun.}\ }\textbf {\bibinfo {volume} {185}},\ \bibinfo {pages} {2309} (\bibinfo {year} {2014})}\BibitemShut {NoStop}%
\bibitem [{\citenamefont {Pizzi}\ \emph {et~al.}(2020)\citenamefont {Pizzi}, \citenamefont {Vitale}, \citenamefont {Arita}, \citenamefont {Bl\"{u}gel}, \citenamefont {Freimuth}, \citenamefont {G{\'{e}}ranton}, \citenamefont {Gibertini}, \citenamefont {Gresch}, \citenamefont {Johnson}, \citenamefont {Koretsune}, \citenamefont {Iba{\~{n}}ez-Azpiroz}, \citenamefont {Lee}, \citenamefont {Lihm}, \citenamefont {Marchand}, \citenamefont {Marrazzo}, \citenamefont {Mokrousov}, \citenamefont {Mustafa}, \citenamefont {Nohara}, \citenamefont {Nomura}, \citenamefont {Paulatto}, \citenamefont {Ponc{\'{e}}}, \citenamefont {Ponweiser}, \citenamefont {Qiao}, \citenamefont {Th\"{o}le}, \citenamefont {Tsirkin}, \citenamefont {Wierzbowska}, \citenamefont {Marzari}, \citenamefont {Vanderbilt}, \citenamefont {Souza}, \citenamefont {Mostofi},\ and\ \citenamefont {Yates}}]{pizzi.vitale.20}%
  \BibitemOpen
  \bibfield  {author} {\bibinfo {author} {\bibfnamefont {G.}~\bibnamefont {Pizzi}}, \bibinfo {author} {\bibfnamefont {V.}~\bibnamefont {Vitale}}, \bibinfo {author} {\bibfnamefont {R.}~\bibnamefont {Arita}}, \bibinfo {author} {\bibfnamefont {S.}~\bibnamefont {Bl\"{u}gel}}, \bibinfo {author} {\bibfnamefont {F.}~\bibnamefont {Freimuth}}, \bibinfo {author} {\bibfnamefont {G.}~\bibnamefont {G{\'{e}}ranton}}, \bibinfo {author} {\bibfnamefont {M.}~\bibnamefont {Gibertini}}, \bibinfo {author} {\bibfnamefont {D.}~\bibnamefont {Gresch}}, \bibinfo {author} {\bibfnamefont {C.}~\bibnamefont {Johnson}}, \bibinfo {author} {\bibfnamefont {T.}~\bibnamefont {Koretsune}}, \bibinfo {author} {\bibfnamefont {J.}~\bibnamefont {Iba{\~{n}}ez-Azpiroz}}, \bibinfo {author} {\bibfnamefont {H.}~\bibnamefont {Lee}}, \bibinfo {author} {\bibfnamefont {J.-M.}\ \bibnamefont {Lihm}}, \bibinfo {author} {\bibfnamefont {D.}~\bibnamefont {Marchand}}, \bibinfo {author} {\bibfnamefont {A.}~\bibnamefont {Marrazzo}}, \bibinfo {author} {\bibfnamefont
  {Y.}~\bibnamefont {Mokrousov}}, \bibinfo {author} {\bibfnamefont {J.~I.}\ \bibnamefont {Mustafa}}, \bibinfo {author} {\bibfnamefont {Y.}~\bibnamefont {Nohara}}, \bibinfo {author} {\bibfnamefont {Y.}~\bibnamefont {Nomura}}, \bibinfo {author} {\bibfnamefont {L.}~\bibnamefont {Paulatto}}, \bibinfo {author} {\bibfnamefont {S.}~\bibnamefont {Ponc{\'{e}}}}, \bibinfo {author} {\bibfnamefont {T.}~\bibnamefont {Ponweiser}}, \bibinfo {author} {\bibfnamefont {J.}~\bibnamefont {Qiao}}, \bibinfo {author} {\bibfnamefont {F.}~\bibnamefont {Th\"{o}le}}, \bibinfo {author} {\bibfnamefont {S.~S.}\ \bibnamefont {Tsirkin}}, \bibinfo {author} {\bibfnamefont {M.}~\bibnamefont {Wierzbowska}}, \bibinfo {author} {\bibfnamefont {N.}~\bibnamefont {Marzari}}, \bibinfo {author} {\bibfnamefont {D.}~\bibnamefont {Vanderbilt}}, \bibinfo {author} {\bibfnamefont {I.}~\bibnamefont {Souza}}, \bibinfo {author} {\bibfnamefont {A.~A.}\ \bibnamefont {Mostofi}},\ and\ \bibinfo {author} {\bibfnamefont {J.~R.}\ \bibnamefont {Yates}},\ }\bibfield
  {title} {\bibinfo {title} {{\sc Wannier90} as a community code: new features and applications},\ }\href {https://doi.org/10.1088/1361-648x/ab51ff} {\bibfield  {journal} {\bibinfo  {journal} {J. Phys.: Condens. Matter}\ }\textbf {\bibinfo {volume} {32}},\ \bibinfo {pages} {165902} (\bibinfo {year} {2020})}\BibitemShut {NoStop}%
\bibitem [{\citenamefont {Sancho}\ \emph {et~al.}(1985)\citenamefont {Sancho}, \citenamefont {Sancho}, \citenamefont {Sancho},\ and\ \citenamefont {Rubio}}]{sancho.sancho.85}%
  \BibitemOpen
  \bibfield  {author} {\bibinfo {author} {\bibfnamefont {M.~P.~L.}\ \bibnamefont {Sancho}}, \bibinfo {author} {\bibfnamefont {J.~M.~L.}\ \bibnamefont {Sancho}}, \bibinfo {author} {\bibfnamefont {J.~M.~L.}\ \bibnamefont {Sancho}},\ and\ \bibinfo {author} {\bibfnamefont {J.}~\bibnamefont {Rubio}},\ }\bibfield  {title} {\bibinfo {title} {Highly convergent schemes for the calculation of bulk and surface {Green} functions},\ }\href {https://doi.org/10.1088/0305-4608/15/4/009} {\bibfield  {journal} {\bibinfo  {journal} {J. Phys. F: Met. Phys.}\ }\textbf {\bibinfo {volume} {15}},\ \bibinfo {pages} {851} (\bibinfo {year} {1985})}\BibitemShut {NoStop}%
\bibitem [{\citenamefont {Wu}\ \emph {et~al.}(2018)\citenamefont {Wu}, \citenamefont {Zhang}, \citenamefont {Song}, \citenamefont {Troyer},\ and\ \citenamefont {Soluyanov}}]{wu.zhang.18}%
  \BibitemOpen
  \bibfield  {author} {\bibinfo {author} {\bibfnamefont {Q.~S.}\ \bibnamefont {Wu}}, \bibinfo {author} {\bibfnamefont {S.~N.}\ \bibnamefont {Zhang}}, \bibinfo {author} {\bibfnamefont {H.-F.}\ \bibnamefont {Song}}, \bibinfo {author} {\bibfnamefont {M.}~\bibnamefont {Troyer}},\ and\ \bibinfo {author} {\bibfnamefont {A.~A.}\ \bibnamefont {Soluyanov}},\ }\bibfield  {title} {\bibinfo {title} {{\sc WannierTools}: An open-source software package for novel topological materials},\ }\href {https://doi.org/10.1016/j.cpc.2017.09.033} {\bibfield  {journal} {\bibinfo  {journal} {Comput. Phys. Commun.}\ }\textbf {\bibinfo {volume} {224}},\ \bibinfo {pages} {405} (\bibinfo {year} {2018})}\BibitemShut {NoStop}%
\bibitem [{\citenamefont {He}\ \emph {et~al.}(2021)\citenamefont {He}, \citenamefont {Helbig}, \citenamefont {Verstraete},\ and\ \citenamefont {Bousquet}}]{he.helbig.21}%
  \BibitemOpen
  \bibfield  {author} {\bibinfo {author} {\bibfnamefont {X.}~\bibnamefont {He}}, \bibinfo {author} {\bibfnamefont {N.}~\bibnamefont {Helbig}}, \bibinfo {author} {\bibfnamefont {M.~J.}\ \bibnamefont {Verstraete}},\ and\ \bibinfo {author} {\bibfnamefont {E.}~\bibnamefont {Bousquet}},\ }\bibfield  {title} {\bibinfo {title} {{TB2J}: A python package for computing magnetic interaction parameters},\ }\href {https://doi.org/10.1016/j.cpc.2021.107938} {\bibfield  {journal} {\bibinfo  {journal} {Comput. Phys. Commun.}\ }\textbf {\bibinfo {volume} {264}},\ \bibinfo {pages} {107938} (\bibinfo {year} {2021})}\BibitemShut {NoStop}%
\bibitem [{Note1()}]{Note1}%
  \BibitemOpen
  \bibinfo {note} {See Supplemental Material at [URL will be inserted by publisher] for additional theoretical results, including Ref.\cite {cococcioni.gironcoli.05}.}\BibitemShut {Stop}%
\bibitem [{\citenamefont {Otrokov}\ \emph {et~al.}(2019)\citenamefont {Otrokov}, \citenamefont {Klimovskikh}, \citenamefont {Bentmann}, \citenamefont {Estyunin}, \citenamefont {Zeugner}, \citenamefont {Aliev}, \citenamefont {Ga{\ss}}, \citenamefont {Wolter}, \citenamefont {Koroleva}, \citenamefont {Shikin}, \citenamefont {Blanco-Rey}, \citenamefont {Hoffmann}, \citenamefont {Rusinov}, \citenamefont {Vyazovskaya}, \citenamefont {Eremeev}, \citenamefont {Koroteev}, \citenamefont {Kuznetsov}, \citenamefont {Freyse}, \citenamefont {S{\'a}nchez-Barriga}, \citenamefont {Amiraslanov}, \citenamefont {Babanly}, \citenamefont {Mamedov}, \citenamefont {Abdullayev}, \citenamefont {Zverev}, \citenamefont {Alfonsov}, \citenamefont {Kataev}, \citenamefont {B{\"u}chner}, \citenamefont {Schwier}, \citenamefont {Kumar}, \citenamefont {Kimura}, \citenamefont {Petaccia}, \citenamefont {Di~Santo}, \citenamefont {Vidal}, \citenamefont {Schatz}, \citenamefont {Ki{\ss}ner}, \citenamefont {{\"U}nzelmann}, \citenamefont {Min},
  \citenamefont {Moser}, \citenamefont {Peixoto}, \citenamefont {Reinert}, \citenamefont {Ernst}, \citenamefont {Echenique}, \citenamefont {Isaeva},\ and\ \citenamefont {Chulkov}}]{otrokov.klimoviskih.19}%
  \BibitemOpen
  \bibfield  {author} {\bibinfo {author} {\bibfnamefont {M.~M.}\ \bibnamefont {Otrokov}}, \bibinfo {author} {\bibfnamefont {I.~I.}\ \bibnamefont {Klimovskikh}}, \bibinfo {author} {\bibfnamefont {H.}~\bibnamefont {Bentmann}}, \bibinfo {author} {\bibfnamefont {D.}~\bibnamefont {Estyunin}}, \bibinfo {author} {\bibfnamefont {A.}~\bibnamefont {Zeugner}}, \bibinfo {author} {\bibfnamefont {Z.~S.}\ \bibnamefont {Aliev}}, \bibinfo {author} {\bibfnamefont {S.}~\bibnamefont {Ga{\ss}}}, \bibinfo {author} {\bibfnamefont {A.~U.~B.}\ \bibnamefont {Wolter}}, \bibinfo {author} {\bibfnamefont {A.~V.}\ \bibnamefont {Koroleva}}, \bibinfo {author} {\bibfnamefont {A.~M.}\ \bibnamefont {Shikin}}, \bibinfo {author} {\bibfnamefont {M.}~\bibnamefont {Blanco-Rey}}, \bibinfo {author} {\bibfnamefont {M.}~\bibnamefont {Hoffmann}}, \bibinfo {author} {\bibfnamefont {I.~P.}\ \bibnamefont {Rusinov}}, \bibinfo {author} {\bibfnamefont {A.~Y.}\ \bibnamefont {Vyazovskaya}}, \bibinfo {author} {\bibfnamefont {S.~V.}\ \bibnamefont {Eremeev}},
  \bibinfo {author} {\bibfnamefont {Y.~M.}\ \bibnamefont {Koroteev}}, \bibinfo {author} {\bibfnamefont {V.~M.}\ \bibnamefont {Kuznetsov}}, \bibinfo {author} {\bibfnamefont {F.}~\bibnamefont {Freyse}}, \bibinfo {author} {\bibfnamefont {J.}~\bibnamefont {S{\'a}nchez-Barriga}}, \bibinfo {author} {\bibfnamefont {I.~R.}\ \bibnamefont {Amiraslanov}}, \bibinfo {author} {\bibfnamefont {M.~B.}\ \bibnamefont {Babanly}}, \bibinfo {author} {\bibfnamefont {N.~T.}\ \bibnamefont {Mamedov}}, \bibinfo {author} {\bibfnamefont {N.~A.}\ \bibnamefont {Abdullayev}}, \bibinfo {author} {\bibfnamefont {V.~N.}\ \bibnamefont {Zverev}}, \bibinfo {author} {\bibfnamefont {A.}~\bibnamefont {Alfonsov}}, \bibinfo {author} {\bibfnamefont {V.}~\bibnamefont {Kataev}}, \bibinfo {author} {\bibfnamefont {B.}~\bibnamefont {B{\"u}chner}}, \bibinfo {author} {\bibfnamefont {E.~F.}\ \bibnamefont {Schwier}}, \bibinfo {author} {\bibfnamefont {S.}~\bibnamefont {Kumar}}, \bibinfo {author} {\bibfnamefont {A.}~\bibnamefont {Kimura}}, \bibinfo {author}
  {\bibfnamefont {L.}~\bibnamefont {Petaccia}}, \bibinfo {author} {\bibfnamefont {G.}~\bibnamefont {Di~Santo}}, \bibinfo {author} {\bibfnamefont {R.~C.}\ \bibnamefont {Vidal}}, \bibinfo {author} {\bibfnamefont {S.}~\bibnamefont {Schatz}}, \bibinfo {author} {\bibfnamefont {K.}~\bibnamefont {Ki{\ss}ner}}, \bibinfo {author} {\bibfnamefont {M.}~\bibnamefont {{\"U}nzelmann}}, \bibinfo {author} {\bibfnamefont {C.~H.}\ \bibnamefont {Min}}, \bibinfo {author} {\bibfnamefont {S.}~\bibnamefont {Moser}}, \bibinfo {author} {\bibfnamefont {T.~R.~F.}\ \bibnamefont {Peixoto}}, \bibinfo {author} {\bibfnamefont {F.}~\bibnamefont {Reinert}}, \bibinfo {author} {\bibfnamefont {A.}~\bibnamefont {Ernst}}, \bibinfo {author} {\bibfnamefont {P.~M.}\ \bibnamefont {Echenique}}, \bibinfo {author} {\bibfnamefont {A.}~\bibnamefont {Isaeva}},\ and\ \bibinfo {author} {\bibfnamefont {E.~V.}\ \bibnamefont {Chulkov}},\ }\bibfield  {title} {\bibinfo {title} {Prediction and observation of an antiferromagnetic topological insulator},\ }\href
  {https://doi.org/10.1038/s41586-019-1840-9} {\bibfield  {journal} {\bibinfo  {journal} {Nature}\ }\textbf {\bibinfo {volume} {576}},\ \bibinfo {pages} {416} (\bibinfo {year} {2019})}\BibitemShut {NoStop}%
\bibitem [{\citenamefont {Klimovskikh}\ \emph {et~al.}(2020)\citenamefont {Klimovskikh}, \citenamefont {Otrokov}, \citenamefont {Estyunin}, \citenamefont {Eremeev}, \citenamefont {Filnov}, \citenamefont {Koroleva}, \citenamefont {Shevchenko}, \citenamefont {Voroshnin}, \citenamefont {Rybkin}, \citenamefont {Rusinov}, \citenamefont {Blanco-Rey}, \citenamefont {Hoffmann}, \citenamefont {Aliev}, \citenamefont {Babanly}, \citenamefont {Amiraslanov}, \citenamefont {Abdullayev}, \citenamefont {Zverev}, \citenamefont {Kimura}, \citenamefont {Tereshchenko}, \citenamefont {Kokh}, \citenamefont {Petaccia}, \citenamefont {Di~Santo}, \citenamefont {Ernst}, \citenamefont {Echenique}, \citenamefont {Mamedov}, \citenamefont {Shikin},\ and\ \citenamefont {Chulkov}}]{klimovskikh.otrokov.20}%
  \BibitemOpen
  \bibfield  {author} {\bibinfo {author} {\bibfnamefont {I.~I.}\ \bibnamefont {Klimovskikh}}, \bibinfo {author} {\bibfnamefont {M.~M.}\ \bibnamefont {Otrokov}}, \bibinfo {author} {\bibfnamefont {D.}~\bibnamefont {Estyunin}}, \bibinfo {author} {\bibfnamefont {S.~V.}\ \bibnamefont {Eremeev}}, \bibinfo {author} {\bibfnamefont {S.~O.}\ \bibnamefont {Filnov}}, \bibinfo {author} {\bibfnamefont {A.}~\bibnamefont {Koroleva}}, \bibinfo {author} {\bibfnamefont {E.}~\bibnamefont {Shevchenko}}, \bibinfo {author} {\bibfnamefont {V.}~\bibnamefont {Voroshnin}}, \bibinfo {author} {\bibfnamefont {A.~G.}\ \bibnamefont {Rybkin}}, \bibinfo {author} {\bibfnamefont {I.~P.}\ \bibnamefont {Rusinov}}, \bibinfo {author} {\bibfnamefont {M.}~\bibnamefont {Blanco-Rey}}, \bibinfo {author} {\bibfnamefont {M.}~\bibnamefont {Hoffmann}}, \bibinfo {author} {\bibfnamefont {Z.~S.}\ \bibnamefont {Aliev}}, \bibinfo {author} {\bibfnamefont {M.~B.}\ \bibnamefont {Babanly}}, \bibinfo {author} {\bibfnamefont {I.~R.}\ \bibnamefont {Amiraslanov}},
  \bibinfo {author} {\bibfnamefont {N.~A.}\ \bibnamefont {Abdullayev}}, \bibinfo {author} {\bibfnamefont {V.~N.}\ \bibnamefont {Zverev}}, \bibinfo {author} {\bibfnamefont {A.}~\bibnamefont {Kimura}}, \bibinfo {author} {\bibfnamefont {O.~E.}\ \bibnamefont {Tereshchenko}}, \bibinfo {author} {\bibfnamefont {K.~A.}\ \bibnamefont {Kokh}}, \bibinfo {author} {\bibfnamefont {L.}~\bibnamefont {Petaccia}}, \bibinfo {author} {\bibfnamefont {G.}~\bibnamefont {Di~Santo}}, \bibinfo {author} {\bibfnamefont {A.}~\bibnamefont {Ernst}}, \bibinfo {author} {\bibfnamefont {P.~M.}\ \bibnamefont {Echenique}}, \bibinfo {author} {\bibfnamefont {N.~T.}\ \bibnamefont {Mamedov}}, \bibinfo {author} {\bibfnamefont {A.~M.}\ \bibnamefont {Shikin}},\ and\ \bibinfo {author} {\bibfnamefont {E.~V.}\ \bibnamefont {Chulkov}},\ }\bibfield  {title} {\bibinfo {title} {Tunable {3D/2D} magnetism in the {(MnBi$_{2}$Te$_{4}$)(Bi$_{2}$Te$_{3}$)$_{m}$} topological insulators family},\ }\href {https://doi.org/10.1038/s41535-020-00255-9} {\bibfield
  {journal} {\bibinfo  {journal} {npj Quantum Materials}\ }\textbf {\bibinfo {volume} {5}},\ \bibinfo {pages} {54} (\bibinfo {year} {2020})}\BibitemShut {NoStop}%
\bibitem [{\citenamefont {Kobia\l{}ka}\ \emph {et~al.}(2022)\citenamefont {Kobia\l{}ka}, \citenamefont {Sternik},\ and\ \citenamefont {Ptok}}]{kobialka.sternik.22}%
  \BibitemOpen
  \bibfield  {author} {\bibinfo {author} {\bibfnamefont {A.}~\bibnamefont {Kobia\l{}ka}}, \bibinfo {author} {\bibfnamefont {M.}~\bibnamefont {Sternik}},\ and\ \bibinfo {author} {\bibfnamefont {A.}~\bibnamefont {Ptok}},\ }\bibfield  {title} {\bibinfo {title} {Dynamical properties of the magnetic topological insulator {$T$Bi$_{2}$Te$_{4}$} ({$T=$Mn, Fe}): Phonons dispersion, {Raman} active modes, and chiral phonons study},\ }\href {https://doi.org/10.1103/PhysRevB.105.214304} {\bibfield  {journal} {\bibinfo  {journal} {Phys. Rev. B}\ }\textbf {\bibinfo {volume} {105}},\ \bibinfo {pages} {214304} (\bibinfo {year} {2022})}\BibitemShut {NoStop}%
\bibitem [{\citenamefont {Eremeev}\ \emph {et~al.}(2022)\citenamefont {Eremeev}, \citenamefont {Otrokov}, \citenamefont {Ernst},\ and\ \citenamefont {Chulkov}}]{eremeev.otrokov.22}%
  \BibitemOpen
  \bibfield  {author} {\bibinfo {author} {\bibfnamefont {S.~V.}\ \bibnamefont {Eremeev}}, \bibinfo {author} {\bibfnamefont {M.~M.}\ \bibnamefont {Otrokov}}, \bibinfo {author} {\bibfnamefont {A.}~\bibnamefont {Ernst}},\ and\ \bibinfo {author} {\bibfnamefont {E.~V.}\ \bibnamefont {Chulkov}},\ }\bibfield  {title} {\bibinfo {title} {Magnetic ordering and topology in {Mn$_{2}$Bi$_{2}$Te$_{5}$} and {Mn$_{2}$Sb$_{2}$Te$_{5}$} van der {Waals} materials},\ }\href {https://doi.org/10.1103/PhysRevB.105.195105} {\bibfield  {journal} {\bibinfo  {journal} {Phys. Rev. B}\ }\textbf {\bibinfo {volume} {105}},\ \bibinfo {pages} {195105} (\bibinfo {year} {2022})}\BibitemShut {NoStop}%
\bibitem [{\citenamefont {Cococcioni}\ and\ \citenamefont {de~Gironcoli}(2005)}]{cococcioni.gironcoli.05}%
  \BibitemOpen
  \bibfield  {author} {\bibinfo {author} {\bibfnamefont {M.}~\bibnamefont {Cococcioni}}\ and\ \bibinfo {author} {\bibfnamefont {S.}~\bibnamefont {de~Gironcoli}},\ }\bibfield  {title} {\bibinfo {title} {Linear response approach to the calculation of the effective interaction parameters in the {LDA+U} method},\ }\href {https://doi.org/10.1103/PhysRevB.71.035105} {\bibfield  {journal} {\bibinfo  {journal} {Phys. Rev. B}\ }\textbf {\bibinfo {volume} {71}},\ \bibinfo {pages} {035105} (\bibinfo {year} {2005})}\BibitemShut {NoStop}%
\bibitem [{\citenamefont {Moore}\ \emph {et~al.}(2024)\citenamefont {Moore}, \citenamefont {Horton}, \citenamefont {Linscott}, \citenamefont {Ganose}, \citenamefont {Siron}, \citenamefont {O'Regan},\ and\ \citenamefont {Persson}}]{moore.horton.24}%
  \BibitemOpen
  \bibfield  {author} {\bibinfo {author} {\bibfnamefont {G.~C.}\ \bibnamefont {Moore}}, \bibinfo {author} {\bibfnamefont {M.~K.}\ \bibnamefont {Horton}}, \bibinfo {author} {\bibfnamefont {E.}~\bibnamefont {Linscott}}, \bibinfo {author} {\bibfnamefont {A.~M.}\ \bibnamefont {Ganose}}, \bibinfo {author} {\bibfnamefont {M.}~\bibnamefont {Siron}}, \bibinfo {author} {\bibfnamefont {D.~D.}\ \bibnamefont {O'Regan}},\ and\ \bibinfo {author} {\bibfnamefont {K.~A.}\ \bibnamefont {Persson}},\ }\bibfield  {title} {\bibinfo {title} {High-throughput determination of {Hubbard} {$U$} and {Hund} {$J$} values for transition metal oxides via the linear response formalism},\ }\href {https://doi.org/10.1103/PhysRevMaterials.8.014409} {\bibfield  {journal} {\bibinfo  {journal} {Phys. Rev. Mater.}\ }\textbf {\bibinfo {volume} {8}},\ \bibinfo {pages} {014409} (\bibinfo {year} {2024})}\BibitemShut {NoStop}%
\bibitem [{\citenamefont {Pajda}\ \emph {et~al.}(2001)\citenamefont {Pajda}, \citenamefont {Kudrnovsk\'y}, \citenamefont {Turek}, \citenamefont {Drchal},\ and\ \citenamefont {Bruno}}]{pajda.kudrnovsky.01}%
  \BibitemOpen
  \bibfield  {author} {\bibinfo {author} {\bibfnamefont {M.}~\bibnamefont {Pajda}}, \bibinfo {author} {\bibfnamefont {J.}~\bibnamefont {Kudrnovsk\'y}}, \bibinfo {author} {\bibfnamefont {I.}~\bibnamefont {Turek}}, \bibinfo {author} {\bibfnamefont {V.}~\bibnamefont {Drchal}},\ and\ \bibinfo {author} {\bibfnamefont {P.}~\bibnamefont {Bruno}},\ }\bibfield  {title} {\bibinfo {title} {Ab initio calculations of exchange interactions, spin-wave stiffness constants, and {Curie} temperatures of {Fe}, {Co}, and {Ni}},\ }\href {https://doi.org/10.1103/PhysRevB.64.174402} {\bibfield  {journal} {\bibinfo  {journal} {Phys. Rev. B}\ }\textbf {\bibinfo {volume} {64}},\ \bibinfo {pages} {174402} (\bibinfo {year} {2001})}\BibitemShut {NoStop}%
\bibitem [{\citenamefont {Chen}\ \emph {et~al.}(2022)\citenamefont {Chen}, \citenamefont {Mao}, \citenamefont {Chung}, \citenamefont {Stone}, \citenamefont {Kolesnikov}, \citenamefont {Wang}, \citenamefont {Murai}, \citenamefont {Gao}, \citenamefont {Delaire},\ and\ \citenamefont {Dai}}]{chen.mao.22}%
  \BibitemOpen
  \bibfield  {author} {\bibinfo {author} {\bibfnamefont {L.}~\bibnamefont {Chen}}, \bibinfo {author} {\bibfnamefont {C.}~\bibnamefont {Mao}}, \bibinfo {author} {\bibfnamefont {J.-H.}\ \bibnamefont {Chung}}, \bibinfo {author} {\bibfnamefont {M.~B.}\ \bibnamefont {Stone}}, \bibinfo {author} {\bibfnamefont {A.~I.}\ \bibnamefont {Kolesnikov}}, \bibinfo {author} {\bibfnamefont {X.}~\bibnamefont {Wang}}, \bibinfo {author} {\bibfnamefont {N.}~\bibnamefont {Murai}}, \bibinfo {author} {\bibfnamefont {B.}~\bibnamefont {Gao}}, \bibinfo {author} {\bibfnamefont {O.}~\bibnamefont {Delaire}},\ and\ \bibinfo {author} {\bibfnamefont {P.}~\bibnamefont {Dai}},\ }\bibfield  {title} {\bibinfo {title} {Anisotropic magnon damping by zero-temperature quantum fluctuations in ferromagnetic {CrGeTe$_{3}$}},\ }\href {https://doi.org/10.1038/s41467-022-31612-w} {\bibfield  {journal} {\bibinfo  {journal} {Nat. Commun.}\ }\textbf {\bibinfo {volume} {13}},\ \bibinfo {pages} {4037} (\bibinfo {year} {2022})}\BibitemShut {NoStop}%
\bibitem [{\citenamefont {Zakeri}\ \emph {et~al.}(2024)\citenamefont {Zakeri}, \citenamefont {von Faber},\ and\ \citenamefont {Ernst}}]{zakeri.vonfaber.24}%
  \BibitemOpen
  \bibfield  {author} {\bibinfo {author} {\bibfnamefont {K.}~\bibnamefont {Zakeri}}, \bibinfo {author} {\bibfnamefont {A.}~\bibnamefont {von Faber}},\ and\ \bibinfo {author} {\bibfnamefont {A.}~\bibnamefont {Ernst}},\ }\bibfield  {title} {\bibinfo {title} {Magnons and fundamental magnetic interactions in a ferromagnetic monolayer: The case of the ni monolayer},\ }\href {https://doi.org/10.1103/PhysRevB.109.L180406} {\bibfield  {journal} {\bibinfo  {journal} {Phys. Rev. B}\ }\textbf {\bibinfo {volume} {109}},\ \bibinfo {pages} {L180406} (\bibinfo {year} {2024})}\BibitemShut {NoStop}%
\bibitem [{\citenamefont {Zhuo}\ \emph {et~al.}(2022)\citenamefont {Zhuo}, \citenamefont {Li},\ and\ \citenamefont {Manchon}}]{zhuo.li.22}%
  \BibitemOpen
  \bibfield  {author} {\bibinfo {author} {\bibfnamefont {F.}~\bibnamefont {Zhuo}}, \bibinfo {author} {\bibfnamefont {H.}~\bibnamefont {Li}},\ and\ \bibinfo {author} {\bibfnamefont {A.}~\bibnamefont {Manchon}},\ }\bibfield  {title} {\bibinfo {title} {Topological thermal {Hall} effect and magnonic edge states in kagome ferromagnets with bond anisotropy},\ }\href {https://doi.org/10.1088/1367-2630/ac51a8} {\bibfield  {journal} {\bibinfo  {journal} {New J. Phys.}\ }\textbf {\bibinfo {volume} {24}},\ \bibinfo {pages} {023033} (\bibinfo {year} {2022})}\BibitemShut {NoStop}%
\bibitem [{\citenamefont {Wu}\ \emph {et~al.}(2015)\citenamefont {Wu}, \citenamefont {Yang}, \citenamefont {Le}, \citenamefont {Fan},\ and\ \citenamefont {Hu}}]{wu.yang.15}%
  \BibitemOpen
  \bibfield  {author} {\bibinfo {author} {\bibfnamefont {X.}~\bibnamefont {Wu}}, \bibinfo {author} {\bibfnamefont {F.}~\bibnamefont {Yang}}, \bibinfo {author} {\bibfnamefont {C.}~\bibnamefont {Le}}, \bibinfo {author} {\bibfnamefont {H.}~\bibnamefont {Fan}},\ and\ \bibinfo {author} {\bibfnamefont {J.}~\bibnamefont {Hu}},\ }\bibfield  {title} {\bibinfo {title} {Triplet ${p}_{z}$-wave pairing in quasi-one-dimensional {${A}_{2}$Cr$_{3}$As$_{3}$} superconductors ({$A=$K, Rb, Cs})},\ }\href {https://doi.org/10.1103/PhysRevB.92.104511} {\bibfield  {journal} {\bibinfo  {journal} {Phys. Rev. B}\ }\textbf {\bibinfo {volume} {92}},\ \bibinfo {pages} {104511} (\bibinfo {year} {2015})}\BibitemShut {NoStop}%
\bibitem [{\citenamefont {Jiang}\ \emph {et~al.}(2015)\citenamefont {Jiang}, \citenamefont {Cao},\ and\ \citenamefont {Cao}}]{jiang.cao.15}%
  \BibitemOpen
  \bibfield  {author} {\bibinfo {author} {\bibfnamefont {H.}~\bibnamefont {Jiang}}, \bibinfo {author} {\bibfnamefont {G.}~\bibnamefont {Cao}},\ and\ \bibinfo {author} {\bibfnamefont {C.}~\bibnamefont {Cao}},\ }\bibfield  {title} {\bibinfo {title} {Electronic structure of quasi-one-dimensional superconductor {K$_{2}$Cr$_{3}$As$_{3}$} from first-principles calculations},\ }\href {https://doi.org/10.1038/srep16054} {\bibfield  {journal} {\bibinfo  {journal} {Scie. Rep.}\ }\textbf {\bibinfo {volume} {5}},\ \bibinfo {pages} {16054} (\bibinfo {year} {2015})}\BibitemShut {NoStop}%
\bibitem [{\citenamefont {Yang}\ \emph {et~al.}(2019)\citenamefont {Yang}, \citenamefont {Feng}, \citenamefont {Lu}, \citenamefont {Wang},\ and\ \citenamefont {Chen}}]{yang.feng.19}%
  \BibitemOpen
  \bibfield  {author} {\bibinfo {author} {\bibfnamefont {Y.}~\bibnamefont {Yang}}, \bibinfo {author} {\bibfnamefont {S.-Q.}\ \bibnamefont {Feng}}, \bibinfo {author} {\bibfnamefont {H.-Y.}\ \bibnamefont {Lu}}, \bibinfo {author} {\bibfnamefont {W.-S.}\ \bibnamefont {Wang}},\ and\ \bibinfo {author} {\bibfnamefont {Z.-P.}\ \bibnamefont {Chen}},\ }\bibfield  {title} {\bibinfo {title} {Electronic structures of newly discovered quasi-one-dimensional superconductors {$A_{2}$Mo$_{3}$As$_{3}$} ({$A=$K, Rb, Cs})},\ }\href {https://doi.org/10.1007/s10948-019-5054-z} {\bibfield  {journal} {\bibinfo  {journal} {J. Supercond. Nov. Magn.}\ }\textbf {\bibinfo {volume} {32}},\ \bibinfo {pages} {2421} (\bibinfo {year} {2019})}\BibitemShut {NoStop}%
\bibitem [{\citenamefont {Xu}\ \emph {et~al.}(2020{\natexlab{b}})\citenamefont {Xu}, \citenamefont {Wu}, \citenamefont {Zhi}, \citenamefont {Lei}, \citenamefont {Duan}, \citenamefont {Ning}, \citenamefont {Cao},\ and\ \citenamefont {Chen}}]{xu.wu.20}%
  \BibitemOpen
  \bibfield  {author} {\bibinfo {author} {\bibfnamefont {C.}~\bibnamefont {Xu}}, \bibinfo {author} {\bibfnamefont {N.}~\bibnamefont {Wu}}, \bibinfo {author} {\bibfnamefont {G.-X.}\ \bibnamefont {Zhi}}, \bibinfo {author} {\bibfnamefont {B.-H.}\ \bibnamefont {Lei}}, \bibinfo {author} {\bibfnamefont {X.}~\bibnamefont {Duan}}, \bibinfo {author} {\bibfnamefont {F.}~\bibnamefont {Ning}}, \bibinfo {author} {\bibfnamefont {C.}~\bibnamefont {Cao}},\ and\ \bibinfo {author} {\bibfnamefont {Q.}~\bibnamefont {Chen}},\ }\bibfield  {title} {\bibinfo {title} {Coexistence of nontrivial topological properties and strong ferromagnetic fluctuations in quasi-one-dimensional {$A_{2}$Cr$_{3}$As$_{3}$}},\ }\href {https://doi.org/10.1038/s41524-020-0294-9} {\bibfield  {journal} {\bibinfo  {journal} {npj Comput. Mater.}\ }\textbf {\bibinfo {volume} {6}},\ \bibinfo {pages} {30} (\bibinfo {year} {2020}{\natexlab{b}})}\BibitemShut {NoStop}%
\bibitem [{\citenamefont {Taddei}\ \emph {et~al.}(2023)\citenamefont {Taddei}, \citenamefont {Lei}, \citenamefont {Susner}, \citenamefont {Zhai}, \citenamefont {Bullard}, \citenamefont {Sanjeewa}, \citenamefont {Zheng}, \citenamefont {Sefat}, \citenamefont {Chi}, \citenamefont {dela Cruz}, \citenamefont {Singh},\ and\ \citenamefont {Lv}}]{taddei.lei.23}%
  \BibitemOpen
  \bibfield  {author} {\bibinfo {author} {\bibfnamefont {K.~M.}\ \bibnamefont {Taddei}}, \bibinfo {author} {\bibfnamefont {B.-H.}\ \bibnamefont {Lei}}, \bibinfo {author} {\bibfnamefont {M.~A.}\ \bibnamefont {Susner}}, \bibinfo {author} {\bibfnamefont {H.-F.}\ \bibnamefont {Zhai}}, \bibinfo {author} {\bibfnamefont {T.~J.}\ \bibnamefont {Bullard}}, \bibinfo {author} {\bibfnamefont {L.~D.}\ \bibnamefont {Sanjeewa}}, \bibinfo {author} {\bibfnamefont {Q.}~\bibnamefont {Zheng}}, \bibinfo {author} {\bibfnamefont {A.~S.}\ \bibnamefont {Sefat}}, \bibinfo {author} {\bibfnamefont {S.}~\bibnamefont {Chi}}, \bibinfo {author} {\bibfnamefont {C.}~\bibnamefont {dela Cruz}}, \bibinfo {author} {\bibfnamefont {D.~J.}\ \bibnamefont {Singh}},\ and\ \bibinfo {author} {\bibfnamefont {B.}~\bibnamefont {Lv}},\ }\bibfield  {title} {\bibinfo {title} {Gapless spin excitations in the superconducting state of a quasi-one-dimensional spin-triplet superconductor},\ }\href {https://doi.org/10.1103/PhysRevB.107.L180504} {\bibfield  {journal}
  {\bibinfo  {journal} {Phys. Rev. B}\ }\textbf {\bibinfo {volume} {107}},\ \bibinfo {pages} {L180504} (\bibinfo {year} {2023})}\BibitemShut {NoStop}%
\bibitem [{\citenamefont {Zhang}\ \emph {et~al.}(2019)\citenamefont {Zhang}, \citenamefont {Jia}, \citenamefont {Wang}, \citenamefont {Li}, \citenamefont {Duan}, \citenamefont {Li}, \citenamefont {Zhao}, \citenamefont {Cao}, \citenamefont {Dai}, \citenamefont {Deng}, \citenamefont {Zhang}, \citenamefont {Feng}, \citenamefont {Yu}, \citenamefont {Liu}, \citenamefont {Hu}, \citenamefont {Zhu},\ and\ \citenamefont {Jin}}]{zhang.jia.19}%
  \BibitemOpen
  \bibfield  {author} {\bibinfo {author} {\bibfnamefont {J.}~\bibnamefont {Zhang}}, \bibinfo {author} {\bibfnamefont {Y.}~\bibnamefont {Jia}}, \bibinfo {author} {\bibfnamefont {X.}~\bibnamefont {Wang}}, \bibinfo {author} {\bibfnamefont {Z.}~\bibnamefont {Li}}, \bibinfo {author} {\bibfnamefont {L.}~\bibnamefont {Duan}}, \bibinfo {author} {\bibfnamefont {W.}~\bibnamefont {Li}}, \bibinfo {author} {\bibfnamefont {J.}~\bibnamefont {Zhao}}, \bibinfo {author} {\bibfnamefont {L.}~\bibnamefont {Cao}}, \bibinfo {author} {\bibfnamefont {G.}~\bibnamefont {Dai}}, \bibinfo {author} {\bibfnamefont {Z.}~\bibnamefont {Deng}}, \bibinfo {author} {\bibfnamefont {S.}~\bibnamefont {Zhang}}, \bibinfo {author} {\bibfnamefont {S.}~\bibnamefont {Feng}}, \bibinfo {author} {\bibfnamefont {R.}~\bibnamefont {Yu}}, \bibinfo {author} {\bibfnamefont {Q.}~\bibnamefont {Liu}}, \bibinfo {author} {\bibfnamefont {J.}~\bibnamefont {Hu}}, \bibinfo {author} {\bibfnamefont {J.}~\bibnamefont {Zhu}},\ and\ \bibinfo {author} {\bibfnamefont
  {C.}~\bibnamefont {Jin}},\ }\bibfield  {title} {\bibinfo {title} {A new quasi-one-dimensional compound {Ba$_{3}$TiTe$_{5}$} and superconductivity induced by pressure},\ }\href {https://doi.org/10.1038/s41427-019-0158-2} {\bibfield  {journal} {\bibinfo  {journal} {NPG Asia Mater.}\ }\textbf {\bibinfo {volume} {11}},\ \bibinfo {pages} {60} (\bibinfo {year} {2019})}\BibitemShut {NoStop}%
\bibitem [{\citenamefont {Okubo}\ \emph {et~al.}(2003)\citenamefont {Okubo}, \citenamefont {Yamada}, \citenamefont {Thamizhavel}, \citenamefont {Kirita}, \citenamefont {Inada}, \citenamefont {Settai}, \citenamefont {Harima}, \citenamefont {Takegahara}, \citenamefont {Galatanu}, \citenamefont {Yamamoto},\ and\ \citenamefont {\={O}nuki}}]{okubo.yamada.03}%
  \BibitemOpen
  \bibfield  {author} {\bibinfo {author} {\bibfnamefont {T.}~\bibnamefont {Okubo}}, \bibinfo {author} {\bibfnamefont {M.}~\bibnamefont {Yamada}}, \bibinfo {author} {\bibfnamefont {A.}~\bibnamefont {Thamizhavel}}, \bibinfo {author} {\bibfnamefont {S.}~\bibnamefont {Kirita}}, \bibinfo {author} {\bibfnamefont {Y.}~\bibnamefont {Inada}}, \bibinfo {author} {\bibfnamefont {R.}~\bibnamefont {Settai}}, \bibinfo {author} {\bibfnamefont {H.}~\bibnamefont {Harima}}, \bibinfo {author} {\bibfnamefont {K.}~\bibnamefont {Takegahara}}, \bibinfo {author} {\bibfnamefont {A.}~\bibnamefont {Galatanu}}, \bibinfo {author} {\bibfnamefont {E.}~\bibnamefont {Yamamoto}},\ and\ \bibinfo {author} {\bibfnamefont {Y.}~\bibnamefont {\={O}nuki}},\ }\bibfield  {title} {\bibinfo {title} {Unique {Fermi} surfaces with quasi-one-dimensional character in {CeRh$_{3}$B$_{2}$} and {LaRh$_{3}$B$_{2}$}},\ }\href {https://doi.org/10.1088/0953-8984/15/46/L02} {\bibfield  {journal} {\bibinfo  {journal} {J. Phys.: Condens. Matter}\ }\textbf {\bibinfo
  {volume} {15}},\ \bibinfo {pages} {L721} (\bibinfo {year} {2003})}\BibitemShut {NoStop}%
\bibitem [{\citenamefont {Song}\ \emph {et~al.}(2020)\citenamefont {Song}, \citenamefont {Li}, \citenamefont {Xu}, \citenamefont {Wu}, \citenamefont {Qian}, \citenamefont {Chen}, \citenamefont {Biswas}, \citenamefont {Xu},\ and\ \citenamefont {Sun}}]{song.li.20}%
  \BibitemOpen
  \bibfield  {author} {\bibinfo {author} {\bibfnamefont {Z.}~\bibnamefont {Song}}, \bibinfo {author} {\bibfnamefont {B.}~\bibnamefont {Li}}, \bibinfo {author} {\bibfnamefont {C.}~\bibnamefont {Xu}}, \bibinfo {author} {\bibfnamefont {S.}~\bibnamefont {Wu}}, \bibinfo {author} {\bibfnamefont {B.}~\bibnamefont {Qian}}, \bibinfo {author} {\bibfnamefont {T.}~\bibnamefont {Chen}}, \bibinfo {author} {\bibfnamefont {P.~K.}\ \bibnamefont {Biswas}}, \bibinfo {author} {\bibfnamefont {X.}~\bibnamefont {Xu}},\ and\ \bibinfo {author} {\bibfnamefont {J.}~\bibnamefont {Sun}},\ }\bibfield  {title} {\bibinfo {title} {Pressure engineering of the {Dirac} fermions in quasi-one-dimensional {Tl$_{2}$Mo$_{6}$Se$_{6}$}},\ }\href {https://doi.org/10.1088/1361-648X/ab73a8} {\bibfield  {journal} {\bibinfo  {journal} {J. Phys.: Condens. Matter}\ }\textbf {\bibinfo {volume} {32}},\ \bibinfo {pages} {215402} (\bibinfo {year} {2020})}\BibitemShut {NoStop}%
\bibitem [{\citenamefont {He}\ \emph {et~al.}(2024)\citenamefont {He}, \citenamefont {Li}, \citenamefont {Ge}, \citenamefont {Zeng}, \citenamefont {Song}, \citenamefont {Zou}, \citenamefont {Wang}, \citenamefont {Li}, \citenamefont {Ding}, \citenamefont {Dai}, \citenamefont {Cao}, \citenamefont {Zhang}, \citenamefont {Xu},\ and\ \citenamefont {Luo}}]{he.li.24}%
  \BibitemOpen
  \bibfield  {author} {\bibinfo {author} {\bibfnamefont {X.}~\bibnamefont {He}}, \bibinfo {author} {\bibfnamefont {Y.}~\bibnamefont {Li}}, \bibinfo {author} {\bibfnamefont {Y.}~\bibnamefont {Ge}}, \bibinfo {author} {\bibfnamefont {H.}~\bibnamefont {Zeng}}, \bibinfo {author} {\bibfnamefont {S.-J.}\ \bibnamefont {Song}}, \bibinfo {author} {\bibfnamefont {S.}~\bibnamefont {Zou}}, \bibinfo {author} {\bibfnamefont {Z.}~\bibnamefont {Wang}}, \bibinfo {author} {\bibfnamefont {Y.}~\bibnamefont {Li}}, \bibinfo {author} {\bibfnamefont {W.}~\bibnamefont {Ding}}, \bibinfo {author} {\bibfnamefont {J.}~\bibnamefont {Dai}}, \bibinfo {author} {\bibfnamefont {G.-H.}\ \bibnamefont {Cao}}, \bibinfo {author} {\bibfnamefont {X.-X.}\ \bibnamefont {Zhang}}, \bibinfo {author} {\bibfnamefont {G.}~\bibnamefont {Xu}},\ and\ \bibinfo {author} {\bibfnamefont {Y.}~\bibnamefont {Luo}},\ }\href@noop {} {\bibinfo {title} {Unconventional {Hall} effects in a quasi-kagome {Kondo} {Weyl} semimetal candidate {Ce$_3$TiSb$_5$}}} (\bibinfo {year}
  {2024}),\ \Eprint {https://arxiv.org/abs/arXiv:2408.04438} {arXiv:2408.04438} \BibitemShut {NoStop}%
\bibitem [{\citenamefont {Basak}\ and\ \citenamefont {Ptok}(2023)}]{basak.ptok.23}%
  \BibitemOpen
  \bibfield  {author} {\bibinfo {author} {\bibfnamefont {S.}~\bibnamefont {Basak}}\ and\ \bibinfo {author} {\bibfnamefont {A.}~\bibnamefont {Ptok}},\ }\bibfield  {title} {\bibinfo {title} {Theoretical study of dynamical and electronic properties of noncentrosymmetric superconductor {NbReSi}},\ }\href {https://doi.org/10.3390/ma16010078} {\bibfield  {journal} {\bibinfo  {journal} {Materials}\ }\textbf {\bibinfo {volume} {16}},\ \bibinfo {pages} {78} (\bibinfo {year} {2023})}\BibitemShut {NoStop}%
\bibitem [{\citenamefont {Mofazzel~Hosen}\ \emph {et~al.}(2017)\citenamefont {Mofazzel~Hosen}, \citenamefont {Dimitri}, \citenamefont {Belopolski}, \citenamefont {Maldonado}, \citenamefont {Sankar}, \citenamefont {Dhakal}, \citenamefont {Dhakal}, \citenamefont {Cole}, \citenamefont {Oppeneer}, \citenamefont {Kaczorowski}, \citenamefont {Chou}, \citenamefont {Hasan}, \citenamefont {Durakiewicz},\ and\ \citenamefont {Neupane}}]{mofazzelhosen.dimitri.17}%
  \BibitemOpen
  \bibfield  {author} {\bibinfo {author} {\bibfnamefont {M.}~\bibnamefont {Mofazzel~Hosen}}, \bibinfo {author} {\bibfnamefont {K.}~\bibnamefont {Dimitri}}, \bibinfo {author} {\bibfnamefont {I.}~\bibnamefont {Belopolski}}, \bibinfo {author} {\bibfnamefont {P.}~\bibnamefont {Maldonado}}, \bibinfo {author} {\bibfnamefont {R.}~\bibnamefont {Sankar}}, \bibinfo {author} {\bibfnamefont {N.}~\bibnamefont {Dhakal}}, \bibinfo {author} {\bibfnamefont {G.}~\bibnamefont {Dhakal}}, \bibinfo {author} {\bibfnamefont {T.}~\bibnamefont {Cole}}, \bibinfo {author} {\bibfnamefont {P.~M.}\ \bibnamefont {Oppeneer}}, \bibinfo {author} {\bibfnamefont {D.}~\bibnamefont {Kaczorowski}}, \bibinfo {author} {\bibfnamefont {F.}~\bibnamefont {Chou}}, \bibinfo {author} {\bibfnamefont {M.~Z.}\ \bibnamefont {Hasan}}, \bibinfo {author} {\bibfnamefont {T.}~\bibnamefont {Durakiewicz}},\ and\ \bibinfo {author} {\bibfnamefont {M.}~\bibnamefont {Neupane}},\ }\bibfield  {title} {\bibinfo {title} {Tunability of the topological nodal-line semimetal
  phase in {ZrSi$X$}-type materials ({$X=$S, Se, Te})},\ }\href {https://doi.org/10.1103/PhysRevB.95.161101} {\bibfield  {journal} {\bibinfo  {journal} {Phys. Rev. B}\ }\textbf {\bibinfo {volume} {95}},\ \bibinfo {pages} {161101(R)} (\bibinfo {year} {2017})}\BibitemShut {NoStop}%
\bibitem [{\citenamefont {Yu}\ \emph {et~al.}(2015)\citenamefont {Yu}, \citenamefont {Weng}, \citenamefont {Fang}, \citenamefont {Dai},\ and\ \citenamefont {Hu}}]{yu.weng.15}%
  \BibitemOpen
  \bibfield  {author} {\bibinfo {author} {\bibfnamefont {R.}~\bibnamefont {Yu}}, \bibinfo {author} {\bibfnamefont {H.}~\bibnamefont {Weng}}, \bibinfo {author} {\bibfnamefont {Z.}~\bibnamefont {Fang}}, \bibinfo {author} {\bibfnamefont {X.}~\bibnamefont {Dai}},\ and\ \bibinfo {author} {\bibfnamefont {X.}~\bibnamefont {Hu}},\ }\bibfield  {title} {\bibinfo {title} {Topological node-line semimetal and {Dirac} semimetal state in antiperovskite {Cu$_{3}$PdN}},\ }\href {https://doi.org/10.1103/PhysRevLett.115.036807} {\bibfield  {journal} {\bibinfo  {journal} {Phys. Rev. Lett.}\ }\textbf {\bibinfo {volume} {115}},\ \bibinfo {pages} {036807} (\bibinfo {year} {2015})}\BibitemShut {NoStop}%
\bibitem [{\citenamefont {Liu}\ \emph {et~al.}(2018{\natexlab{b}})\citenamefont {Liu}, \citenamefont {Lou}, \citenamefont {Guo}, \citenamefont {Wang}, \citenamefont {Sun}, \citenamefont {Li}, \citenamefont {Thirupathaiah}, \citenamefont {Fedorov}, \citenamefont {Shen}, \citenamefont {Liu}, \citenamefont {Lei},\ and\ \citenamefont {Wang}}]{liu.lou.18}%
  \BibitemOpen
  \bibfield  {author} {\bibinfo {author} {\bibfnamefont {Z.}~\bibnamefont {Liu}}, \bibinfo {author} {\bibfnamefont {R.}~\bibnamefont {Lou}}, \bibinfo {author} {\bibfnamefont {P.}~\bibnamefont {Guo}}, \bibinfo {author} {\bibfnamefont {Q.}~\bibnamefont {Wang}}, \bibinfo {author} {\bibfnamefont {S.}~\bibnamefont {Sun}}, \bibinfo {author} {\bibfnamefont {C.}~\bibnamefont {Li}}, \bibinfo {author} {\bibfnamefont {S.}~\bibnamefont {Thirupathaiah}}, \bibinfo {author} {\bibfnamefont {A.}~\bibnamefont {Fedorov}}, \bibinfo {author} {\bibfnamefont {D.}~\bibnamefont {Shen}}, \bibinfo {author} {\bibfnamefont {K.}~\bibnamefont {Liu}}, \bibinfo {author} {\bibfnamefont {H.}~\bibnamefont {Lei}},\ and\ \bibinfo {author} {\bibfnamefont {S.}~\bibnamefont {Wang}},\ }\bibfield  {title} {\bibinfo {title} {Experimental observation of {Dirac} nodal links in centrosymmetric semimetal {TiB$_{2}$}},\ }\href {https://doi.org/10.1103/PhysRevX.8.031044} {\bibfield  {journal} {\bibinfo  {journal} {Phys. Rev. X}\ }\textbf {\bibinfo {volume}
  {8}},\ \bibinfo {pages} {031044} (\bibinfo {year} {2018}{\natexlab{b}})}\BibitemShut {NoStop}%
\bibitem [{\citenamefont {V.}\ \emph {et~al.}(2024)\citenamefont {V.}, \citenamefont {Pradhan},\ and\ \citenamefont {Kanchana}}]{anusree.sonali.24}%
  \BibitemOpen
  \bibfield  {author} {\bibinfo {author} {\bibfnamefont {A.~C.}\ \bibnamefont {V.}}, \bibinfo {author} {\bibfnamefont {S.~S.}\ \bibnamefont {Pradhan}},\ and\ \bibinfo {author} {\bibfnamefont {V.}~\bibnamefont {Kanchana}},\ }\bibfield  {title} {\bibinfo {title} {Coexistence of electron and phonon topology in conjunction with quantum transport device modeling},\ }\href {https://doi.org/10.1088/1361-648X/ad1a5b} {\bibfield  {journal} {\bibinfo  {journal} {J. Phys.: Condens. Matter}\ }\textbf {\bibinfo {volume} {36}},\ \bibinfo {pages} {155501} (\bibinfo {year} {2024})}\BibitemShut {NoStop}%
\bibitem [{\citenamefont {Zhou}\ \emph {et~al.}(2024)\citenamefont {Zhou}, \citenamefont {Yang}, \citenamefont {Zhang},\ and\ \citenamefont {Zhang}}]{zhou.yang.24}%
  \BibitemOpen
  \bibfield  {author} {\bibinfo {author} {\bibfnamefont {L.}~\bibnamefont {Zhou}}, \bibinfo {author} {\bibfnamefont {F.}~\bibnamefont {Yang}}, \bibinfo {author} {\bibfnamefont {S.}~\bibnamefont {Zhang}},\ and\ \bibinfo {author} {\bibfnamefont {T.}~\bibnamefont {Zhang}},\ }\bibfield  {title} {\bibinfo {title} {Chemical rules for stacked kagome and honeycomb topological semimetals},\ }\href {https://doi.org/10.1002/adma.202309803} {\bibfield  {journal} {\bibinfo  {journal} {Adv. Mater.}\ }\textbf {\bibinfo {volume} {36}},\ \bibinfo {pages} {2309803} (\bibinfo {year} {2024})}\BibitemShut {NoStop}%
\bibitem [{\citenamefont {Rosmus}\ \emph {et~al.}(2022)\citenamefont {Rosmus}, \citenamefont {Olszowska}, \citenamefont {Bukowski}, \citenamefont {Starowicz}, \citenamefont {Piekarz},\ and\ \citenamefont {Ptok}}]{rosmus.olszowska.22}%
  \BibitemOpen
  \bibfield  {author} {\bibinfo {author} {\bibfnamefont {M.}~\bibnamefont {Rosmus}}, \bibinfo {author} {\bibfnamefont {N.}~\bibnamefont {Olszowska}}, \bibinfo {author} {\bibfnamefont {Z.}~\bibnamefont {Bukowski}}, \bibinfo {author} {\bibfnamefont {P.}~\bibnamefont {Starowicz}}, \bibinfo {author} {\bibfnamefont {P.}~\bibnamefont {Piekarz}},\ and\ \bibinfo {author} {\bibfnamefont {A.}~\bibnamefont {Ptok}},\ }\bibfield  {title} {\bibinfo {title} {Electronic band structure and surface states in dirac semimetal {LaAgSb$_{2}$}},\ }\href {https://doi.org/10.3390/ma15207168} {\bibfield  {journal} {\bibinfo  {journal} {Materials}\ }\textbf {\bibinfo {volume} {15}},\ \bibinfo {pages} {7168} (\bibinfo {year} {2022})}\BibitemShut {NoStop}%
\bibitem [{\citenamefont {Lu}\ \emph {et~al.}(2022)\citenamefont {Lu}, \citenamefont {Ran}, \citenamefont {Li}, \citenamefont {Xu}, \citenamefont {Hu}, \citenamefont {Yin}, \citenamefont {Wang}, \citenamefont {Zhang}, \citenamefont {Luo}, \citenamefont {Xu},\ and\ \citenamefont {Qian}}]{lu.ran.22}%
  \BibitemOpen
  \bibfield  {author} {\bibinfo {author} {\bibfnamefont {Q.}~\bibnamefont {Lu}}, \bibinfo {author} {\bibfnamefont {Z.}~\bibnamefont {Ran}}, \bibinfo {author} {\bibfnamefont {Y.}~\bibnamefont {Li}}, \bibinfo {author} {\bibfnamefont {C.}~\bibnamefont {Xu}}, \bibinfo {author} {\bibfnamefont {J.}~\bibnamefont {Hu}}, \bibinfo {author} {\bibfnamefont {X.}~\bibnamefont {Yin}}, \bibinfo {author} {\bibfnamefont {G.}~\bibnamefont {Wang}}, \bibinfo {author} {\bibfnamefont {W.}~\bibnamefont {Zhang}}, \bibinfo {author} {\bibfnamefont {W.}~\bibnamefont {Luo}}, \bibinfo {author} {\bibfnamefont {X.}~\bibnamefont {Xu}},\ and\ \bibinfo {author} {\bibfnamefont {D.}~\bibnamefont {Qian}},\ }\bibfield  {title} {\bibinfo {title} {Topologically protected surface states in {TaPdTe$_{5}$}},\ }\href {https://doi.org/10.1007/s44214-022-00009-7} {\bibfield  {journal} {\bibinfo  {journal} {Quantum Front.}\ }\textbf {\bibinfo {volume} {1}},\ \bibinfo {pages} {9} (\bibinfo {year} {2022})}\BibitemShut {NoStop}%
\bibitem [{\citenamefont {Souza}\ \emph {et~al.}(2023)\citenamefont {Souza}, \citenamefont {Ale~Crivillero}, \citenamefont {Dawczak-D\c{e}bicki}, \citenamefont {Ptok}, \citenamefont {Pagliuso},\ and\ \citenamefont {Wirth}}]{souza.criivillero.23}%
  \BibitemOpen
  \bibfield  {author} {\bibinfo {author} {\bibfnamefont {J.~C.}\ \bibnamefont {Souza}}, \bibinfo {author} {\bibfnamefont {M.~V.}\ \bibnamefont {Ale~Crivillero}}, \bibinfo {author} {\bibfnamefont {H.}~\bibnamefont {Dawczak-D\c{e}bicki}}, \bibinfo {author} {\bibfnamefont {A.}~\bibnamefont {Ptok}}, \bibinfo {author} {\bibfnamefont {P.~G.}\ \bibnamefont {Pagliuso}},\ and\ \bibinfo {author} {\bibfnamefont {S.}~\bibnamefont {Wirth}},\ }\bibfield  {title} {\bibinfo {title} {Tuning the topological character of half-heusler systems: A comparative study on {Y$T$Bi} ({$T=$Pd, Pt})},\ }\href {https://doi.org/10.1103/PhysRevB.108.165154} {\bibfield  {journal} {\bibinfo  {journal} {Phys. Rev. B}\ }\textbf {\bibinfo {volume} {108}},\ \bibinfo {pages} {165154} (\bibinfo {year} {2023})}\BibitemShut {NoStop}%
\bibitem [{\citenamefont {Momma}\ and\ \citenamefont {Izumi}(2011)}]{momma.izumi.11}%
  \BibitemOpen
  \bibfield  {author} {\bibinfo {author} {\bibfnamefont {K.}~\bibnamefont {Momma}}\ and\ \bibinfo {author} {\bibfnamefont {F.}~\bibnamefont {Izumi}},\ }\bibfield  {title} {\bibinfo {title} {{{\sc vesta3} for three-dimensional visualization of crystal, volumetric and morphology data}},\ }\href {https://doi.org/10.1107/S0021889811038970} {\bibfield  {journal} {\bibinfo  {journal} {J. Appl. Crystallogr.}\ }\textbf {\bibinfo {volume} {44}},\ \bibinfo {pages} {1272} (\bibinfo {year} {2011})}\BibitemShut {NoStop}%
\bibitem [{\citenamefont {Kokalj}(1999)}]{kokalj.99}%
  \BibitemOpen
  \bibfield  {author} {\bibinfo {author} {\bibfnamefont {A.}~\bibnamefont {Kokalj}},\ }\bibfield  {title} {\bibinfo {title} {Xcrysden--a new program for displaying crystalline structures and electron densities},\ }\href {https://doi.org/10.1016/S1093-3263(99)00028-5} {\bibfield  {journal} {\bibinfo  {journal} {J. Mol. Graph. Model.}\ }\textbf {\bibinfo {volume} {17}},\ \bibinfo {pages} {176} (\bibinfo {year} {1999})}\BibitemShut {NoStop}%
\end{thebibliography}%

%%%%%%%%%%%%%%%%%%%%%%%%%%%%%%%%%%%
%%%%%%%%%%%%%%%%%%%%%%%%%%%%%%%%%%%
%%%%%%%%%%%%%%%%%%%%%%%%%%%%%%%%%%%

\clearpage
\newpage

\onecolumngrid

\begin{center}
  \textbf{\Large Supplemental Material}\\[.3cm]
  \textbf{\large Correlation stabilized ferromagnetic MnRuAs with distorted kagome lattice}\\[.3cm]
  %%%%%%
  Anusree C V$^{1}$, Andrzej Ptok$^{2}$, Pawe\l{} Sobieszczyk$^{2}$, G. Vaitheeswaran$^{3}$, V. Kanchana$^{1}$ \\[.2cm]
  %%%%%%
  {\itshape
    $^{1}$Department of Physics, Indian Institute of Technology Hyderabad, Kandi, Sangareddy 502285, Telangana, India \\[.2cm]
    $^{2}$Institute of Nuclear Physics, Polish Academy of Sciences, W. E. Radzikowskiego 152, PL-31342 Krak\'{o}w, Poland \\[.2cm]
    \mbox{$^{3}$School of Physics, University of Hyderabad, Prof. C. R. Rao Road, Gachibowli, Hyderabad 500-046, Telangana, India} \\[.2cm]
  }
  (Dated: \today)
\\[0.3cm]
\end{center}

\setcounter{equation}{0}
\renewcommand{\theequation}{S\arabic{equation}}
\setcounter{figure}{0}
\renewcommand{\thefigure}{S\arabic{figure}}
\setcounter{section}{0}
\renewcommand{\thesection}{S\arabic{section}}
\setcounter{table}{0}
\renewcommand{\thetable}{S\arabic{table}}
\setcounter{page}{1}

%%%%%%%%%%%%%%%%%%%%%%%%%%%%%%%%%%%
%%%%%%%%%%%%%%%%%%%%%%%%%%%%%%%%%%%
%%%%%%%%%%%%%%%%%%%%%%%%%%%%%%%%%%%

\begin{figure}[!t]
\begin{center}
\includegraphics[width=\linewidth]{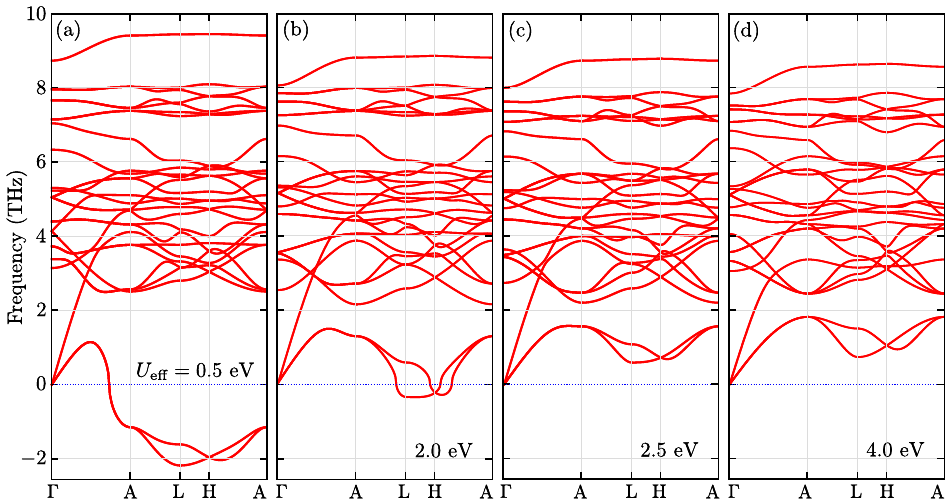}
\caption{
The evolution of the imaginary soft mode in the phonon dispersion curves with $U_\text{eff}$. For relatively small $U_\text{eff}$, the imaginary soft mode (represented as negative frequencies) is realized along the A--L--H--A plane.
\label{fig.smphband}}
\end{center}
\end{figure}

\begin{figure}[!b]
\begin{center}
\includegraphics[width=\linewidth]{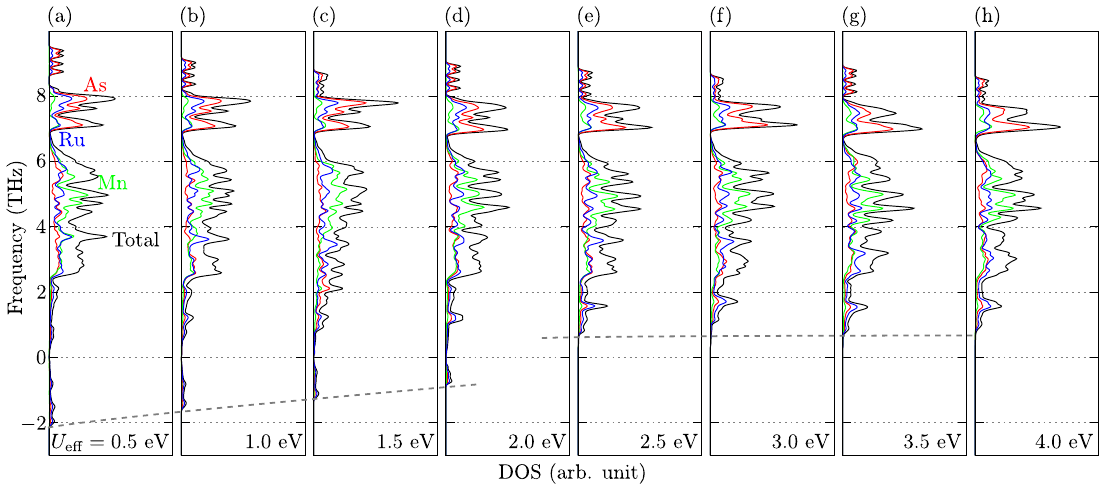}
\caption{
The phonon density of states (DOS) for several values of $U_\text{eff}$ (as labeled). The total DOS is represented by the black line, while the partial DOS for Mn, Ru, and As are shown as green, blue, and red lines, respectively. The imaginary soft mode disperses between $U_\text{eff} = 2$~eV and $2.5$~eV.
\label{fig.smphdos}}
\end{center}
\end{figure}

\begin{figure}[!t]
\begin{center}
\includegraphics[width=0.55\linewidth]{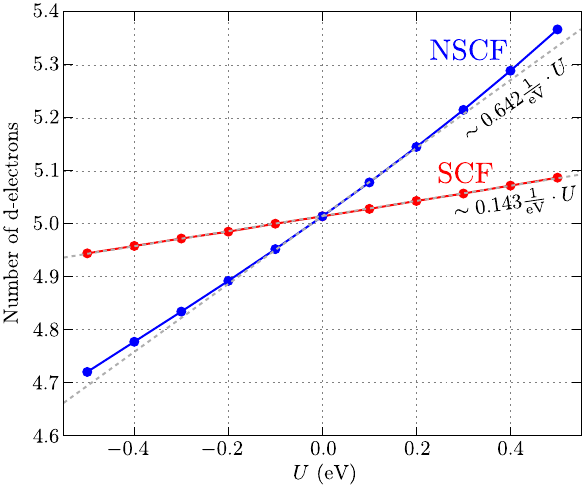}
\caption{
The estimation of the $U_{\text{eff}}$ within linear response technique.
Here, we determine the $U$ parameter for the DFT+U treatment of Mn $d$-electrons in MnRuAs, following the linear response approach of Cococcioni \textit{et al.}~\cite{cococcioni.gironcoli.05}. The calculation employs a $2 \times 2 \times 2$ supercell to evaluate the response function, $\chi = \lim_{U \rightarrow 0} \partial N / \partial U$, in both self-consistent (SCF) and non-self-consistent (NSCF) frameworks, denoted as $\chi_\text{SCF}$ and $\chi_\text{NSCF}$, respectively. This function characterizes the response of the $d$-electrons on a Mn site to an additional spherical potential $U$. As shown, in the $U \rightarrow 0$ limit, the data points align well with a linear fit (indicated by dashed grey lines). Consequently, the $U$ parameter is calculated as $U = \chi_\text{SCF}^{-1} - \chi_\text{NSCF}^{-1} = 1/0.143 \; \text{eV} - 1/0.642 \; \text{eV} \approx 5.44$~eV.
\label{fig.lin_resp}}
\end{center}
\end{figure}

\begin{figure}[!b]
\begin{center}
\includegraphics[width=\linewidth]{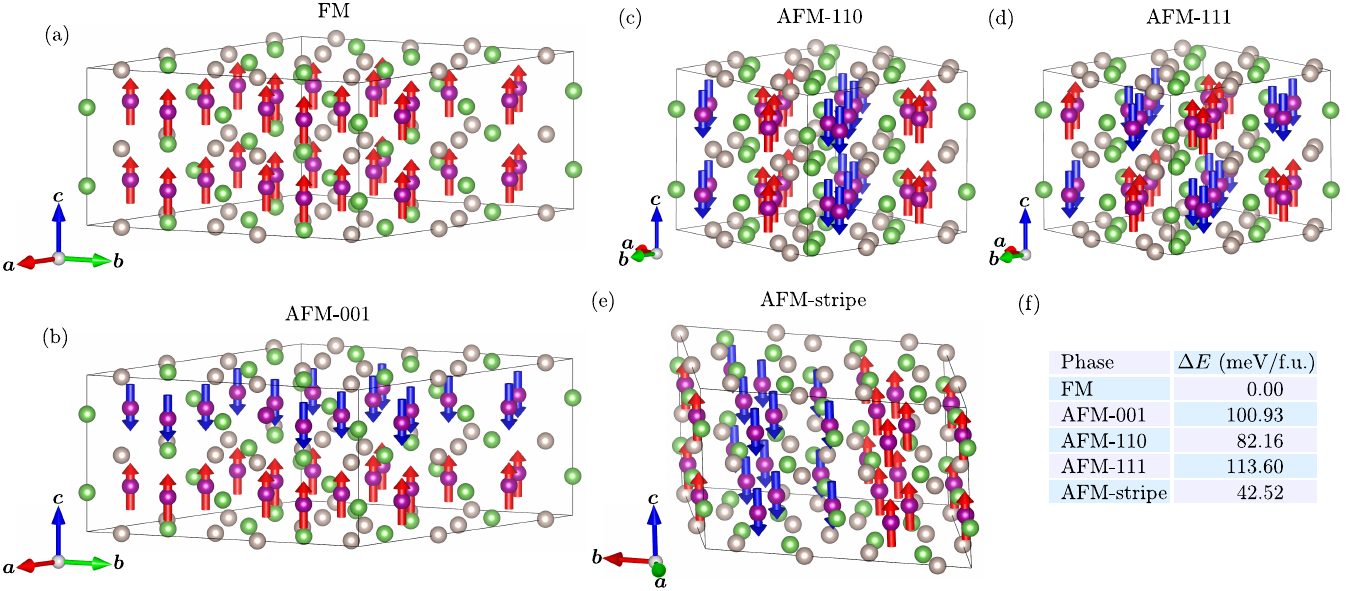}
\caption{
{ The comparison of energies for the different magnetic configurations studied}:
(a) ordinary ferromagnetic state,
(b) antiferromagnetic state along (001) direction,
(c) antiferromagnetic state along (110) direction,
(d) antiferromagnetic state along (111) direction,
(e) antiferromagnetic stripes along (010) direction.
The energy of system for all cases was calculated in $2 \times 2 \times 2$ supercell.
The energy differences presented in Table (f) are referenced to the magnetic ground state (in the absence of spin--orbit coupling), which corresponds to the ferromagnetic phase, and is set to ``zero''.
\label{fig.ene_mag}}
\end{center}
\end{figure}

\begin{figure}[!t]
\begin{center}
\includegraphics[width=0.7\linewidth]{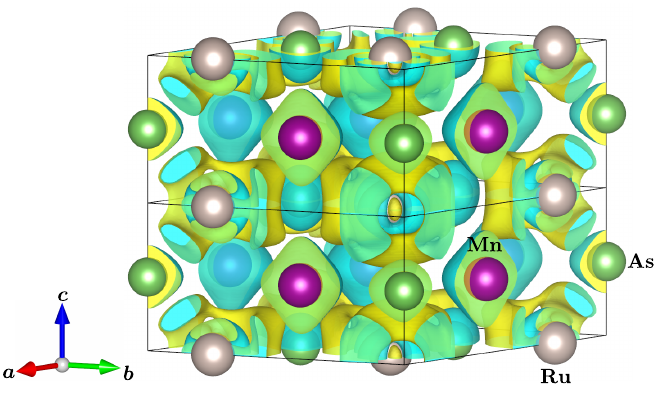}
\caption{
The three-dimensional plot of the charge density difference.
Results is the unit cell doubled along $c$.
Yellow and cyan regions indicate the electron accumulation and depletion, respectively.
The system exhibits metallic bonds -- the charge is accumulated in whole volume of the system.
This is also supported by relatively small value of the electron localization function (not presented).
\label{fig.bond}}
\end{center}
\end{figure}

\begin{figure}[!b]
\centering
\includegraphics[width=0.6\linewidth]{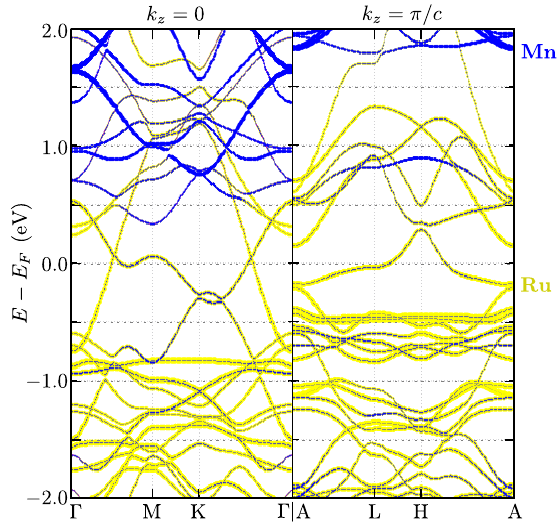}
\caption{
Contribution of Mn (blue) and Ru (yellow) states to the bulk electronic band structure along high-symmetry paths for the $k_z = 0$ and $\pi / c$ planes.
\label{fig.orb}
}
\end{figure}

\begin{figure}[!t]
\centering
\includegraphics[width=0.7\linewidth]{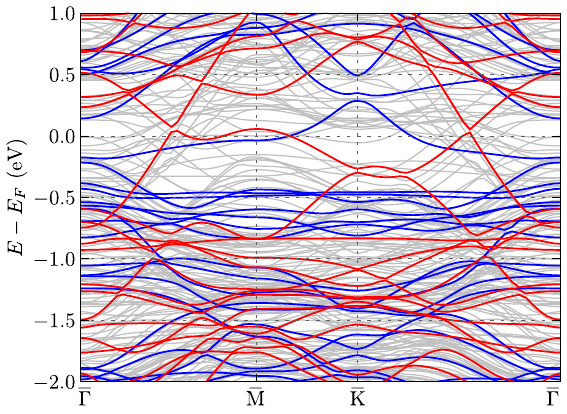}
\caption{
The bulk states of MnRuAs projected onto the (001) surface, obtained from exact DFT calculations. The red and blue lines highlight the projections for $k_{z} = 0$ and $\pm \pi/c$, respectively.
\label{fig.bandkz_bulk}
}
\end{figure}

\begin{figure}[!b]
\centering
\includegraphics[width=0.7\linewidth]{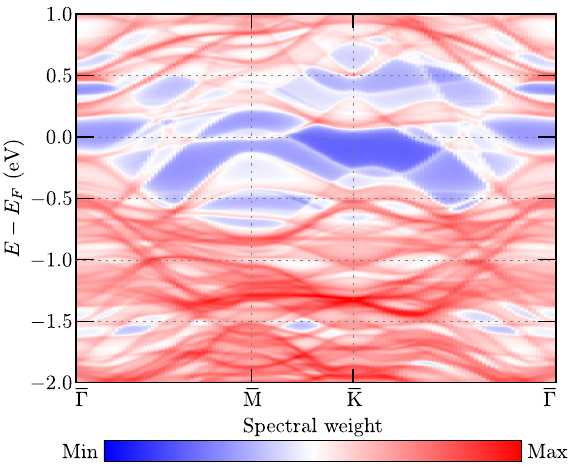}
\caption{
The bulk states contribution to the surface Green function for the MnRuAs (001) surface.
\label{fig.spec_bulk}
}
\end{figure}

\begin{figure}[!t]
\centering
\includegraphics[width=\linewidth]{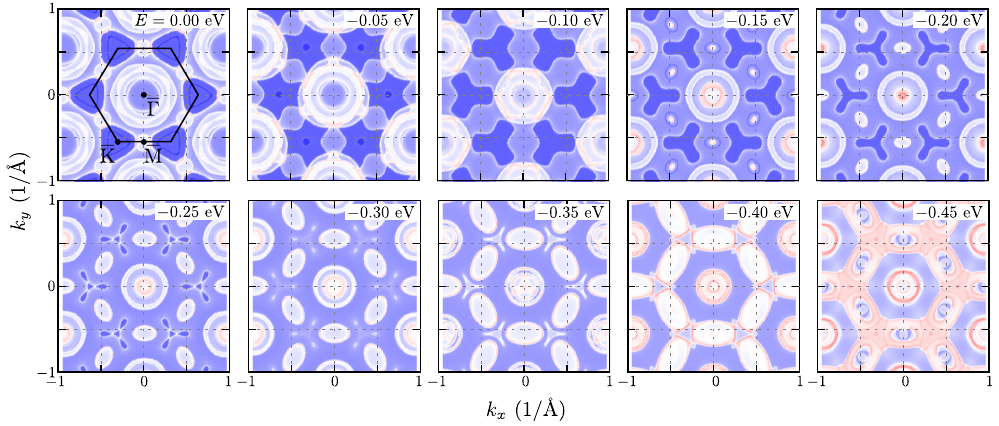}
\caption{
The bulk states contribution to the constant energy contour of the surface Green function for the MnRuAs (001) surface.
\label{fig.arc_bulk}
}
\end{figure}

In this Supplemental Material, we present additional results:
\begin{itemize}
%%%%%%%%%%%%%%%%%%%%%%%%%%
\item Tab.~\ref{tab.hubbers} -- The role of $U_{\text{eff}}$ on the system parameters.
%%%%%%%%%%%%%%%%%%%%%%%%%%
\item Fig.~\ref{fig.smphband} -- The evolution of the imaginary soft modes in the phonon dispersion curve for various values of $U_{\text{eff}}$.
%%%%%%%%%%%%%%%%%%%%%%%%%%
\item Fig.~\ref{fig.smphdos} -- The phonon density of states for different values of $U_{\text{eff}}$.
%%%%%%%%%%%%%%%%%%%%%%%%%%
\item Fig.~\ref{fig.lin_resp} -- The estimation of the $U_{\text{eff}}$ within the linear response technique.
%%%%%%%%%%%%%%%%%%%%%%%%%%
\item Fig.~\ref{fig.ene_mag} -- The comparison of the energy states for various magnetic orders.
%%%%%%%%%%%%%%%%%%%%%%%%%%
\item Fig.~\ref{fig.bond} -- The charge density difference plot.
%%%%%%%%%%%%%%%%%%%%%%%%%%
\item Fig.~\ref{fig.orb} -- The contribution of Mn and Ru states in the bulk electronic band structure.
%%%%%%%%%%%%%%%%%%%%%%%%%%
\item Fig.~\ref{fig.bandkz_bulk} -- The bulk states projected on the (001) surface obtained from exact DFT calculations.
%%%%%%%%%%%%%%%%%%%%%%%%%%
\item Fig.~\ref{fig.spec_bulk} -- The bulk state contribution to the surface Green functions.
%%%%%%%%%%%%%%%%%%%%%%%%%%
\item Fig.~\ref{fig.arc_bulk} -- The bulk state contribution to the constant energy contour of the surface Green functions.
%%%%%%%%%%%%%%%%%%%%%%%%%%
\end{itemize}

\begin{table}[!b]
\caption{
\label{tab.hubbers}
The table summarizing lattice parameters and Mn magnetic moments obtained from system optimization within the DFT+U framework, highlighting the role of the $U_{\text{eff}}$ parameter.}
\begin{ruledtabular}
\begin{tabular}{cccc}
U (eV) & a (\AA) & c (\AA) & Mn mag. mom. ($\mu_\text{B}$) \\
\hline
0.0 & 6.528 & 3.601 & 3.610 \\
0.5 & 6.539 & 3.611 & 3.727 \\
1.0 & 6.551 & 3.622 & 3.834 \\
1.5 & 6.562 & 3.632 & 3.929 \\
2.0 & 6.574 & 3.640 & 4.014 \\
2.5 & 6.587 & 3.647 & 4.090 \\
3.0 & 6.599 & 3.655 & 4.162 \\
3.5 & 6.613 & 3.662 & 4.227 \\
4.0 & 6.623 & 3.671 & 4.285 \\
4.5 & 6.634 & 3.678 & 4.337 \\
5.0 & 6.643 & 3.686 & 4.383 \\
\end{tabular}
\end{ruledtabular}
\end{table}

%\nocite{*}
%\bibliography{biblio.bib}

\end{document}